\documentclass[12pt,a4paper]{article}
\usepackage[latin1]{inputenc}
\usepackage{amsmath}
\usepackage{amsmath} 
\usepackage{amsfonts}
\usepackage{amssymb}
\usepackage{float}
\usepackage{cite}
\usepackage{breqn}
\usepackage{graphics}
\DeclareGraphicsExtensions{.pdf,.png,.jpg}
\usepackage[colorlinks=true,linkcolor=blue,citecolor=red]{hyperref}

\makeatletter
\newcommand{\mathleft}{\@fleqntrue\@mathmargin0pt}
\newcommand{\mathcenter}{\@fleqnfalse}
\makeatother
\newcommand*{\SavedEqref}{}
\let\SavedEqref\eqref
\renewcommand*{\eqref}[1]{%
  \begingroup
    \hypersetup{
      linkcolor=linkequation,
      linkbordercolor=linkequation,
    }%
    \SavedEqref{#1}%
  \endgroup
}

\usepackage[a4paper, width = 160mm, top = 30mm, bottom = 25mm, 
bindingoffset = 0mm]{geometry}

\def\beq{\begin{equation}}
\def\eeq{\end{equation}}
\def\bea{\begin{eqnarray}}
\def\eea{\end{eqnarray}}

\begin{document}

\begin{center}
	{\Large \bf Polyadic Cantor potential of minimum lacunarity: \\ Special case of super periodic generalized unified Cantor potential}
\vspace{1.3cm}		
		
{\sf  Mohammad Umar\footnote[1]{e-mail address:\ \ aliphysics110@gmail.com, \ \ opz238433@opc.iitd.ac.in},  Mohammad Hasan\footnote[2]{e-mail address:\ \ mhasan@isro.gov.in, \ \ mohammadhasan786@gmail.com}, Vibhav Narayan Singh\footnote[3]{e-mail address:\ \ vibhavn.singh13@bhu.ac.in, \ \ vibhav.ecc123@gmail.com}, Bhabani Prasad Mandal\footnote[4]{e-mail address:\ \ bhabani.mandal@gmail.com, \ \ bhabani@bhu.ac.in  }}
		
\bigskip

{\em $^{1}$Optics and Photonics Centre, Indian Institute of Technology, Delhi-110016, INDIA. \\ 
$^{2}$Indian Space Research Organisation,
Bangalore-560094, INDIA. \\
$^{3,4}$Department of Physics,
Banaras Hindu University,
Varanasi-221005, INDIA. \\}

\bigskip	

%%%%%%%%%%%%%%%%%%%%%%%%%%%%%%%%%%%%%%% Abstract %%%%%%%%%%%%%%%%%%%%%%%%%%%%%%%%%%%%
\noindent {\bf Abstract}		
\end{center}
\noindent
To bridge the fractal and non-fractal potentials we introduce the concept of generalized unified Cantor potential (GUCP) with the key parameter $N$ which represents the potential count at the stage $S=1$. This system is characterized by total span $L$, stages $S$, scaling parameter $\rho$ and two real numbers $\mu$ and $\nu$. Notably, the polyadic Cantor potential (PCP) system with minimal lacunarity is a specific instance within the GUCP paradigm. Employing the super periodic potential (SPP) formalism, we formulated a closed-form expression for transmission probability $T_{S}(k, N)$ using the $q$-Pochhammer symbol and investigated the features of non-relativistic quantum tunneling through this potential configuration. We show that GUCP system exhibits sharp transmission resonances, differing from traditional quantum systems. Our analysis reveals saturation in the transmission profile with evolving stages $S$ and establishes a significant scaling relationship between reflection probability and wave vector $k$
through analytical derivations.
\medskip
\vspace{1in}
\newpage
	
%%%%%%%%%%%%%%%%%%%%%%%%%%%%%%%%% Introduction %%%%%%%%%%%%%%%%%%%%%%%%%%%%%%%%%%%%

\section{Introduction}
Quantum tunneling, a seminal concept introduced in the realm of quantum mechanics in 1928 \cite{nordheim, gurney}, has continuously held the interest of the scientific community. 
This phenomenon, essential in understanding quantum behavior, has witnessed considerable advancements due to the collective efforts of researchers, who have significantly deepened our understanding of its underlying principles \cite{condon1931quantum, wigner1955lower, bohm2012quantum, albeverio2012solvable, razavy2013quantum}. These efforts have been focused on exploring how matter waves interact with various potential barriers, a study driven by the theoretical implications, experimental applications, and practical significance of tunneling. The study of quantum tunneling has evolved over time, with a diverse array of potential barriers examined and a multitude of methodologies developed to tackle the broad spectrum of challenges associated with tunneling phenomena \cite{albeverio2012solvable, razavy2013quantum}. Spanning over a century, contributions from numerous scholars have added to the rich tapestry of knowledge surrounding various aspects of quantum tunneling \cite{nordheim, gurney, condon1931quantum, wigner1955lower, bohm2012quantum, albeverio2012solvable, razavy2013quantum, esaki1974long, burstein1969tunneling,giaever1974electron, josephson1974discovery, lauhon2000direct}.  Additionally, the exploration of tunneling characteristics in specific contexts, such as non-Hermitian quantum mechanics (NHQM) \cite{angelopoulou1995non,hasan2020hartman, hasan2020role, longhi2022non, garcia2021unitarity, garcia2021hermitian}, space-fractional quantum mechanics (SFQM) \cite{guo2006some, de2011tunneling, tare2014transmission, hasan2018tunneling, hasan2020tunneling,singh2023quantum} and quaternionic quantum mechanics (QQM) \cite{sobhani2016scattering, hasan2020new, hassanabadi2017relativistic, sobhani2017new, de2010closed, davies1989nonrelativistic, de2006analytic}  has been noteworthy. In these domains, the analytical computation of scattering coefficients for given potential distributions has provided a more profound understanding of transmission features, offering insights that are often more nuanced and informative than those obtained through numerical methods alone.\\
\indent
Over the last three decades, both theoretical and practical investigations into the wave properties and transmission in fractal structures, or fractal potentials, have gained significant attention \cite{konotop1990wave, sun1991wave, bertolotti1994spectral, bertolotti1996transmission, lavrinenko2002propagation, chiadini2003self, hatano2005strong, sangawa2005resonance, esaki2009wave, guerin1996scattering, honda2006rigorous, chuprikov2006new, chuprikov2000transfer, sakaguchi2017scaling, ogawana2018transmission,monsoriu2005transfer,takeda2004localization, chuprikov2000electron,jaggard1990reflection,konotop1991transmission}. These structures exemplify self-similarity \cite{sun1991wave, bertolotti1994spectral, bertolotti1996transmission, lavrinenko2002propagation, chiadini2003self}, a characteristic feature that has led researchers to delve into their unique physical attributes. The exploration of these properties has been a key focus, shedding light on the distinctive aspects inherent to fractal-based systems. Due to experimental observations of theoretical features of tunneling from fractal systems \cite{takeda2004localization, chuprikov2000electron, miyamoto2004smart}, quantum tunneling through fractal potentials has gained much attention.
In our recent research, we investigated quantum tunneling phenomena through the general Smith-Volterra-Cantor (GSVC or SVC-$\rho$) potential system, a non-fractal structure, as detailed in \cite{narayan2023tunneling}. Also, we introduce an expanded formulation of the GSVC potential, named as Smith-Volterra-Cantor of order $n$ and designated as SVC($\rho, n$), where the variable $n$ is a real number \cite{singh2023quantum01}. Subsequently, we developed and introduced a novel potential system, assigned as the UCP system, as reported in \cite{umar2023quantum}. This new system integrates aspects of both fractal and non-fractal systems, encompassing the general Cantor (GC) potential (representative of fractal systems) and the SVC potential (representative of non-fractal systems). \\
\indent
Geometrical entities that maintain self-similarity and uniformity across their smaller components are known as fractals \cite{mandelbrot1982fractal, mandelbrot1984fractal,mandelbrot1987fractals,mandelbrot1998nature,takayasu1990fractals, pietronero2012fractals, barnsley1988fractals, falconer2004fractal, lofstedt2008fractal, barnsley2014fractals}. This defining trait implies that fractals are composed of smaller replicas of their overall form, exhibiting a recursive and repetitive pattern that is central to their structure and identity. The concept of fractals, originally introduced by mathematician Benoit B. Mandelbrot \cite{mandelbrot1982fractal}, emerges from the recursive application of a mathematical operation on a geometric figure, referred to as the initiator. This repeated application, conducted at multiple levels using a process known as the generator, leads to the formation of self-similar structures, known as fractals. At each level of recursion, the resulting sub-components bear a complete resemblance to the entire original object, illustrating the principle of self-similarity. This characteristic of self-similarity implies that fractals are scale-invariant, a property that remains consistent across all scales in mathematical contexts. Moreover, the occurrence and relevance of fractals extend beyond mathematics into the natural world, where the structure of many naturally occurring objects can be effectively approximated using fractal geometry  \cite{mandelbrot1982fractal, hurd1988resource, barnsley1988fractals}.\\
\indent
In the domain of fractal potentials, the Cantor fractal potential stands as the simplest and prototypical member. Its construction commences with a rectangular potential barrier, characterized by a height $V$ and length $L$, identified as the initial stage $S=0$. Progressing to higher stages $S=1,2,3,...., S$, the GC fractal potential is formed by systematically excising a fraction $\frac{1}{\rho}$ of the length of the potential at the preceding stage from the central part of the remaining potential segments at each stage, where $\rho$ is a positive real number greater than 1. The case where $\rho=3$ is particularly notable, as it defines the standard Cantor fractal potential, often simply termed the Cantor potential. Parallel to this, the SVC-$\rho$ potential represents a related potential distribution. This potential diverges from the Cantor model in its methodology: at each stage $S$, a fraction $\frac{1}{\rho^{S}}$ of the length from the preceding stage $S-1$ is removed from the midst of each segment. The standard version of this system, SVC-4, emerges when $\rho =4$. Unlike the Cantor fractal potential, the SVC-$\rho$ system does not classify as a fractal due to its lack of consistent self-similar structure at various stages. The quantum mechanical scattering properties through the Cantor fractal potential have been extensively studied \cite{konotop1990wave, sun1991wave, bertolotti1994spectral, bertolotti1996transmission, lavrinenko2002propagation, chiadini2003self, hatano2005strong, sangawa2005resonance, esaki2009wave, guerin1996scattering, honda2006rigorous, chuprikov2006new, chuprikov2000transfer, sakaguchi2017scaling, ogawana2018transmission,monsoriu2005transfer}. These investigations predominantly employ the transfer matrix method in quantum mechanics to elucidate the scattering coefficients and their related properties. In the context of the Cantor fractal potential system, researchers have discerned a notable scaling behavior of the transmission coefficient in relation to the wave vector $k$, as documented in \cite{guerin1996scattering, esaki2009wave, sakaguchi2017scaling}. Interestingly, this same scaling behavior has been reported in non-fractal systems (SVC-$\rho$ potential) \cite{narayan2023tunneling} as well and has also been identified within the realm of SFQM for GC and SVC-$\rho$ potentials \cite{singh2023quantum}.
\\
\indent
In our previous research, we introduced the concept of the super periodic potential (SPP) \cite{hasan2018super}, which serves as an advanced form of the locally periodic potential. This approach allows for the calculation of reflection and transmission probability for potentials of any given periodicity, provided the transfer matrix for each unit cell potential is known. Our findings demonstrate that the GC and SVC-$\rho$ potential systems are specific examples of the broader SPP framework. Through the application of SPP, we have investigated the transmission coefficients for these potentials, examining them in the contexts of both standard quantum mechanics (QM) \cite{hasan2018super, narayan2023tunneling, umar2023quantum} and in SFQM domain \cite{singh2023quantum}.\\ 
\indent
Recently, we introduced a more general class of Cantor potential system that unifies the GC and SVC-$\rho$ system. We name this potential as UCP or UCP-$\rho$  \cite{umar2023quantum} and provide its transmission coefficient by using the SPP formalism developed earlier. The construction of the UCP-$\rho$ system of stage $S$ is similar to the construction of the SVC-$\rho$ system \cite{narayan2023tunneling} of stage $S$. The UCP-$\rho$ system of stage $S$ is characterized by  four parameters $\mu$, $\nu$, $\rho$ and $S$. The UCP-$\rho$ system is constructed by removing $\frac{1}{\rho^{\mu + \nu S}}$ fraction of the length of the potential segment of stage $S-1$ from the middle of each remaining portion of the potential at every stage $S$. Again here, $\rho$ is a real positive number and $\mu$ and $\nu$ also belongs to the real number ($\mu$, $\nu$ $\in$ $\mathbb{R}$) with condition that $\mu$ and $\nu$ can not be simultaneously zero. 
It is essential to highlight that in the UCP-$\rho$ system, the potential at stage 
$S=1$ comprises just two ($N=2$) segments. This indicates that at the initial stage 
$S=0$, there was only a single removal action executed within the potential. Similarly, at each stage, every potential segment undergoes a division, resulting in two potential segments and entails a single removal operation per segment.\\
\indent
%This article presents an extension of the UCP-$\rho$ system, wherein the limitation of stage $S=1$ to only two potential segments is transcended. Instead, we propose a modified potential structure where stage $S=1$ comprises $N$ potential segments, and which are the result of $N-1$ excise operation.
This article presents an extension of the UCP-$\rho$ system with a modified potential structure, where stage $S=1$ includes $N > 2$ potential segments instead of $N=2$, achieved through $N-1$ excision operations.
%
%In this expanded formulation of the UCP-$\rho$ system, each potential segment, at every subsequent stage, is subject to a division process. This results in the generation of $N$ new potential segments from each original segment, accompanied by $N-1$ removal operations per segment. 
%
We call this expanded version of the UCP-$\rho$ system as GUCP or UCP-$\rho_{N}$ system. \textcolor{black}{Throughout this paper, we will use the term UCP-$\rho_{N}$ to describe this system}. Within this nomenclature, the UCP-$\rho$ system is designated as the UCP-$\rho_{2}$ system. We provide the transmission coefficient from the UCP-$\rho_{N}$ system by using SPP formalism developed earlier. It is also crucial to highlight that $N$ is a member of the set of natural numbers ($N\in \mathbb{N}$). In general, the UCP-$\rho_{N}$ system is constructed by removing a fraction $\frac{1}{\rho^{\mu + \nu S}}$ of the length of the potential segment at stage $S-1$, executed at $N-1$ symmetric locations (from the center of the potential system) within the potential. This process is designed to produce $N$ potential segments from a selected potential bar, leading to stage $S$ containing $N^{S}$ potential segments. In the case where $(\mu, \nu) = (\mathbb{R}, 0)$, the resulting potential is a fractal UCP-$\rho_{N}$ potential, due to its sustained self-similarity across each stage. Conversely, the configuration $(\mu, \nu) = (0, \mathbb{R})$ does not manifest fractal properties and let us call this non-fractal UCP-$\rho_{N}$ potential, as it does not maintain self-similarity in subsequent stages. Interestingly, fractal UCP-$\rho_{N}$ system (FUCP-$\rho_{N}$ system) represents a PCP \cite{jarrendahl1995x,jaggard1997polyadic,jaggard1998scattering,monsoriu2006quantum,villatoro2008tunneling} of minimum lacunarity. Lacunarity \cite{mandelbrot1982fractal, mandelbrot1994fractal, mandelbrot1995measures, lin1986suggested, gefen1983geometric, allain1991characterizing, yasar2005fractal} is one of the fractal descriptors and is explained in detail Section \ref{fucp}.\\
\indent
The polyadic Cantor system \cite{mandelbrot1998nature, falconer2004fractal, barnsley2014fractals} represents an intriguing extension of the classic Cantor set, offering a broader spectrum of fractal configurations. In this system, the process of segment removal at each iteration is not limited to a single middle third, as in the traditional Cantor set, but involves removing multiple segments, allowing for a more diverse range of fractal patterns. This approach results in varying degrees of complexity and lacunarity within the fractal structures, a concept explored in depth by Falconer \cite{falconer2004fractal}. Additionally, the system's unique properties have implications for understanding natural phenomena, as Mandelbrot illustrates in his seminal work \cite{mandelbrot1982fractal}. These various studies underline the polyadic significance of the Cantor system as a tool for exploring the complexities of fractal geometry and its real-world manifestations.\\
\indent
Fractal (and Cantor) geometry is not only significant in studying the transmission through structures with Cantor-like potentials, but it also has an important role in optics. The earliest documented research on the diffraction of electromagnetic waves by fractal structures is believed to originate from the theoretical studies conducted by Berry \cite{berry1979diffractals}. Further, an extensive array of research papers has been published, exploring the interaction between electromagnetic waves and fractal structures. These studies have covered topics such as the optical Fourier transform of fractals \cite{allain1986optical}, and the diffraction phenomena associated with various fractal objects. Such studies have examined diffraction through serrated fractals \cite{kim1991diffraction}, Cantor bars \cite{ sakurada1992fresnel,jaggard1992triadic,ledesma1997scaling}, fractals with modifiable lacunarity \cite{zunino2003fraunhofer}, fractal (metallic) supergratings \cite{skigin2007diffraction} and so on. Also, numerous studies have been conducted where fractal geometry is incorporated into the design of optical lenses, leading to the development of what are known as fractal lenses or fractal zone plates \cite{saavedra2003fractal,monsoriu2004fractal, monsoriu2006fractal, furlan2007white, monsoriu2007devil, remon2009fractal, pu2015devil, cheng2016optical} and also fractal squared zone plate has been studied \cite{calatayud2013fractal}. Furthermore, the polyadic Cantor geometry has been integrated into the domains of optics \cite{calatayud2009polyadic} and acoustics \cite{perez2019fractal,castineira2017polyadic}.\\
\indent
The organization of the paper is structured as follows: Section \ref{cantor} revisits the UCP-$\rho$ system. In Section \ref{gucpn}, we expand upon the UCP-$\rho$ system, introducing the GUCP or UCP-$\rho_{N}$ system, and calculated the lengths of the potential segments for each stage, utilizing the $q$-Pochhammer. Section \ref{ucpspp} delineates how the UCP-$\rho_{N}$ system, at any arbitrary stage $S$, constitutes a specific instance of the rectangular SPP of order $S$. In Section \ref{fucp}, the discussion encompasses fractal descriptors, including fractal dimension and lacunarity within fractals. Utilizing these concepts, we demonstrate that the PCP system is encapsulated within the UCP-$\rho_{N}$ system as a particular case. Section \ref{tucp} articulates the explicit expression for the argument $\Gamma_{q}$ of the Chebyshev polynomial of the second kind, which is essential for calculating the transmission probability for the UCP-$\rho_{N}$ system. Section \ref{tfeatures} offers a graphical and detailed exposition of the transmission features. This section also addresses the phenomenon of transmission probability saturation and elucidates a universal scaling law that the reflection coefficient adheres to for large values of $k$. The paper reaches its conclusion in Section \ref{conclusion}, where the discussions and conclusions are presented.

%%%%%%%%%%%%%%%%%%%%%%%%%%%%%%%%%%%%%%%%%%%%%%%%%%%%%%%%%%%%%%%%%%%%%%%%%%%%%%%%%%%%%

\section{Unified Cantor potential system}
\label{cantor}
\begin{figure}[H]
	\begin{center}
		\includegraphics[scale=0.9]{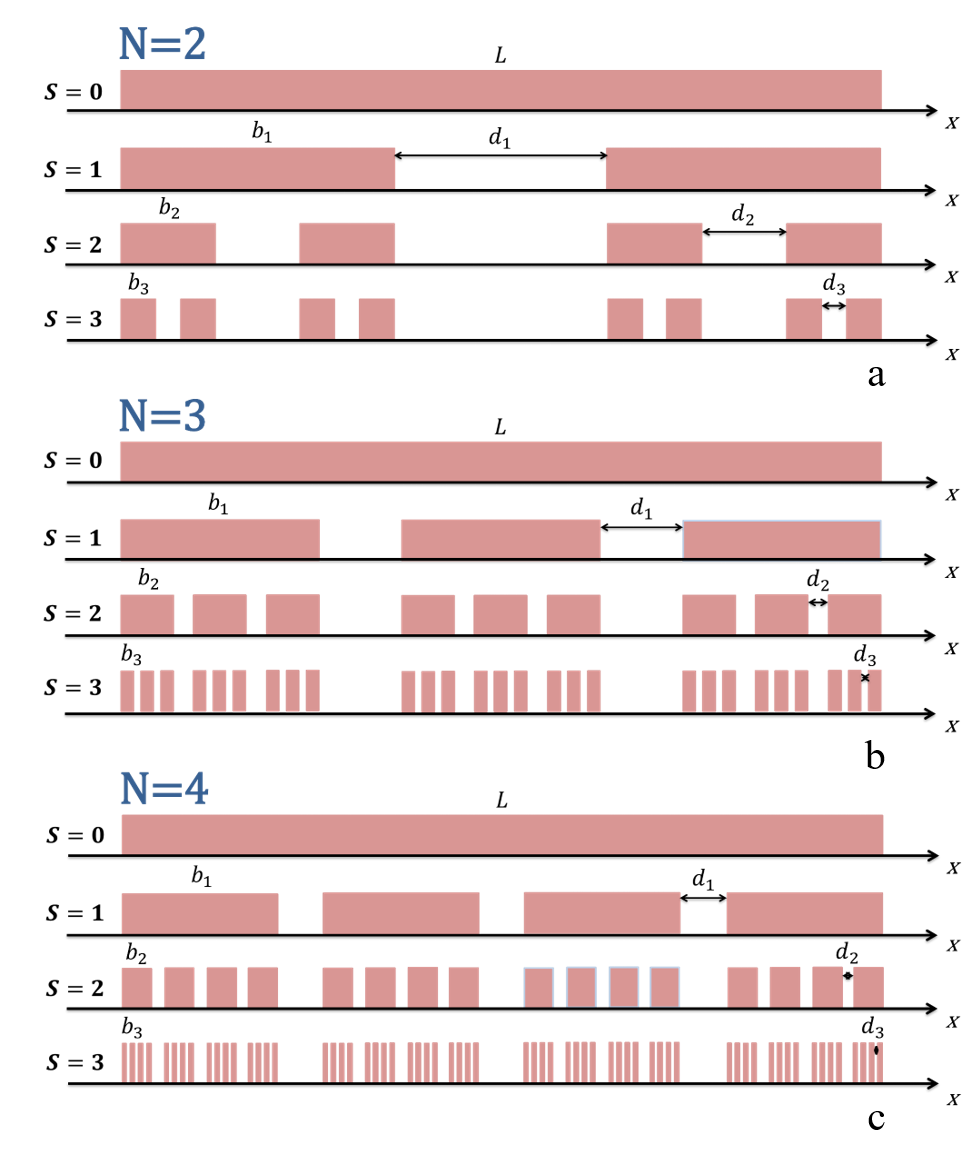} 
		\caption{\it 
			The figure elucidates the formation of the generalized unified Cantor potential systems through the potential segment removal exercise, specifically the (a) UCP-\(\rho_{2}\), (b) UCP-\(\rho_{3}\) and (c) UCP-\(\rho_{4}\) configurations, each distinguished by \(N=2\), \(3\) and \(4\) potential segments at the initial stage \(S=1\) respectively. It visually demonstrates the methodical formation of these systems through various stages of development. For each designated stage \(S\), the notation \(b_{S}\) specifies the length of individual potential segments, providing a quantifiable measure of the structural dimensions of the system at each stage. Furthermore, the outermost gap, labeled \(d_{S}\), is quantitatively described by the ratio \(\frac{b_{S-1}}{\rho^{\mu+\nu S}}\) capturing the dynamic scaling of the outermost gap as a function of parameters \(\rho\), \(\mu\), \(\nu\), and the stage \(S\). The height of the opaque region is V representing the height of the potential.
		}
		\label{fractal_figure}
	\end{center}
\end{figure}

The unified Cantor potential system denoted as UCP-$\rho$, is a concept that arises from the modification of a rectangular barrier potential with finite length $L$ and height $V (\equiv V_{0})$. This system is generated by iteratively removing a specific fraction of the potential at each stage, denoted as $S$, with the fraction being proportional to $\frac{1}{\rho^{\mu + \nu S}}$. Here, $\rho \in \mathbb{R}^{+}$ and $\rho>1$, $\mu$ and $\nu$ are real numbers ($\mu$, $\nu$ $\in$ $\mathbb{R}$), with the constraint that they cannot be simultaneously zero. If both parameters were to attain a simultaneous value of zero the system's inaugural stage $S = 1$, would manifest as a localized empty space, devoid of any potential segments. This distinctive condition would result in absolute transmission, permitting unobstructed passage of particles or waves through this void. Under these circumstances, further progression into subsequent stages would become unattainable. The construction of the UCP-$\rho$ system can be understood as follows:
\begin{itemize}
    \item \textbf{Stage $S=0$:} This stage simply represents a symmetric potential of finite length $L$ and potential height $V \equiv V_{0}$.
    \item \textbf{Stage $S=1$:} At this stage, a fraction $\frac{1}{\rho^{\mu + \nu}}$ of length $L$ is removed from the center of the potential of stage $S=0$. This results in two potential segments, each with a length denoted as $b_{1}$.
    \item \textbf{Stage $S=2$:} In the second stage, a further fraction $\frac{1}{\rho^{\mu + 2\nu}}$ of length $b_{1}$ is removed from the center of the two remaining potential segments created in the previous stage $(S = 1)$. This process divides each segment into smaller segments, and each of these has a length of $b_2$. We have a total of four potential segments at this stage. 
    \item \textbf{Stage $S=3$:} In the second stage, a fraction $\frac{1}{\rho^{\mu + 3\nu}}$ of length $b_{2}$ is removed from the center of the segments created at the previous stage $(S = 2)$ and each of the newly created segments at this stage has a length of $b_3$. This stage has a total of eight potential segments of equal length. 
    \item \textbf{Stage $S=4$:} In the fourth stage, a fraction $\frac{1}{\rho^{\mu + 4\nu}}$ of length $b_{3}$ is removed from the center of the segments created at the previous stage $(S = 3)$ and each of the newly created segments at this stage has a length of $b_4$. This stage has a total of sixteen potential segments of equal length.
\end{itemize}
In general, applying the same process of elimination of the fraction $\frac{1}{\rho^{\mu+\nu S}}$ of the potential length $b_{S-1}$ from the middle of the remaining potential segments of stage $S-1$ constructs the UCP-$\rho$ system of any arbitrary stage $S$ consisting of $2^{S}$ potential segment of equal length $b_{S}$. We find the expression of $b_{S}$ as follows. From Fig. \ref{fractal_figure}a, it is clear that the length of the potential segment at stage $S=1$ is expressed through,
\begin{equation}
b_{1} = \frac{L}{2}\left(1-\frac{1}{\rho^{\mu+\nu}}\right).
\end{equation}
Similarly, the length of the potential segment at stage $S=2$ is expressed through
\begin{equation}
b_{2} = \frac{L}{2^{2}}\left(1-\frac{1}{\rho^{\mu+\nu}}\right)\left(1-\frac{1}{\rho^{\mu+2\nu}}\right).
\end{equation}
Similarly, the length of the potential segment at stage $S=3$ is expressed through
\begin{equation}
b_{3} = \frac{L}{2^{3}}\left(1-\frac{1}{\rho^{\mu+\nu}}\right)\left(1-\frac{1}{\rho^{\mu+2\nu}}\right)\left(1-\frac{1}{\rho^{\mu+3\nu}}\right).
\end{equation}
and so on. 
Similarly, length of the potential segments at stage $S=4$ is expressed through
\begin{equation}
b_{4} = \frac{L}{2^{4}}\left(1-\frac{1}{\rho^{\mu+\nu}}\right)\left(1-\frac{1}{\rho^{\mu+2\nu}}\right)\left(1-\frac{1}{\rho^{\mu+3\nu}}\right)\left(1-\frac{1}{\rho^{\mu+4\nu}}\right).
\end{equation}
In general, length $b_{S}$ of potential segment of any arbitrary stage $G$ can be expressed as
\begin{equation}
b_{S} = \frac{L}{2^{S}}\prod_{j=1}^{S}\left(1-\frac{1}{\rho^{\mu+\nu j}}\right).
\label{b02}
\end{equation}
The product series can be expressed in terms of $q$-Pochhammer symbol as
\begin{equation}
\prod_{j=1}^{S}\left(1-\frac{1}{\rho^{\mu+\nu j}}\right) = \frac{\rho^{\mu}}{\rho^{\mu}-1}\times q\left(\frac{1}{\rho^{\mu}};\frac{1}{\rho^{\nu}}\right)_{S+1},
\end{equation}
where $q$-Pochhammer symbol is defined by  \cite{abramowitz1968handbook},
\begin{equation}
\begin{split}
    q(\alpha;\beta)_{p} & = (1-\alpha)(1-\alpha.\beta)(1- \alpha.\beta^{2}).....(1-\alpha.\beta^{p-1}). \\
    & = \prod_{j=0}^{p-1}(1-\alpha.\beta^{j})
    \label{qp}
\end{split}
\end{equation}
Hence $b_{S}$ in terms of $q$-Pochhammer symbol is expressed through
\begin{equation}
b_{S} = \frac{L\rho^{\mu}}{2^{S}(\rho^{\mu}-1)}\times q\left(\frac{1}{\rho^{\mu}};\frac{1}{\rho^{\nu}}\right)_{S+1}.
\label{b2}
\end{equation}
The equation presented above serves as a general expression for determining the length of the potential segment at each stage in the UCP-$\rho$ system. 
In the ensuing discussion, we will introduce the UCP-$\rho_{N}$ system and develop an expression for the length of the potential segment at each stage, denoted as $b_{S}\equiv (b_{S})_{N}$, for this particular UCP-$\rho_{N}$ configuration.

%%%%%%%%%%%%%%%%%%%%%%%%%%%%%%%%%%%%%%%%%%%%%%%%%%%%%%%%%%%%%%%%%%

\section{Generalized unified Cantor potential: GUCP or UCP-\texorpdfstring{$\rho_{N}$}{rho-N} system}
\label{gucpn}
In section \ref{cantor}, we have discussed the concept of UCP-$\rho$ system, wherein we implement a fractional removal operation involving a factor of $\frac{1}{\rho^{\mu+\nu S}}$ from the central portion of every available potential at each progressive stage, indexed by $S$. Consequently, this process causes each potential segment at every stage to undergo a division into two distinct potential segments, leading to a total of $2^{S}$ potential segments at each respective stage $S$. Now consider the geometry of Fig. \ref{fractal_figure}b, where stage $S=1$ configuration comprises three potential segments of width $b_{1}$ separated by distance $d_{1}$. This particular arrangement is achieved through a fractional removal operation, specifically fraction $\frac {1}{\rho^{\mu+\nu}}$ of the length $L$, executed at two symmetric locations. The important thing in this process is that the removal operation should be done in such a way that the length of the newly created potential segment should be the same. Stage $S=1$ represents that a potential of height $V$ and length $b_{1}$ is periodically repeated three times at a periodic distance $d_{1}$. For the next stage $S=2$, a fraction $\frac {1}{\rho^{\mu+2\nu}}$ of the length $b_{1}$ is eliminated at the two symmetric locations from each potential segment of stage $S=1$ in such a fashion that each newly created potential segments at stage $S=2$ should be of same length $b_{2}$. This stage has a total of 9 potential segments and it is depicted in Fig. \ref{fractal_figure}b. Next, for stage $S=3$, a fraction $\frac {1}{\rho^{\mu+3\nu}}$ of the length $b_{2}$ is eliminated at the two symmetric locations from each potential segment of stage $S=2$ such that each potential segment has length $b_{3}$. There are a total of 27 potential segments at stage $S=3$. In general, applying the same process of elimination of a fraction $\frac{1}{\rho^{\mu + \nu S}}$ of length $b_{S-1}$ at two symmetric locations from each potential segment of stage $S-1$ constructs the UCP-$\rho_{3}$ system of any arbitrary stage $S$ consisting of $3^{S}$ potential segment of equal length $b_{S}$. The nomenclature UCP-$\rho_{3}$ suggests that each potential segment at each stage is undergoing a division into three potential segments. In a similar fashion, the UCP-$\rho$ system discussed in Section \ref{cantor} can be represented as a UCP-$\rho_{2}$ system because in this case, each potential segment at each stage is undergoing a division into two potential segments. The construction of the UCP-$\rho_{4}$ potential system is also illustrated in Fig. \ref{fractal_figure}c. In general, we can construct the UCP-$\rho_{N}$ system where each potential segment at each stage $S$ is undergoing a division into $N$ potential segments. UCP-$\rho_{N}$ system represents a generalized version of UCP-$\rho_{2}$ system and we also call this system a generalized unified Cantor potential (GUCP) system. The UCP-$\rho_{N}$ system comprises a set of $N^{S}$ potential segments of equal length $b_{S}$. It is important to note here that, for all $N$, stage $S=1$ represents a periodic potential system of total $N$ potentials, but this periodicity breaks for stages $S>1$. Stages $S>1$ represent the SPP structure which we will discuss in the next section. Hence a transition from periodic potential to SPP occurs as we transit from stage $S=1$ to stage $S>1$. Next, we will find the length $b_{S}$ of any arbitrary stage $S$ for UCP-$\rho_{N}$ system.

%%%%%%%%%%%%%%%%%%%%%%%%%%%%%%%%%%%%%%%%%%%%%%%%%%%%%%%%%%%%%%%%%%%%

\subsection{Calculation of \texorpdfstring{$b_{S}$}{b-S} for UCP-\texorpdfstring{$\rho_{3}$}{rho-3} system}
The geometry of Fig. \ref{fractal_figure}b implies that the length of the potential segment at stage $S=1$ is expressed through,
\begin{equation}
b_{1} = \frac{L}{3}\left(1-\frac{2}{\rho^{\mu+\nu}}\right).
\end{equation}
Similarly, the length of the potential segment at stage $S=2$ is expressed through
\begin{equation}
b_{2} = \frac{L}{3^{2}}\left(1-\frac{2}{\rho^{\mu+\nu}}\right)\left(1-\frac{2}{\rho^{\mu+2\nu}}\right).
\end{equation}
Similarly, the length of the potential segment at stage $S=3$ is expressed through
\begin{equation}
b_{3} = \frac{L}{3^{3}}\left(1-\frac{2}{\rho^{\mu+\nu}}\right)\left(1-\frac{2}{\rho^{\mu+2\nu}}\right)\left(1-\frac{2}{\rho^{\mu+3\nu}}\right).
\end{equation}
Similarly, the length of the potential segment at stage $S=4$ is expressed through
\begin{equation}
b_{4} = \frac{L}{3^{4}}\left(1-\frac{2}{\rho^{\mu+\nu}}\right)\left(1-\frac{2}{\rho^{\mu+2\nu}}\right)\left(1-\frac{2}{\rho^{\mu+3\nu}}\right)\left(1-\frac{2}{\rho^{\mu+4\nu}}\right).
\end{equation}
In general, length $b_{S}$ of potential segment of any arbitrary stage $S$ for UCP-$\rho_{3}$ system can be expressed as
\begin{equation}
b_{S} = \frac{L}{3^{S}}\prod_{j=1}^{S}\left(1-\frac{2}{\rho^{\mu+\nu j}}\right).
\label{b03}
\end{equation}
In terms of $q$-Pochhammer symbol $b_{S}$ can be expressed through
\begin{equation}
b_{S} = \frac{L\rho^{\mu}}{3^{S}(\rho^{\mu}-2)}\times q\left(\frac{2}{\rho^{\mu}};\frac{1}{\rho^{\nu}}\right)_{S+1}.
\label{b3}
\end{equation}

%%%%%%%%%%%%%%%%%%%%%%%%%%%%%%%%%%%%%%%%%%%%%%%%%%%%%%%%%%%%%%%%%%%%

\subsection{Calculation of \texorpdfstring{$b_{S}$}{b-S} for UCP-\texorpdfstring{$\rho_{4}$}{rho-4} system}
From Fig. \ref{fractal_figure}c, it is evident that the length of the potential segment at stage $S=1$ is expressed through,
\begin{equation}
b_{1} = \frac{L}{4}\left(1-\frac{3}{\rho^{\mu+\nu}}\right).
\end{equation}
Similarly, the length of the potential segment at stage $S=2$ is expressed through
\begin{equation}
b_{2} = \frac{L}{4^{2}}\left(1-\frac{3}{\rho^{\mu+\nu}}\right)\left(1-\frac{3}{\rho^{\mu+2\nu}}\right).
\end{equation}
Similarly, the length of the potential segment at stage $S=3$ is expressed through
\begin{equation}
b_{3} = \frac{L}{4^{3}}\left(1-\frac{3}{\rho^{\mu+\nu}}\right)\left(1-\frac{3}{\rho^{\mu+2\nu}}\right)\left(1-\frac{3}{\rho^{\mu+3\nu}}\right).
\end{equation}
Similarly, the length of the potential segment at stage $S=4$ is expressed through
\begin{equation}
b_{4} = \frac{L}{4^{4}}\left(1-\frac{3}{\rho^{\mu+\nu}}\right)\left(1-\frac{3}{\rho^{\mu+2\nu}}\right)\left(1-\frac{3}{\rho^{\mu+3\nu}}\right)\left(1-\frac{3}{\rho^{\mu+4\nu}}\right).
\end{equation}
In general, length $b_{S}$ of potential segment of any arbitrary stage $G$ for UCP-$\rho_{4}$ system can be expressed as
\begin{equation}
b_{S} = \frac{L}{4^{S}}\prod_{j=1}^{S}\left(1-\frac{3}{\rho^{\mu+\nu j}}\right).
\label{b04}
\end{equation}
In terms of $q$-Pochhammer symbol this $b_{S}$ can be expressed through
\begin{equation}
b_{S} = \frac{L\rho^{\mu}}{4^{S}(\rho^{\mu}-3)}\times q\left(\frac{3}{\rho^{\mu}};\frac{1}{\rho^{\nu}}\right)_{S+1}.
\label{b4}
\end{equation}

\label{b5}
%\end{equation}

%%%%%%%%%%%%%%%%%%%%%%%%%%%%%%%%%%%%%%%%%%%%%%%%%%%%%%%%%%%%%%%%%%%%

%\subsection{Calculation of \texorpdfstring{$b_{S}$}{b-S} for UCP-\texorpdfstring{$\rho_{N}$}{rho-N} system} 
%From, Eqs. (\ref{b02}), (\ref{b03}), (\ref{b04}) and (\ref{b05}), we can express $b_{S}$ for further consecutive GUCP system. The expression of $b_{S}$ for UCP-$\rho_{10}$, UCP-$\rho_{50}$ and UCP-$\rho_{100}$ system are given below
%\begin{equation}
%b_{S} = \frac{L}{10^{S}}\prod_{j=1}^{S}\left(1-\frac{9}{\rho^{\mu+\nu j}}\right),
\label{b010}
%\end{equation}
%\begin{equation}
%b_{S} = \frac{L}{50^{S}}\prod_{j=1}^{S}\left(1-\frac{49}{\rho^{\mu+\nu j}}\right),
\label{b011}
%\end{equation}
%and
%\begin{equation}
%b_{S} = \frac{L}{100^{S}}\prod_{j=1}^{S}\left(1-\frac{99}{\rho^{\mu+\nu j}}\right).
\label{b012}
%\end{equation}
%respectively. In general, we can express $b_{S}$ for UCP-$\rho_{N}$ system as
\noindent
From Eqs. (\ref{b02}), (\ref{b03}) and (\ref{b04}), we can write a general expression of $b_{S}$ for the UCP-$\rho_{N}$ system as 
\begin{equation}
b_{S} = \frac{L}{N^{S}}\prod_{j=1}^{S}\left(1-\frac{N-1}{\rho^{\mu+\nu j}}\right).
\label{b0N}
\end{equation}
Further, in terms of $q$-Pochhammer symbol, the above equation is expressed through
\begin{equation}
b_{S} = \frac{L\rho^{\mu}}{N^{S}(\rho^{\mu}-1-N)}\times q\left(\frac{N-1}{\rho^{\mu}};\frac{1}{\rho^{\nu}}\right)_{S+1}.
\label{bsN}
\end{equation}
where $N$ represents the potential counts at stage $S=1$.

%%%%%%%%%%%%%%%%%%%%%%%%%%%%%%%%%%%%%%%%%%%%%%%%%%%%%%%%%%%%%%%%%%%%%%%%%%%%%%%

\section{Generalized unified Cantor potential as a special case of super periodic potential}
\label{ucpspp}
In the context of the concept of SPP introduced in reference \cite{hasan2018super}, we briefly expound upon this concept for the sake of comprehensiveness within the scope of this paper. Commencing with a unit cell potential denoted as $V$, we construct a periodic system by repetitively placing the unit cell system in a consecutive manner, with a specified finite repetition count denoted as $N_{1}$, and at regularly spaced intervals denoted as $r_{1}$. 
Here, $r_{1}$ signifies the distance between the starting positions of two consecutive unit cells. The resulting periodic potential is denoted as $V_{1}=(V, N_{1},r_{1})$. Subsequently, call the system $V_{1}$ as a new unit cell and periodically repeat this by introducing another repetition count denoted as $N_{2}$ and another regular spacing interval denoted as $r_{2}$, resulting in the formation of the system $V_{2}=(V_{1}, N_{2},r_{2})$. Now, call the system $V_{2}$ as a new unit cell and iterate this periodic repetition $N_{3}$ times at consecutive distances $r_{3}$ to obtain another unit cell system $V_{3}=(V_{2}, N_{3},r_{3})$ which is further periodically repeated in a similar manner to obtain $V_{4}=(V_{3}, N_{4},r_{4})$. This sequence of periodic repetitions can extend to an arbitrary finite number of iterations denoted as $S$, resulting in the establishment of the super periodic potential system of order $S$, designated as $V_{S}=(V_{S-1}, N_{S}, r_{S})$. This hierarchical approach allows for the generation of increasingly complex SPP structures through successive iterations of unit cell repetition and spacing adjustments. \\
\indent
Next, we will demonstrate that the UCP-$\rho_{N}$ system at stage $S$ can be regarded as a specific instance of the rectangular SPP of order $S$.  Consequently, it becomes evident that the term stage $S$ within the UCP-$\rho_{N}$ framework and the term order $S$ within the realm of rectangular SPP are synonymous, denoting the same underlying concept. As discussed earlier, the UCP-$\rho_{N}$ system can be obtained by removing the fraction $\frac{1}{\rho^{\mu+\nu S}}$ of the length of the available potential at stage $S-1$ from $N-1$ symmetric locations (symmetry should be from the center of the potential) in such a fashion that each potential segment has length $b_{S}$. 
\begin{figure}[h! tbp]
	\begin{center}
		\includegraphics[scale=0.96]{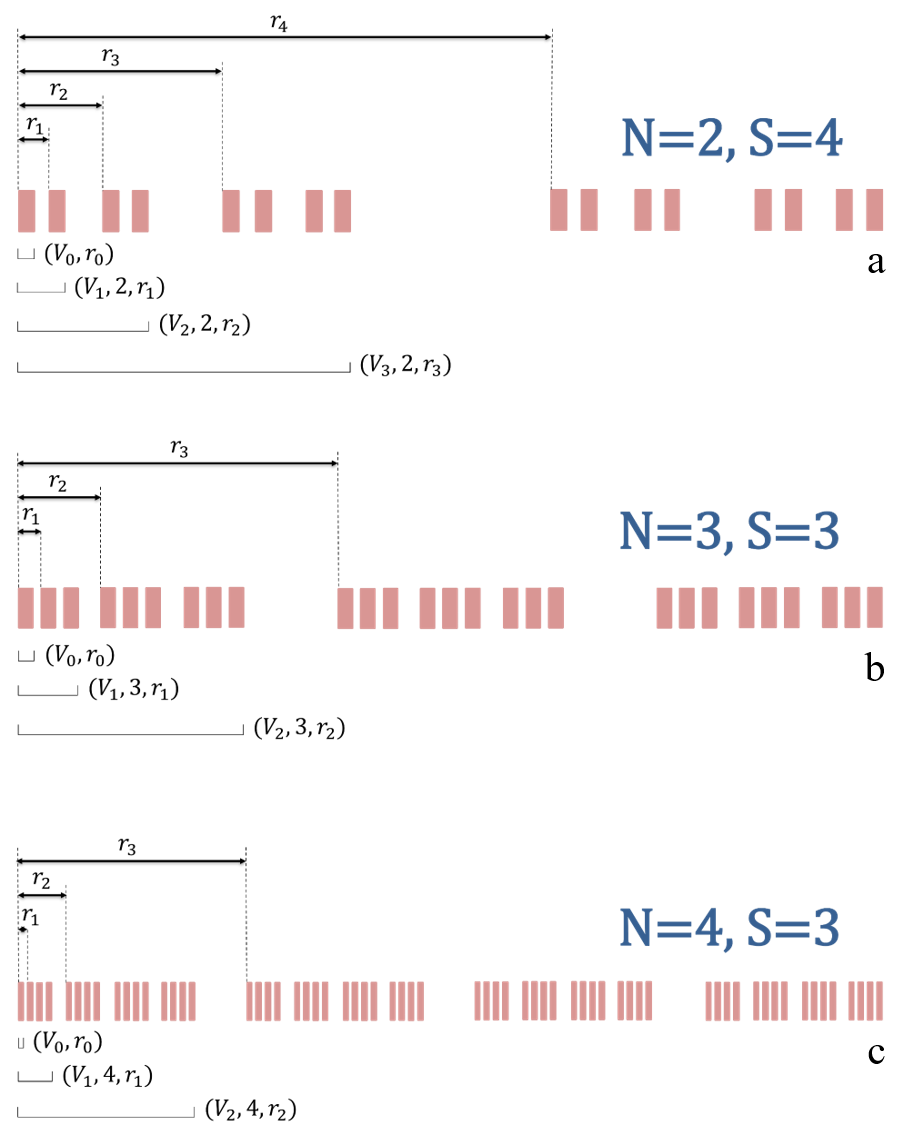} 
		\caption{\it Illustration of the super periodic development of the (a) UCP-$\rho_{2}$ system of stage $S=4$, (b) UCP-$\rho_{3}$ system of stage $S=3$ and (c) UCP-$\rho_{4}$ system of stage $S=3$. The foundational element (unit cell), termed the initiator, is represented by a rectangular barrier with a height $V$, while the construction process itself is driven by the principle of super periodicity. $r_{1}$, $r_{2}$, $r_{3}$ and $r_{4}$ are the super periodic distances that regulate the arrangement of the potentials within the system. These distances are instrumental in dictating the precise arrangement potentials, thereby guiding the Cantor-like construction of the UCP-$\rho_{N}$ systems.}
		\label{construction}
	\end{center}
\end{figure}
Fig. \ref{construction}a, \ref{construction}b and \ref{construction}c illustrates the super periodic construction of UCP-$\rho_{2}$ system of stage $S=4$, UCP-$\rho_{3}$ system of stage $S=3$ and UCP-$\rho_{4}$ system of stage $S=3$ repectively. First, consider the geometry of first of Fig. \ref{construction}a, here a potential of height $V$ and width $b_{S=4}$ is periodically repeated with count $N_{1}=2$ at a distance $r_{1}>b_{4}$. The resultant system of these potential barriers is further periodically repeated with count $N_{2}=2$ at distance $r_{2}$ thereby generating a system of four rectangular potentials. This system as a whole is further periodically repeated with count $N_{3}=2$ at a distance of $r_{3}$. This represents UCP-$\rho_{2}$ system of stage $S=4$. Hence, with the appropriate choice of length $b_{S}$ of the potential segment, we can construct a UCP-$\rho_{2}$ system of any stage $S$. In short UCP-$\rho_{2}$ system can be obtained through the following sequence of operation: start with potential $V$, obtain $V_{1}=(V, 2, r_{1})$, then obtain $V_{2}=(V_{1}, 2, r_{2})$, then $V_{3}=(V_{2}, 2, r_{3})$,....., $V_{S}=(V_{S-1}, 2, r_{S})$. The generated system $V_{S}$ will be our $S^{th}$ stage UCP-$\rho_{2}$ system. Next, consider the geometry of Fig. \ref{construction}b. Here, a UCP-$\rho_{3}$ system of stage $S=3$ is created by employing a sequence of repetitive operations. Initially, we start with a potential barrier of height $V$ and width $b_{S=3}$. This potential is then periodically repeated with a count of $N_{1}=3$ at a distance $r_{1}>b_{3}$.
The resulting configuration represents three periodically repeated potential barriers separated by $r_{1}$. Subsequently, this system as a whole is further periodically repeated with a count of $N_{2}=3$ at a distance of $r_{2}$, generating a system comprising nine rectangular potentials. This system as a whole is further periodically repeated with count $N_{3}=3$ at a distance of $r_{3}$. Now this system represents the UCP-$\rho_{3}$ system of stage $S=3$. Hence, again with the appropriate choice of length $b_{S}$ of the potential segment, we can construct UCP-$\rho_{3}$ of any stage $S$ through the super periodic manner. In summary, a UCP-$\rho_{3}$ system can be obtained by following this sequence of operations: begin with the potential $V$, obtain $V_{1}=(V, 3, r_{1})$, then obtain $V_{2}=(V_{1}, 3, r_{2})$, then $V_{3}=(V_{2}, 3, r_{3})$ and continue this process up to stage $S$ to obtain $V_{S}=(V_{S-1}, 3, r_{S})$. The generated system $V_{S}$ will represent the UCP-$\rho_{3}$ system. Next, Fig. \ref{construction}c illustrated the construction of UCP-$\rho_{4}$ system of stage $S=3$. Following the same sequence of steps, we can construct the UCP-$\rho_{4}$ system of any arbitrary stage $S$: start with potential $V$, then create $V_{1}=(V, 4, r_{1})$, proceed to obtain $V_{2}=(V_{1}, 4, r_{2})$, then form $V_{3}=(V_{2}, 4, r_{3})$, and so on, until you reach $V_{S}=(V_{S-1}, 4, r_{S})$. The resulting system, denoted as $V_{S}$, will represent the UCP-$\rho_{4}$ system at the $S^{th}$ stage. In general, proceeding through the identical sequence of procedural steps, a UCP-$\rho_{N}$ system at any arbitrary stage $S$ can be systematically constructed.\\
\indent
By leveraging the geometric framework inherent in the UCP-$\rho_{N}$ system and from the geometries of Fig. \ref{fractal_figure} and \ref{construction}, it can be easily substantiated that,
\begin{equation}
r_{1} = b_{S} + \frac{b_{S-1}}{\rho^{\mu+\nu S}},
\end{equation}
\begin{equation}
r_{2} = b_{S-1} + \frac{b_{S-2}}{\rho^{\mu+\nu (S-1)}},
\end{equation}
\begin{equation}
r_{3} = b_{S-2} + \frac{b_{S-3}}{\rho^{\mu+\nu (S-2)}},
\end{equation}
\begin{equation}
r_{4} = b_{S-3} + \frac{b_{S-4}}{\rho^{\mu+\nu (S-3)}}.
\end{equation}
In general,
\begin{equation}
r_{q} = b_{S+1-q} + \frac{b_{S-q}}{\rho^{\mu+ \nu (S+1-q)}}.
\end{equation}
Using Eq. (\ref{b0N}) in above, we arrive at
\begin{equation}
r_{q}(N)\equiv r_{q} = \frac{L}{N^{S+1-q}}\prod_{j=1}^{S+1-q}\left(1-\frac{N-1}{\rho^{\mu+\nu j}}\right)+\frac{L}{N^{S-q}\rho^{\mu+\nu(S+1-q)}}\prod_{j=1}^{S-q}\left(1-\frac{N-1}{\rho^{\mu+\nu j}}\right).
\end{equation}
After simplification, this can be expressed as,
\begin{equation}
r_{q}(N) = \frac{L}{N^{S+1-q}}\left( 1+\frac{1}{\rho^{\mu+\nu(S+1-q)}}\right)\prod_{j=1}^{S-q}\left(1-\frac{N-1}{\rho^{\mu+\nu j}}\right).
\label{s_m}
\end{equation}
Further in terms of $q$-Pochhammer symbol $r_{q}$ can be expressed as
\begin{equation}
r_{q}(N) = \frac{L\rho^{\mu}}{N^{S+1-q}(\rho^{\mu}-1-N)}\left( 1+\frac{1}{\rho^{\mu+\nu(S+1-q)}}\right)\times q\left(\frac{N-1}{\rho^{\mu}};\frac{1}{\rho^{\nu}}\right)_{S+1-q}.
\label{s_m1}
\end{equation}
Thus with the knowledge of $r_{q}$, $q=1,2,....,S$ and starting from a rectangular potential barrier of width $b_{S}$ and height $V$, the $S^{th}$ stage UCP-$\rho_{N}$ system can be obtained through the following sequence of operation: $V_{1}= (V,N,r_{1})$, then obtain $V_{2}= (V_{1},N,r_{2})$, then $V_{3}= (V_{2},N,r_{3})$$,.....$, $V_{S}= (V_{S-1},N,r_{S})$.  The generated system $V_{S}$ will be our $S^{th}$ stage UCP-$\rho_{N}$ system constructed through SPP formalism.

%%%%%%%%%%%%%%%%%%%%%%%%%%%%%%%%%%%%%%%%%%%%%%%%%%%%%%%%%%%%%%%%%%%%%%%%%%%%%%%%%%%%

\section{Fractal UCP-\texorpdfstring{$\rho_{N}$}{rho-N} system: Polyadic Cantor potential of minimum lacunarity}
\label{fucp}
The polyadic Cantor potential (PCP) system, a subset of fractal potential systems, expands upon the foundational Cantor potential framework. Within this broader context, the GC system emerges as a particular example of the versatility of the PCP system. Similarly, the PCP system, characterized by the minimum lacunarity, represents a specific example within the UCP-$\rho_{N}$ system. A key aspect of the taxonomy of the PCP system is its association with a parameter known as lacunarity, a parameter that profoundly influences its structural property. To fully appreciate the scope of this discussion, a solid grasp of the fractal dimension and lacunarity is essential, as these concepts are central to the detailed examination of the PCP system and its unique properties.

%%%%%%%%%%%%%%%%%%%%%%%%%%%%%%%%%%%%%%%%%%%%%%%%%%%%%%%%%%%%%%%%%%%%%%%
 
\subsection{Fractal dimension}
Fractals are characterized by unique properties, with self-similarity being a prominent feature. This property denotes that fractals contain self-copies, which can be defined recursively. In essence, an object is deemed self-similar when its constituent parts mirror the same shape or structure as the whole, albeit potentially occurring on different scales and with slight deformations. Notably, the self-similarity observed in natural elements often has a finite limit, disappearing after a certain number of iterations, rather than extending infinitely. In scientific terms, the self-similarity property in natural structures is indicative of the underlying fractal nature, marked by a recursive generation process. The term \textit{iteration} in the context of fractals refers to the successive stages of development, each unveiling a refined iteration of the fractal structure. It is noteworthy that the intricate patterns observed in nature, exhibiting self-similarity, are subject to limitations in their recursive complexity, reaching a discernible endpoint.\\
\indent
The fractal dimension \cite{mandelbrot1982fractal, barcellos1984fractal, theiler1990estimating,mandelbrot1967long,fernandez2014fractal} is a non-integer index that quantifies the complexity of a fractal object, describing how the detail in a pattern changes with the scale at which it is measured. Unlike topological dimensions, which are integer values describing the most basic characteristics of a geometrical shape, the fractal dimension conveys more nuanced information about the scaling of the shape and self-similarity properties. The fractal dimension, a concept rooted in mathematical metrics, diverges from traditional topological dimensions by allowing for noninteger values. While lines, squares, and cubes possess topological dimensions of one, two, and three respectively, the notion of fractal dimension, as introduced by Mandelbrot building upon the earlier contributions of Hausdorff \cite{schleicher2007hausdorff}, extends the possibility of geometric objects having dimensions between these integer values. This concept serves not only as a departure from traditional topological dimensions but also as a measure of the intricate space-filling capacity inherent in fractal patterns. The groundbreaking research of Mandelbrot opened up a new realm of understanding, revealing the existence of geometric entities with dimensions that transcend the conventional boundaries of whole numbers. The fractal dimension thus emerges as a nuanced and dynamic metric, capturing the intricate complexity and spatial intricacies of fractal patterns. Mathematically, fractal dimension $D$ can be calculated as\cite{monsoriu2006quantum, villatoro2008tunneling}
\begin{equation}
    D= \frac{\ln N}{\ln \zeta},
    \label{dimesnion}
\end{equation}
where $N$ is the number of the segments at stage $S=1$ (generator stage) and parameter $\zeta=\frac{1}{\rho}<1$ $(\rho \in \mathbb{R}^{+})$ serves as the scale factor governing the length contraction in each subsequent iteration. For general Cantor set $(N=2, \rho)$, Eq. (\ref{dimesnion}) is simplified as
\begin{equation}
    D= \frac{\ln 2}{\ln \zeta}=-\frac{\ln 2}{\ln \rho}.
    \label{dimesnion01}
\end{equation}
Further, Eq. (\ref{dimesnion}) can be expressed as a function of the number of available number of gaps $(N_{g}=N-1)$ at stage $S=1$ as
\begin{equation}
    D= -\frac{\ln (N_{g}+1)}{\ln \zeta}.
    \label{dimesnion04}
\end{equation}
There is another definition for the fractal dimension $D$ of generalized Cantor set as \cite{cherny2010scattering}
\begin{equation}
D=\frac{\ln2}{\ln\left(\frac{1-\zeta}{2}\right)}.
\label{dimension02}
\end{equation}
Eqs. (\ref{dimesnion01}) and (\ref{dimension02}) are two distinct equations to define the fractal dimension for the general Cantor set and both equations are only equal for $\rho=3$. The present discourse aims to establish that fractal construction characterized by different iteration counts, denoted by $N$, can indeed exhibit the same fractal dimensions and for this purpose, we will utilize Eq. (\ref{dimesnion}). For the standard Cantor (SC) potential $(N=2$, $\rho =3)$ fractal dimension $D=0.6309$ and for GC potentials like $(N, \rho)$ $=$ $(2,4)$, $(2,4.5)$, $(2,5)$, the fractal dimension $D$ is 0.50, 0.4607 and 0.4302 respectively.
\begin{figure}[H]
\begin{center}
\includegraphics[scale=0.37]{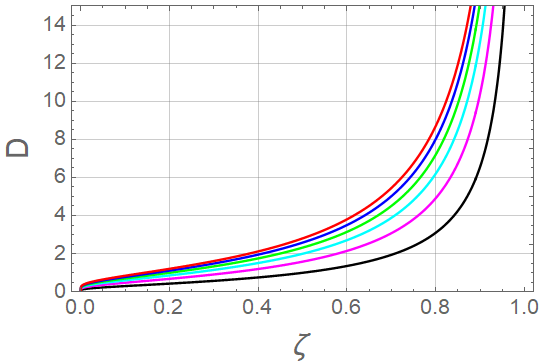}
\includegraphics[scale=0.37]{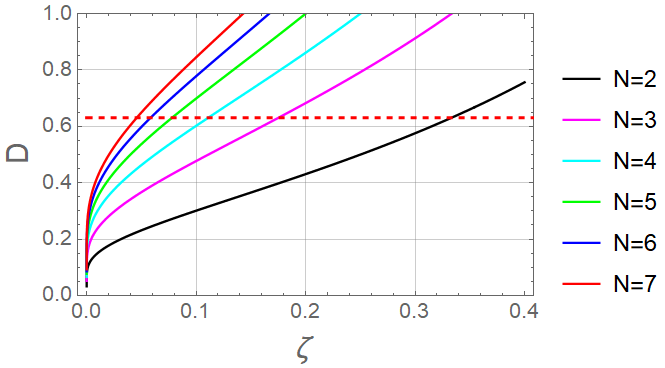} 
\caption{\it 
Graphs depicting the relationship between the fractal dimension $D$ and the scaling parameter $\zeta$ are presented for various values of $N$. The visual representation illustrates that different sets of fractals can yield the same fractal dimension. The second graph provides a closer look at the region where $D$ ranges from $0$ to $1$ and $\zeta$ ranges from $0$ to $0.4$. A red dashed line in this graph represents the fractal dimension value $D=0.6309$ for the standard Cantor set with parameters $(N,\rho)=(2,3)$. The points of intersection between this dashed line and the plots for different values of $N$ indicate fractals that share the same fractal dimension as the standard Cantor set.}
\label{dimension_figure}
\end{center}
\end{figure}
While fractals are commonly characterized by their dimensions, such characterization alone does not provide a comprehensive description of fractals. A notable aspect is the existence of multiple fractals with identical fractal dimensions, contingent upon their specific $N$ and $\zeta$ (or $\rho$) values. This is exemplified in Fig. \ref{dimension_figure}, which depicts the variation of the fractal dimension $D$ in relation to the scaling parameter $\zeta$ for different values of $N$ (potential counts at stage $S=1$). The figure illustrates that distinct sets of fractals can yield identical fractal dimensions. The second figure represents a magnified view of the first figure and focuses on the range where $D$ spans from 0 to 1, and $\zeta$ ranges from 0 to 0.4. A red dashed line in the second figure denotes the fractal dimension value $D=0.6309$ corresponding to the standard Cantor set with parameters $(N=2$, $\rho =3)$. The points of intersection between this dashed line and plots corresponding to different $N$ values highlight fractals that share the same fractal dimension as the standard Cantor set.

%%%%%%%%%%%%%%%%%%%%%%%%%%%%%%%%%%%%%%%%%%%%%%%%%%%%%%%%%%%%%%%%%%%%%%%%%%%

\subsection{Lacunarity in fractals}
\label{lacunarity}
It has been discussed in the last subsection that the dimension of fractals commonly serves as a primary characterization, this singular aspect does not offer a comprehensive depiction of fractals. The limitation arises from the observation that numerous fractals can exhibit identical fractal dimensions, contingent upon their specific $N$ and $\zeta$ (or $\rho$) values. Consequently, the exploration of alternative fractal descriptors becomes imperative to augment the understanding of fractal structures. In this context, one such alternative descriptor is lacunarity, which provides additional information about the spatial distribution and arrangement of features within fractals.\\
\indent
In the realm of mathematics, symmetries manifest as invariances, i.e., indicating a lack of change across various operations, such as spatial translations. Mandelbrot first introduced the concept of lacunarity, a parameter designed to quantify the level of translational invariance or homogeneity within a fractal \cite{mandelbrot1982fractal, mandelbrot1994fractal, mandelbrot1995measures}. Lacunarity serves as a metric for how fractal patterns occupy space, with those exhibiting larger voids or lower translational invariance characterized by high lacunarity. Conversely, fractals demonstrating greater homogeneity or nearing translational invariance are associated with low lacunarity. Consequently, even fractals sharing similar construction procedures and identical fractal dimensions may exhibit distinct lacunarity values, dependent on their heterogeneity levels. The lacunarity parameter, therefore, acts as a descriptor of the textural intricacies of a fractal pattern, facilitating the differentiation of sets possessing identical fractal dimensions but distinctive textures. Next, we will discuss the PCP, paving the way for a more elucidated understanding of the concept of lacunarity and also we will discuss how UCP-$\rho_{N}$ potential is a type of polyadic Cantor potential of minimum lacunarity.
\begin{figure}[H]
\begin{center}
\includegraphics[scale=0.56]{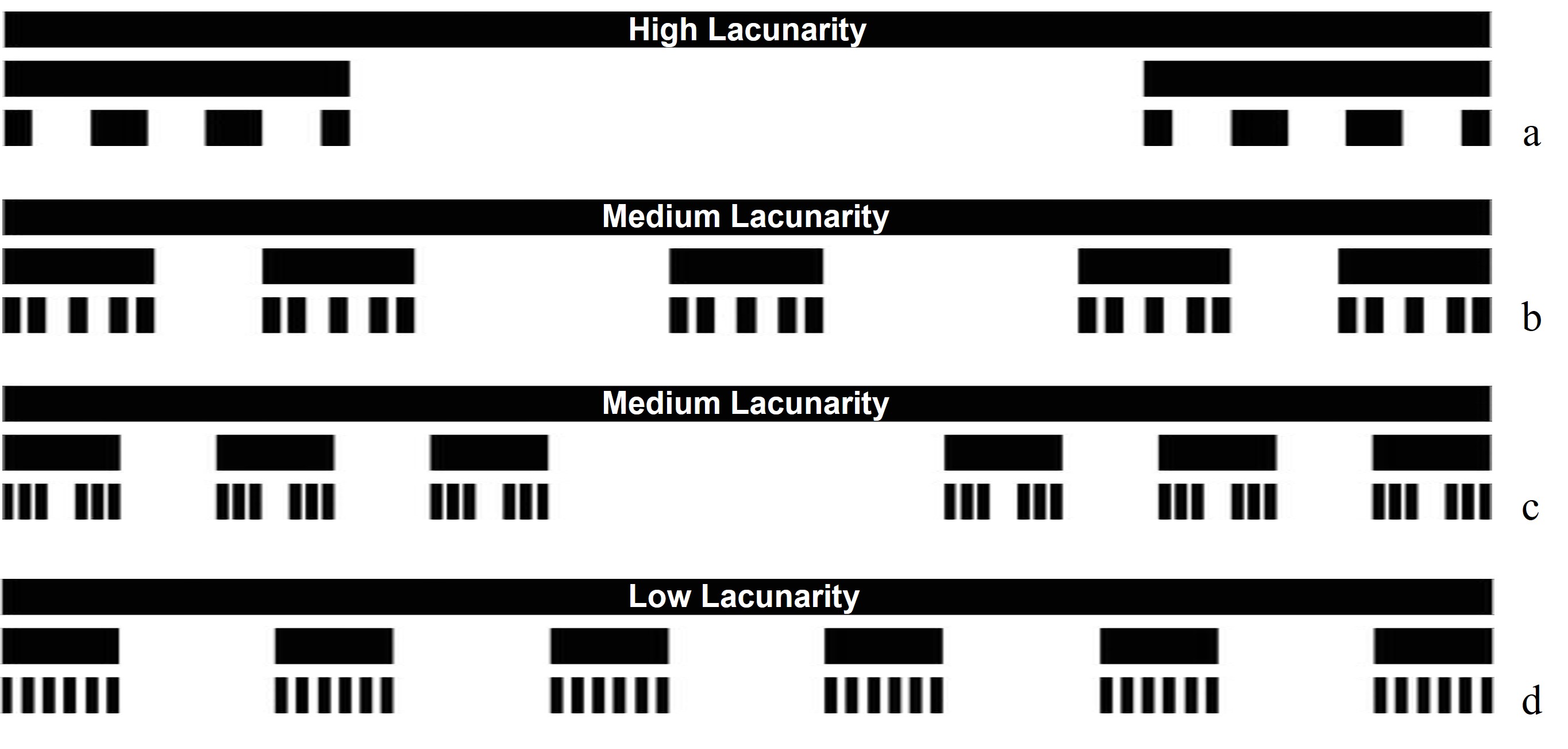} \caption{\textit{In the presented figure, the polyadic Cantor fractal potential system is depicted, categorizing into high, medium, and low lacunarity configurations. This representation predominantly consists of structures with $N=6$ at the initial stage $S=1$, with the sole exception of Fig. b where $N=5$ at the same stage. The high lacunarity structure is characterized by the absence of an outermost gap, while the low lacunarity structure features a symmetrical design where central and adjacent gaps are of identical width. In the medium lacunarity structure, a distinct asymmetry in gap sizes is observed at stage $S=1$,  highlighting the heterogeneity of the fractal potential system. The geometry of the figure is taken from} \cite{castineira2017polyadic}.}
\label{polyadiccantor}
\end{center}
\end{figure}
\indent
GC and SVC-$\rho$ potentials have two ($N=2$) self-similar potentials at stage $S=1$ while the PCP system (sometimes also referred to as Cantor superlattice) is a generalization of the Cantor potential where stage $S=1$ can have more than two self-similar potentials. Some polyadic Cantor structures are shown in the Fig. \ref{polyadiccantor}. A symmetrical PCP system is characterized by three parameters, the number of self-similar potentials $N$, the scaling factor $\zeta$, and the width of the outermost gap $\varepsilon$ at the first stage $S=1$. Much literature refers to the outermost gap at the first stage in fractals as lacunarity parameter and this parameter is sufficient to realize the lacunarity for the polyadic Cantor structure \cite{jaggard1998scattering, monsoriu2004fractal, monsoriu2006quantum, villatoro2008tunneling} (remember that lacunarity parameter $\varepsilon$ is different from lacunarity). Hence, according to the definition of the UCP-$\rho_{N}$ system, the lacunarity parameter will be $\varepsilon=d_{1}$ for each $N$ i.e. $\varepsilon=(d_{1})_{N}$. It is to be noted that the fractal dimension of all PCP systems is independent of the lacunarity parameter $\varepsilon$ as it is another fractal descriptor. Three parameters $(\text{namely }N, \zeta, \varepsilon)$ of a PCP system must satisfy certain constraints to avoid overlapping between the potentials and also to maintain the self-similarity. The maximum value of the scaling factor depends on the value of $N$, such that
\begin{equation}
0<\zeta<\zeta_{max}=\frac{1}{N}.\nonumber
\end{equation}
It is imperative to note that in the discussion of lacunarity, the potential length at stage $S=0$ has been designated as a unit length $(L=1)$. Next, for each $N$ and $\zeta$, there are two extreme values of $\varepsilon$: $\varepsilon_{min}$ and $\varepsilon_{max}$ (see Fig. \ref{epsilon}). The first one is $\varepsilon\equiv \varepsilon_{min}=0$, for which the highest lacunar fractal is obtained, that is, one with the largest possible gap. This implies that the outermost gap (lateral gap) is removed. Thus for an even order (even $N$)
\begin{figure}[H]
\begin{center}
\includegraphics[scale=0.50]{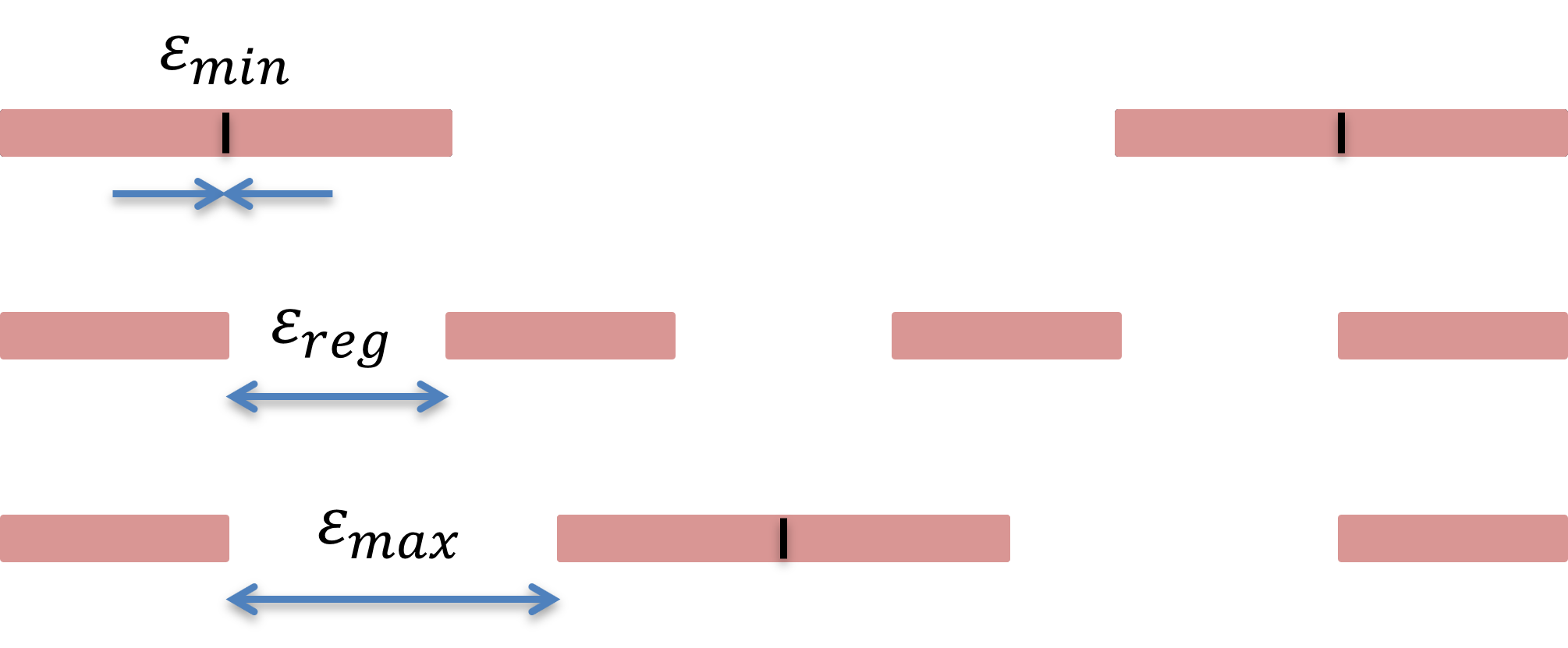}
\caption{\it Illustration of the polyadic Cantor fractal potential of $\varepsilon_{min}=0$ (high or maximum lacunarity fractal), $\varepsilon_{reg}$ (low or minimum lacunarity fractal) and $\varepsilon_{max}$ (medium lacunarity fractal). It is important to note that all layer represents stage $S=1$ with $N=4$. Specifically, in the depiction of the second layer, which is illustrative of minimal lacunarity, there is a noteworthy feature where both the central and adjacent gaps maintain equal widths.}
\label{epsilon}
\end{center}
\end{figure}
\noindent
Cantor potential a central gap remains in the middle and the two identical segments are placed at both sides of the central gap as depicted in Fig. \ref{polyadiccantor}a. This structure is an example of the high (maximum) lacunarity. For an odd order (odd $N$) there is a segment in the center with two identical adjacent gaps. Then two identical segments without lateral gaps are found on both sides of the adjacent gaps as in the even $N$ case, again showing a structure of maximum lacunarity. The other extreme value of lacunarity parameter is $\varepsilon \equiv \varepsilon_{max}$. For even $N$ (Fig. \ref{polyadiccantor}c) the central gap (let us say $g_{c}$) just adjacent to the central segment  has a width equal to 
\begin{equation}
    g_{c}=1-\zeta N,
\end{equation}
and for odd $N$, both large gaps (let us say $g$) surrounding the central rectangular potential (Fig. \ref{polyadiccantor}b) is 
\begin{equation}
    g=\frac{1}{2}(1-\zeta N)=\frac{g_{c}}{2}.
\end{equation} 
In this case, the lacunarity parameter $\varepsilon$ is expressed through \cite{monsoriu2006quantum}
\begin{equation}
\begin{aligned}
\varepsilon \equiv \varepsilon_{max} &= \frac{g_{c}}{N-2},
& \text{even \textit{N}} \\
& = \frac{g_{c}}{N-3},
& \text{odd \textit{N}}
\end{aligned}
\end{equation}
The PCP systems discussed so far have different central and adjacent gaps. Moving on, let us consider the structure shown in Fig. \ref{polyadiccantor}d. There is a normal periodic potential distribution at stage $S=1$, where all gaps including both central and adjacent ones, are of the same width resulting in the lowest possible lacunarity in fractal structure, which produces minimum lacunarity (also evident in Fig. \ref{epsilon}), is expressed through \cite{monsoriu2006quantum}
\begin{equation}
\varepsilon\equiv\varepsilon_{reg} = \frac{g_{c}}{N-1}
\label{45}
\end{equation}
To clarify, the inequality $0=\varepsilon_{min}<\varepsilon_{reg}<\varepsilon_{max}$ is essential and this is well illustrated graphically in Fig. \ref{epsilon}.\\
\begin{figure}[H]
\begin{center}
\includegraphics[scale=0.095]{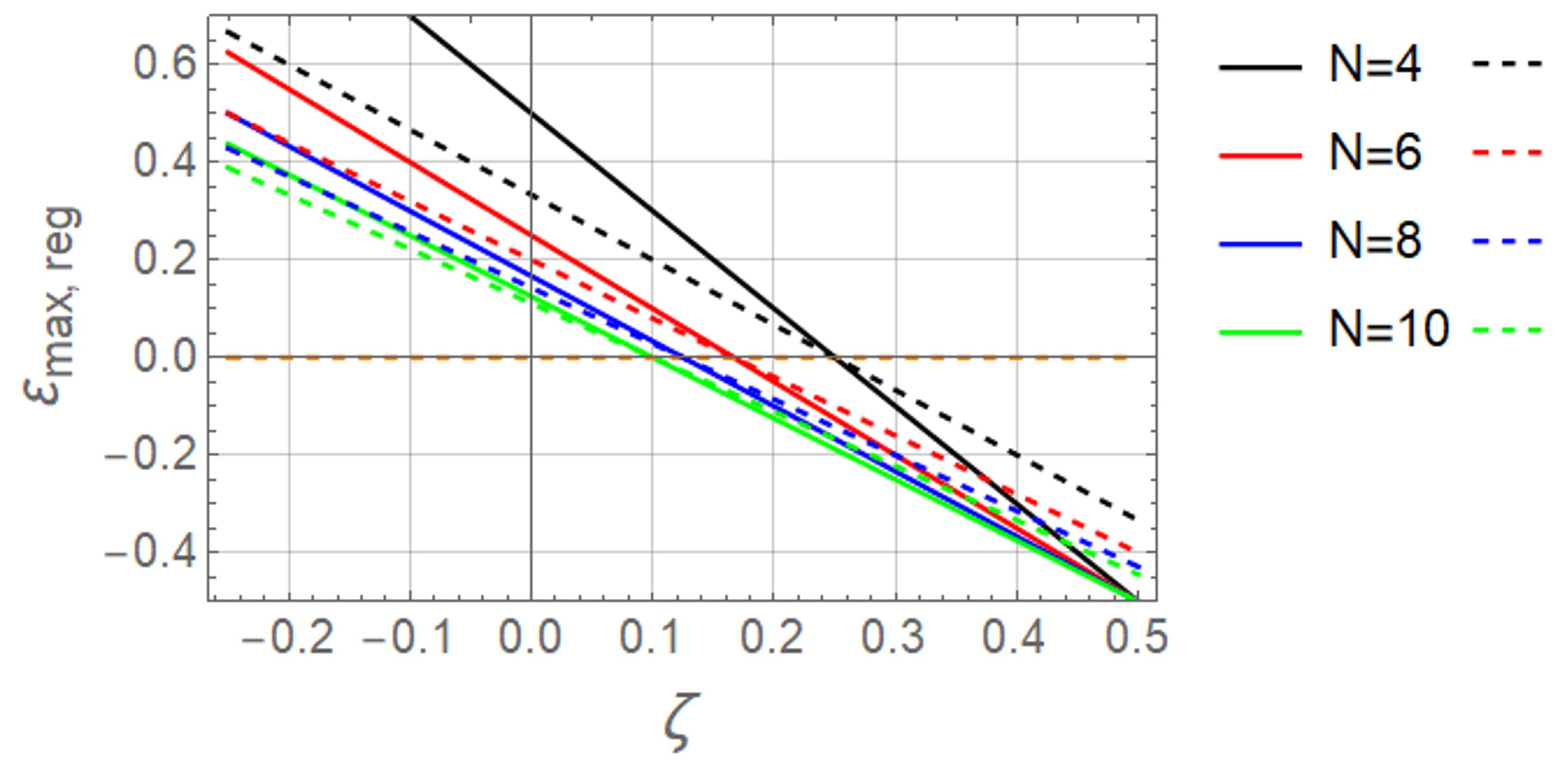} a\\
\includegraphics[scale=0.095]{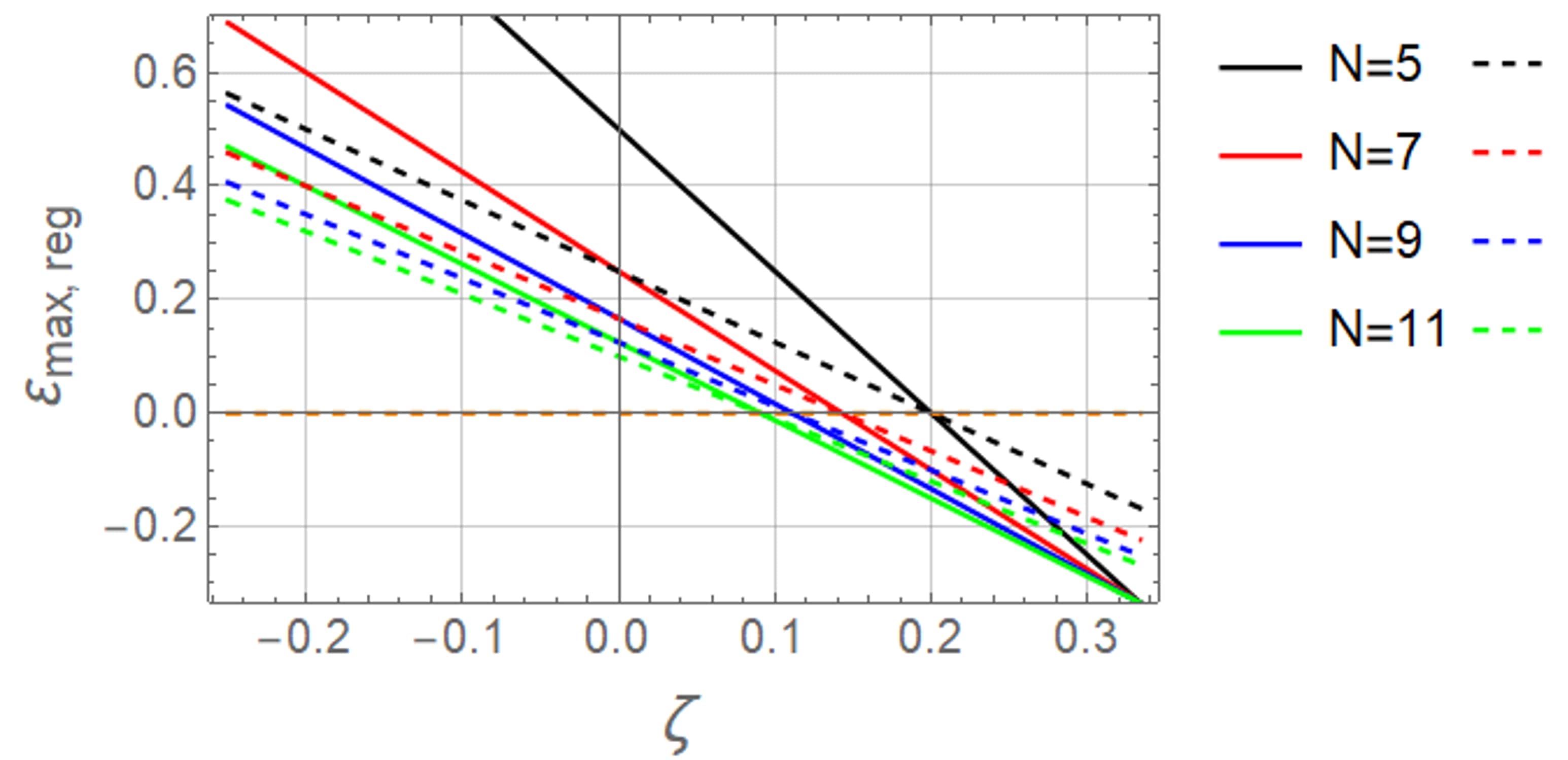} b
\caption{\it 
The presented graphs illustrate the variations of the lacunarity parameter $\varepsilon_{max}$, represented by a dashed curve, and $\varepsilon_{reg}$, depicted as a continuous curve, in relation to the scaling factor $\zeta$. These variations are shown separately for (a) even values of $N$ and (b) odd values of $N$. In both figures, the orange dashed line consistently represents $\varepsilon_{min}=0$. Notably, for a given $N$ and scaling factor $\zeta$, the two extremities in the lacunarity parameter $\varepsilon$, specifically $\varepsilon_{max}$ and $\varepsilon_{min}$, are distinctly observed. for certain fixed values of $N$ and $\zeta$, the lacunarity parameters $\varepsilon_{max}$ and $\varepsilon_{reg}$ can take on negative values. Since $\varepsilon$ represents the width of the outermost gap, negative values are not feasible. Thus, it becomes necessary to consider $\varepsilon$ as its absolute value, $\lvert \varepsilon \rvert$, in cases where it is negative.}
\label{lacunarity_figure02}
\end{center}
\end{figure}
Fig. \ref{lacunarity_figure02} provides a visualization of how the lacunarity parameter behaves under varying conditions. Two distinct types of lacunarity parameters are examined: $\varepsilon_{max}$, which is visualized through a dashed curve, and $\varepsilon_{reg}$, represented by a continuous curve. These parameters are analyzed in the context of their relationship with the scaling factor $\zeta$. The analysis is bifurcated based on the parity of $N$, with one set of plots (a) focusing on even values of $N$ and another set (b) examining odd values of $N$. In both sets of plots, an orange dashed line is consistently used to denote the value of $\varepsilon_{min}=0$. This inclusion is critical as it provides a baseline for comparing the other lacunarity parameters. The graphical representation clearly delineates the two extreme manifestations of the lacunarity parameter $\varepsilon$ for a fixed value of $N$ and the scaling factor $\zeta$. These extremes are encapsulated by $\varepsilon_{max}$, highlighting the medium degree of lacunarity, and $\varepsilon_{min}$, representing the highest degree of lacunarity. This contrast is essential for understanding the range and scope of lacunarity within the studied system. It is important to highlight under certain conditions, specifically for some fixed values of $N$ and $\zeta$, the lacunarity parameters $\varepsilon_{max}$ and $\varepsilon_{reg}$ can assume negative values. However, considering that $\varepsilon$ fundamentally represents the width of the outermost gap, the occurrence of negative values for $\varepsilon$ is not physically meaningful. Therefore, in instances where $\varepsilon$ manifests as a negative quantity, it is essential to interpret $\varepsilon$ in terms of its absolute value, denoted as $\lvert \varepsilon \rvert$, to maintain physical relevance and accuracy in the context of lacunarity measurement.\\
\indent
Summarizing our discussion on lacunarity: a fractal potential structure with a lacunarity parameter $\varepsilon=0$ is characterized by high lacunarity. In contrast, a potential structure with $\varepsilon = \varepsilon_{reg}$ displays the lowest lacunarity, and one where $\varepsilon = \varepsilon_{max}$ has a medium level of lacunarity. It is essential to clarify that the minimum and maximum values of $\varepsilon$ do not correspond to the minimum and maximum lacunarity in the fractal 
structure. Next we will discuss how fractal UCP-$\rho_{N}$ represents a polyadic Cantor potential of low or minimum lacunarity.

%%%%%%%%%%%%%%%%%%%%%%%%%%%%%%%%%%%%%%%%%%%%%%%%%%%%%%%%%%%%%%%%%%%%%%%%%%%%%%

\subsection{Fractal UCP-\texorpdfstring{$\rho_{N}$}{rho-N} system as polyadic Cantor potential of minimum lacunarity}
The PCP system is an interesting extension of the Cantor potential system, resulting in a fractal geometric PCP system. The fractal geometry of the PCP system signifies that self-similarity is maintained at each stage of the system. 
In contrast, the UCP-$\rho_{N}$ system is characterized by a non-fractal geometric structure, primarily due to the absence of self-similarity across its construction stages. This characteristic arises from the construction methodology of the UCP-$\rho_{N}$ system, which includes the removal of a fraction, specifically $\frac{1}{\rho^{\mu+\nu S}}$, from the length $b_{S-1}$ of the potential at stage $S-1$. The term $\frac{1}{\rho^{\mu+\nu S}}$ is a stage $S$ ($S \in \mathbb{N}$) dependent term, having a distinct value at every stage for constant values of $\rho$, $\mu$ and $\nu$. This variation disrupts the preservation of self-similarity throughout the stages, culminating in a non-fractal potential structure. Therefore in the UCP-$\rho_{N}$ system, the self-similarity at each stage will be preserved only if $v=0$. In this case, the removal of the potential segment is exclusively determined by the term $\frac{1}{\rho^{\mu}}$, which remains constant at each stage given fixed values of $\rho$ and $\mu$. When $\nu=0$, let us call this system a fractal UCP-$\rho_{N}$ system (FUCP-$\rho_{N}$ system). It is crucial to recognize that the construction methodologies for the UCP-$\rho_{N}$ and FUCP-$\rho_{N}$ systems are the same.\\
\indent
Next, as elaborated in Section \ref{lacunarity}, the PCP system, despite sharing a common fractal dimension, is differentiated by different lacunarity: minimum, medium, and maximum. Fig. \ref{polyadiccantor} shows that only the PCP system of minimum lacunarity exhibits a periodic distribution of potentials at stage $S=1$, akin to the FUCP-$\rho_{N}$ system at the same stage implying that the construction process of both the systems are same. In contrast, the PCP systems characterized by medium and maximum lacunarity display non-periodic distributions at this stage hence these systems can not be the same as the FUCP-$\rho_{N}$ system. The periodicity at stage $S=1$ suggests a congruence in the geometric constructions of the PCP system with minimum lacunarity and the FUCP-$\rho_{N}$ system, indicating that both configurations adhere to the same underlying fractal geometry. Therefore, it can be inferred that these systems are fundamentally the same (FUCP-$\rho_{N}$ system $=$ PCP system of minimum lacunarity). Thus, for instance, the configuration depicted in Fig. \ref{polyadiccantor}d can be accurately described as a FUCP-$\rho_{6}$ system. 
From the discussion presented, it is evident that the PCP system of minimum lacunarity is a special case of the UCP-$\rho_{N}$ system. Thus, with these detailed explanations, the discussion in this section reaches its completion. Next, we calculate the tunneling probability from the UCP-$\rho_{N}$ system.
%\\
%Next, the PCP system with minimum lacunarity demonstrates a periodic distribution of potential at stage $S=1$, mirroring the distribution found in the FUCP-$\rho_{N}$ system at the same stage, suggesting that the PCP system of minimum lacunarity is fundamentally same as the FUCP-$\rho_{N}$ system. So it is evident that the PCP system of minimum lacunarity, represents a specific case of the UCP-$\rho_{N}$ system. Thus, with these detailed explanations, the discussion in this section reaches its completion. Next, we calculate the tunneling probability from the UCP-$\rho_{N}$ system. Based on these explanations we can say that PCP system of minimum lacunarity is a special case of the UCP-$\rho_{N}$ system. 
\\

\section{Tunneling probability from UCP-\texorpdfstring{$\rho_{N}$}{rho-N} system}
\label{tucp}
In the above section, it has been shown that the UCP-$\rho_{N}$ system is a special case of SPP system with periodic repetitions $N_{j}=N$, $j \in \{1,2,3,...,S \}$. If the transfer matrix of one-dimensional \textit{initiator} or unit cell potential is known
\beq
 \mathbb{M}(k)= \begin{pmatrix}   \mathbb{M}_{11}(k) & \mathbb{M}_{12}(k) \\ \mathbb{M}_{21}(k) & \mathbb{M}_{22}(k) \end{pmatrix},
\eeq
then the transmission probability of SPP of order $S$ is given by \cite{hasan2018super},
\begin{equation}
		T(N_{1},N_{2},....,N_{S})=\frac{1}{1+\left[|\mathbb{M}_{12}|U_{N_{1}-1}(\Gamma_{1})U_{N_{2}-1}(\Gamma_{2})U_{N_{3}-1}(\Gamma_{3})........U_{N_{S}-1}(\Gamma_{S})\right]^{2}}. 
		\label{t_spp}
	\end{equation}
Where $N_{1}$, $N_{2}$,....,$N_{S}$ is the number of periodicity of the unit cell potential of order $1$, $2$,....,$S$ respectively. $U_{N}(\Gamma)$ is the Chebyshev polynomial of second kind \cite{abramowitz1968handbook}. Various $\Gamma$s appearing in the above equation are the arguments of the Chebyshev polynomial which represents the Bloch phases of the corresponding fully developed periodic systems. The expression for $\Gamma_{q}$, $ q \in \{1,2,3,.....,S \}$ is given by \cite{hasan2018super, narayan2023tunneling}, 
\begin{equation}
    \Gamma_{q}(k)=\widetilde{\Gamma}_{a}(k)+\widetilde{\Gamma}_{b}(k),
\end{equation}
where, $\widetilde{\Gamma}_{a}(k)$ and $\widetilde{\Gamma}_{b}(k)$ are expressed through
\begin{equation}
    \widetilde{\Gamma}_{a}(k) = \vert \mathbb{M}_{22} \vert \cos \left[\tau - k \left \{ \sum_{p=1}^{q-1}(N_{p}-1)r_{p} - r_{q}\right \} \right ] {\prod_{p=1}^{q-1}U_{N_{p}-1}(\Gamma_{p})}
\end{equation}
and 
\begin{equation}
    \widetilde{\Gamma}_{b}(k) = \sum_{h=1}^{q-1}\left [\cos \left\{k\left({\sum_{p=h}^{q-1}N_{p}r_{p}} - {\sum_{p=h+1}^{q}r_{p}}\right)\right\}U_{N_{h}-2}(\Gamma_{h}){\prod_{p=h+1}^{q-1}U_{N_{p}-1}(\Gamma_{p})}\right ]
\end{equation}
respectively. 
Where, $\tau$ represents the argument of $\mathbb{M}_{22}(k)$. The expression is also valid for $q$ = $1$ and $2$ provided we drop the terms when the running variable is more than the upper limit for the summation symbols and take the terms as unity when the running variable is more than the upper limit for the product symbols. Also in the above series $N_{0}=1$ and $r_{0}=0$. As we know, for UCP-$\rho_{N}$ system, $N_{q}=N$, $ q \in \{1,2,3,...,S \}$ therefore $\widetilde{\Gamma}_{a}(k)$ and $\widetilde{\Gamma}_{b}(k)$ can be further simplified as 

\begin{equation}
    \widetilde{\Gamma}_{a}(k) = \vert \mathbb{M}_{22} \vert \cos \big\{\tau-k\chi_{1}(q)\big\} {\prod_{p=1}^{q-1}U_{N-1}(\Gamma_{p})}
\end{equation}
and 
\begin{equation}
    \widetilde{\Gamma}_{b}(k) = \sum_{h=1}^{q-1}\left \{\cos \left\{k\chi_{2}(q,h)\right\}U_{N-2}(\Gamma_{h}){\prod_{p=h+1}^{q-1}U_{N-1}(\Gamma_{p})}\right \}
\end{equation}
respectively. Where $\chi_{1} (q)$ and $\chi_{2}(q,h)$ is given by,
\begin{equation}
\chi_{1}(q) = \left\{\sum_{p = 1}^{q-1}(N-1)r_{p}\right\}-r_{q},
\label{eta1}
\end{equation}
\begin{equation}
\chi_{2}(q,h)  =\left\{\sum_{p = r}^{q}(N-1)r_{p}\right\}-( Nr_{q}-r_{h}).
\end{equation}
Further, it can be easily shown that
\begin{equation}
\chi_{2}(q,h)  =\chi_{1}(q) -\chi_{1}(h).
\label{gamma2}
\end{equation}
Next, let us find the simplified expression for $\chi_{1}(q)$ and $\chi_{2}(q, h)$. It is evident from the geometries of Fig. \ref{fractal_figure} and \ref{construction}, that
\begin{equation}
    r_{1}(N) = (N-1)r_{0}+b_{S} + d_{S}, \nonumber
\end{equation}
\begin{equation}
    r_{2}(N) = (N-1)r_{1} + b_{S} + d_{S-1}, \nonumber
\end{equation}
\begin{equation}
    r_{3}(N) = (N-1)(r_{1} + r_{2}) + b_{S} + d_{S-2}, \nonumber
\end{equation}
\begin{equation}
    r_{4}(N) = (N-1)(r_{1} + r_{2}+r_{3}) + b_{S} + d_{S-3}. \nonumber
\end{equation}
In general, the periodic distance for each stage $S$ can be expressed through
\begin{equation}
r_{n} = \left(\sum_{p = 1}^{n-1}r_{p}\right)+b_{S} + d_{S-n+1}.
\end{equation}
Using above equation in Eq. (\ref{eta1}), $\chi_{1}(q)$ is expressed through,
\begin{equation}
    \chi_{1}(q) = -(b_{S} + d_{S-q+1}),
\label{gamma1}
\end{equation}
which further can be expressed as
\begin{equation}
    \chi_{1}(q) = -\left(b_{S} + \frac{b_{S-q}}{\rho^{\mu+\nu(S-q+1)}}\right)
\label{52}
\end{equation}
which implies that $\chi_{1}(q)$ is always negative. Now using Eq. (\ref{52}) in Eq. (\ref{gamma2}), we get
\begin{equation}
    \chi_{2}(q,h)=\frac{b_{S-h}}{\rho^{\mu+\nu(S-h+1)}}-\frac{b_{S-q}}{\rho^{\mu+\nu(S-q+1)}}.
\end{equation}
Further, above equation can be expressed as
\begin{equation}
    \chi_{2}(q, h) = d_{S-h+1} - d_{S-q+1}.
    \label{gamma22}
\end{equation}
Upon examination of the geometrical representation depicted in Fig. \ref{fractal_figure} and \ref{construction}, it is discernible that in instances when $x>y$ in magnitude, a corollary relationship manifests as $d_{x}<d_{y}$. Concurrently, in scenarios where $h<q$, it logically ensues that $S-h+1>S-q+1$, which inherently implies that $d_{S-h+1}<d_{S-q+1}$. Thus, for conditions satisfying $h<q$, it is inferable that $\chi_{2}(q, h)<0$. The formulations presented in Eqs. (\ref{gamma1}) and (\ref{gamma22}) provide the foundational expressions for $\chi_{1}(q)$ and $\chi_{2}(q, h)$ respectively.\\
\indent
The above discussion completes the calculation for $\Gamma_{q}(k)$. With the knowledge of  $\Gamma_{1}(k)$, $\Gamma_{2}(k)$, $\Gamma_{3}(k)$,.....,$\Gamma_{S}(k)$ one can find the transmission probability by using Eq. (\ref{t_spp}) with $N_{q}=N$. Next, the transfer matrix of \textit{unit cell} rectangular barrier of width $b_{S}$ and height $V$ is \cite{griffiths2001waves},
\begin{subequations}
		\begin{equation}
			\mathbb{M}_{11}=\frac{\cos{\widetilde{k} b_{S}}-i\sigma_{+} \sin{\widetilde{k} b_{S}}}{e^{-ikb_{S}}},
		\end{equation}
		\begin{equation}
			\mathbb{M}_{12}=i\sigma_{-} \sin{\widetilde{k} b_{S}},
		\end{equation}
		\begin{equation}
			\mathbb{M}_{21}=-i\sigma_{-} \sin{\widetilde{k} b_{S}},
		\end{equation}
		\begin{equation}
			\mathbb{M}_{22}=\frac{\cos{\widetilde{k} b_{S}}+i\sigma_{+} \sin{\widetilde{k} b_{S}}}{e^{ikb_{S}}}.
		\end{equation}
	\end{subequations}
	Where,
	\begin{equation}
		\sigma_{\pm}=\frac{1}{2} \left( \frac{k}{\widetilde{k}}\pm \frac{\widetilde{k}}{k}\right), 
		\label{epsilon_plus_minus}
	\end{equation}
	\begin{equation}
		k=\frac{\sqrt{2mE}}{\hbar} \ \ , \ \ \widetilde{k}=\frac{\sqrt{2m(E-V)}}{\hbar}. \nonumber
	\end{equation}
With $N_{q}=N$ and using the knowledge of the transfer matrix of \textit{unit cell} rectangular barrier, we simplify Eq. (\ref{t_spp}) to obtain the final expression of transmission coefficient for UCP-$\rho_{N}$ system as,
\begin{equation}
T_{S}(k, N)\equiv T(k) =\frac{1}{1+\sigma_{-}^{2}\sin^{2}{(\widetilde{k} b_{S})} \left[\prod_{q=1} ^{S}U_{N-1}(\Gamma_{q})\right]^{2}}.
\label{T_s}
\end{equation}

%%%%%%%%%%%%%%%%%%%%%%%%%%%%%%%%%%%%%%%%%%%%%%%%%%%%%%%%%%%%%%%%%%%%%%

\section{Transmission features}
\label{tfeatures}
The analytical derivation of the transmission coefficient for the UCP-$\rho_{N}$ system was outlined in the preceding section. This segment of our study shifts focus towards examining the transmission characteristics across the spectrum of the UCP-$\rho_{N}$ configurations. Given the comprehensive analysis of the  UCP-$\rho_{2}$ system already provided in \cite{umar2023quantum}, our investigation will predominantly concentrate on exploring the transmission properties of the remaining variants within the  UCP-$\rho_{N}$ system. This endeavor aims to elucidate the distinct transmission phenomena and underlying mechanics specific to each variant, thereby broadening our understanding of the comprehensive transmission dynamics of the UCP-$\rho_{N}$ system. Through this inquiry, we seek to delineate how these variants differ in affecting the transmission spectra of the system, thereby providing a clearer picture of this model.
\begin{figure}[H]
\begin{center}
\includegraphics[scale=0.31]{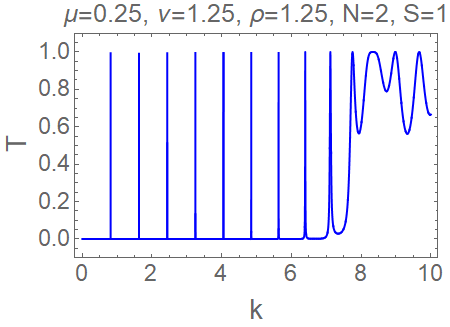} a
\includegraphics[scale=0.31]{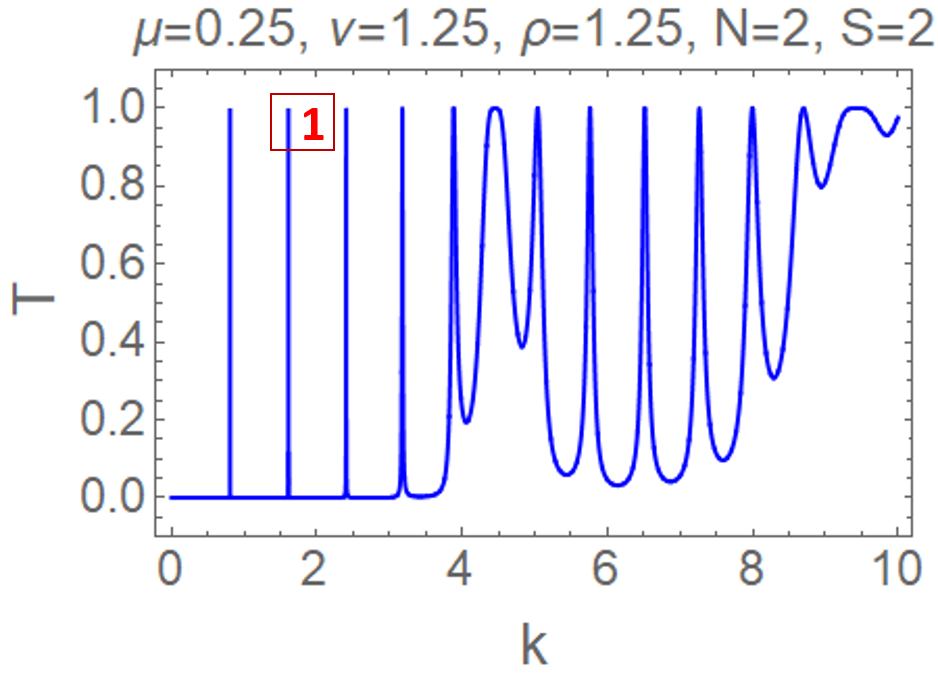} b
\includegraphics[scale=0.31]{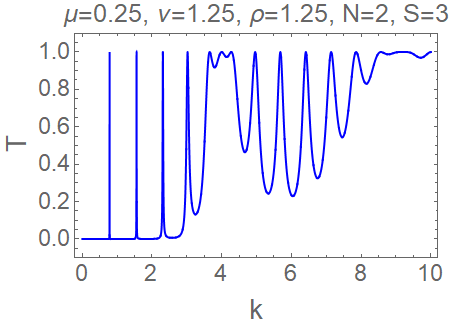} c\\
\includegraphics[scale=0.31]{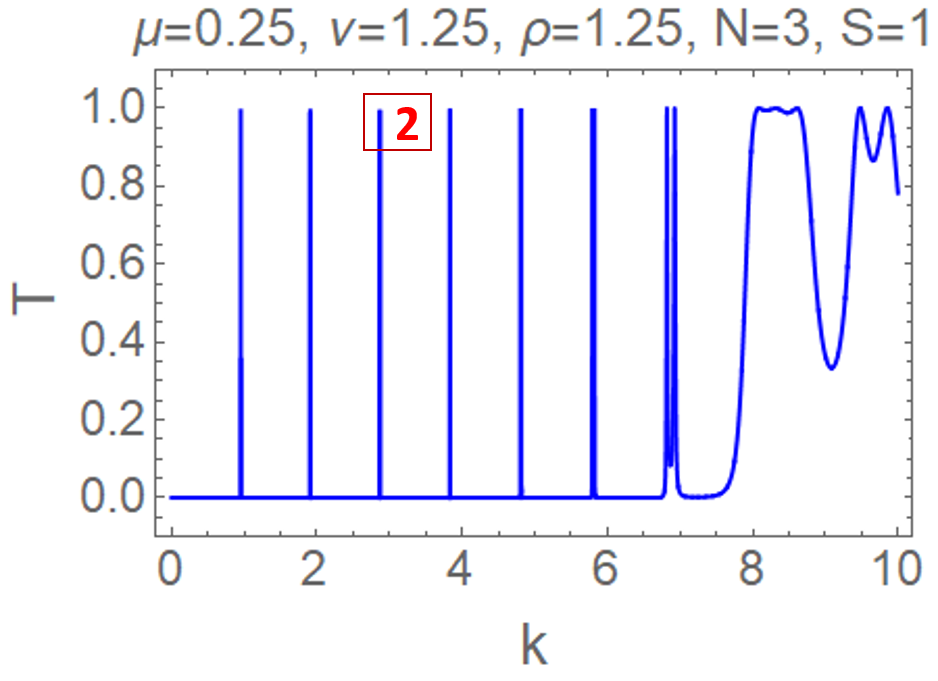} d
\includegraphics[scale=0.31]{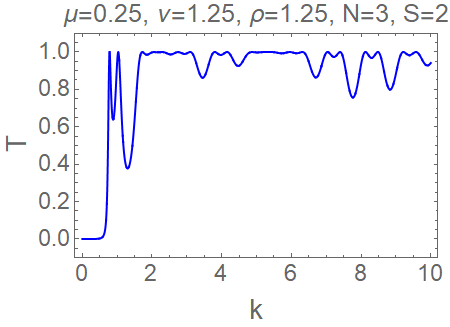} e
\includegraphics[scale=0.31]{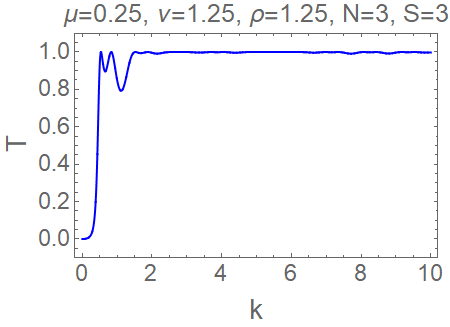} f\\
\includegraphics[scale=0.31]{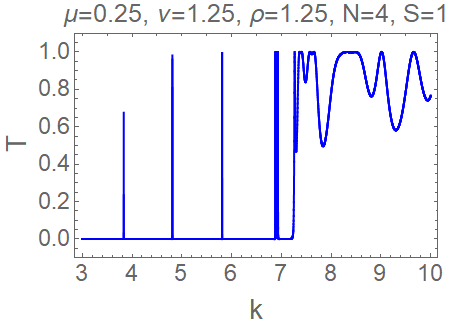} g
\includegraphics[scale=0.31]{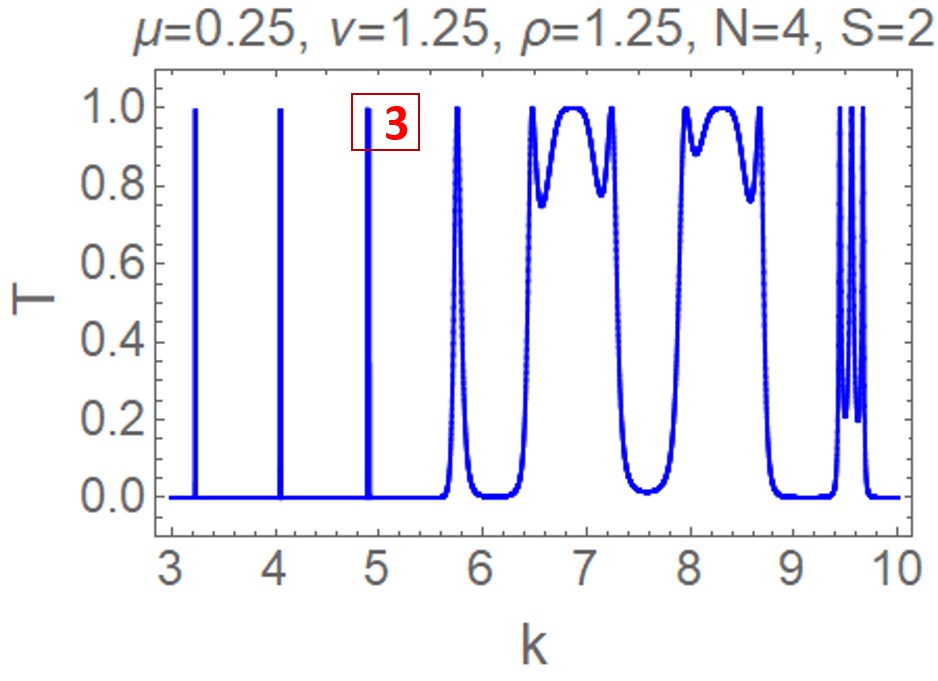} h
\includegraphics[scale=0.31]{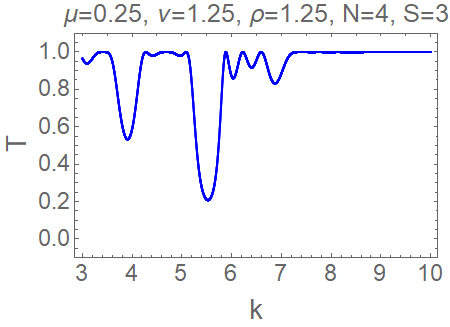} i \\
\includegraphics[scale=0.31]{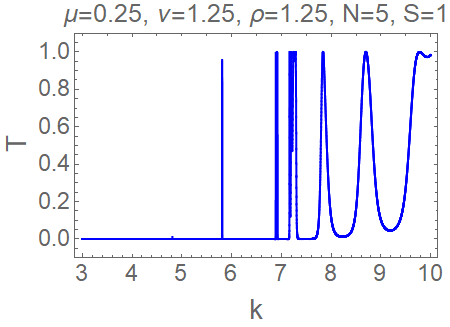} j
\includegraphics[scale=0.31]{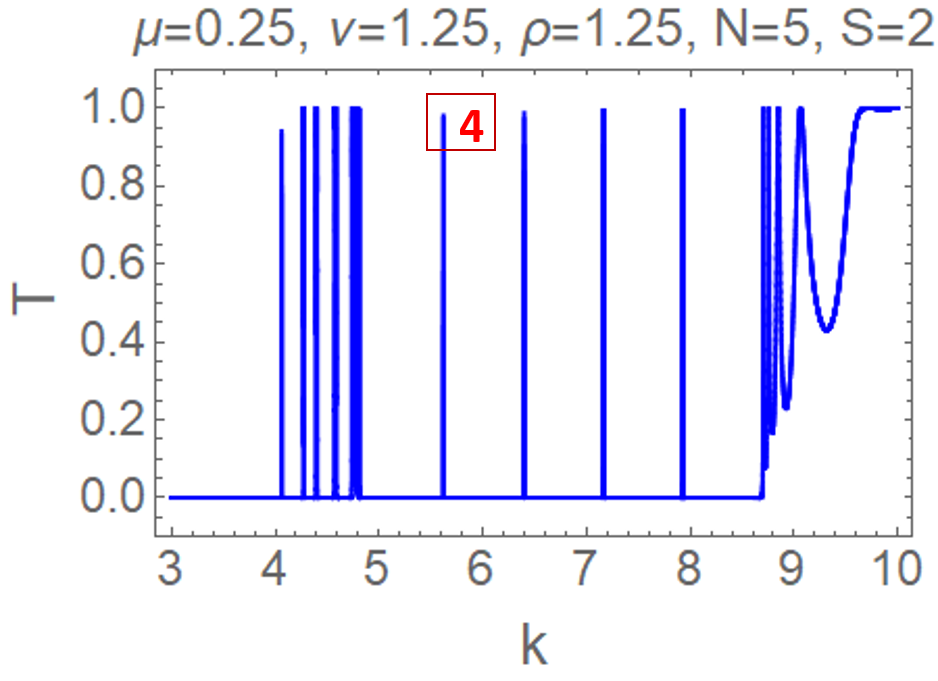} k
\includegraphics[scale=0.31]{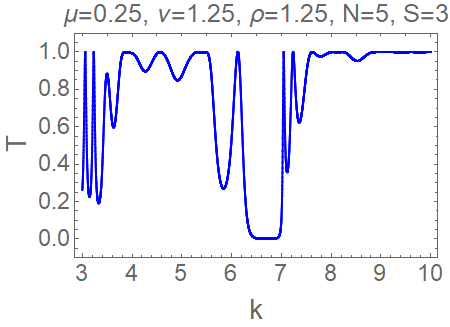} l
\caption{\it Plots showing the transmission profile for the UCP-$\rho_{2}$ (first row), UCP-$\rho_{3}$ (second row), UCP-$\rho_{4}$ (third row) and UCP-$\rho_{5}$ (fourth row) systems of stage $S=1$, $2$ and $3$. These plots are crafted for $L=5$ and $V=25$. Other parameters are mentioned at the top of each figure. Very sharp transmission resonances are observed here. In each row, a peak is designated as $1$, $2$, $3$, and $4$, respectively. This labeling serves to identify these peaks for subsequent magnification and detailed presentation in Fig. $\ref{transmission02}$.}
\label{transmission01}
\end{center}
\end{figure}
The transmission coefficient profiles for the UCP-$\rho_2$, UCP-$\rho_3$, UCP-$\rho_4$ and UCP-$\rho_5$ systems are depicted graphically in Fig. \ref{transmission01}. Within these graphical representations, there is a noticeable emergence of exceptionally sharp resonances, particularly at lower $k$ values. Here, the sharp resonances in the transmission coefficient, $T_{S}(k, N) \equiv T(k)$, are delineated by profound troughs, leading to intervals in the $k$ (null region) dimension where $T(k)$ approaches zero. This phenomenon is of particular interest because, in the context of any Hermitian potential as is the case here, the transmission coefficient is theoretically never precisely zero \cite{christodoulides2018parity}.
A significant characteristic of many transmission resonances within these systems is their acute sharpness, manifesting as abrupt transitions from $T(k)=0$ to $T(k)=1$. 
Detailed analysis of specific pronounced peaks, denoted as
\begin{figure}[H]
\begin{center}
\includegraphics[scale=0.475]{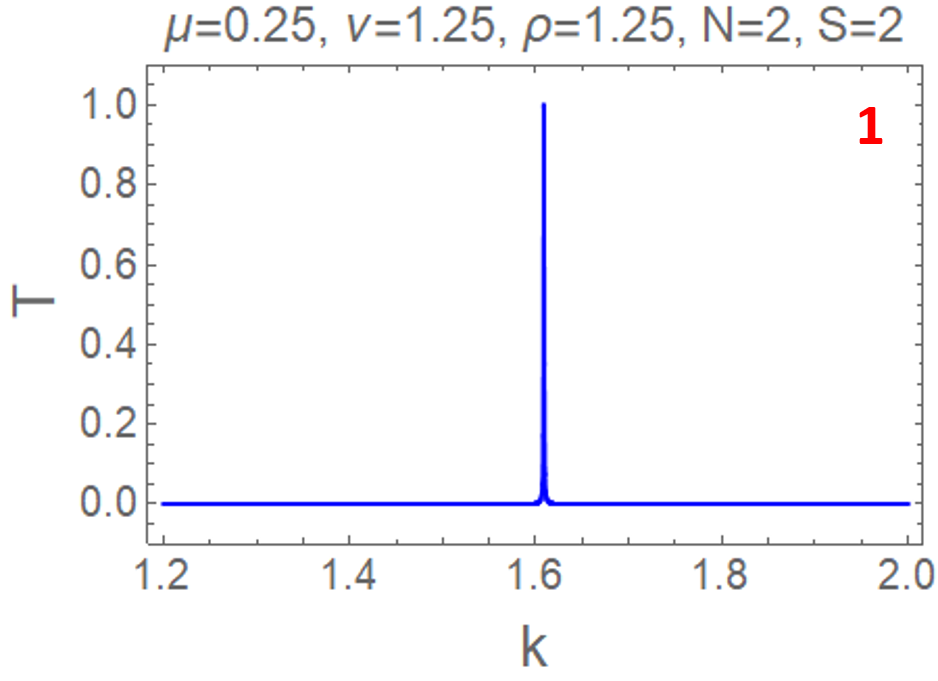} a
\includegraphics[scale=0.475]{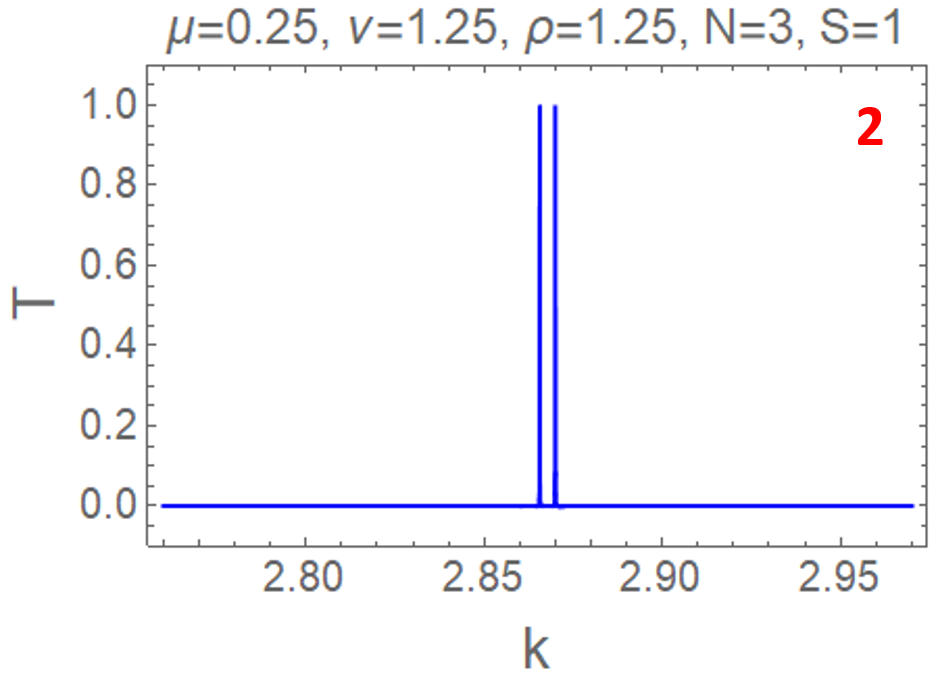} b\\
\includegraphics[scale=0.471]{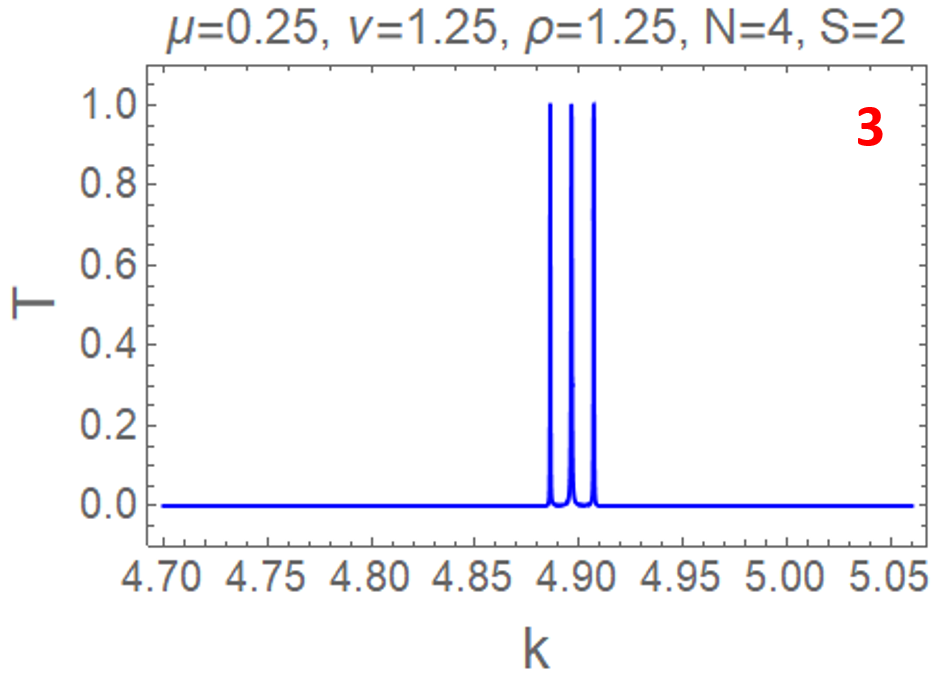} c
\includegraphics[scale=0.485]{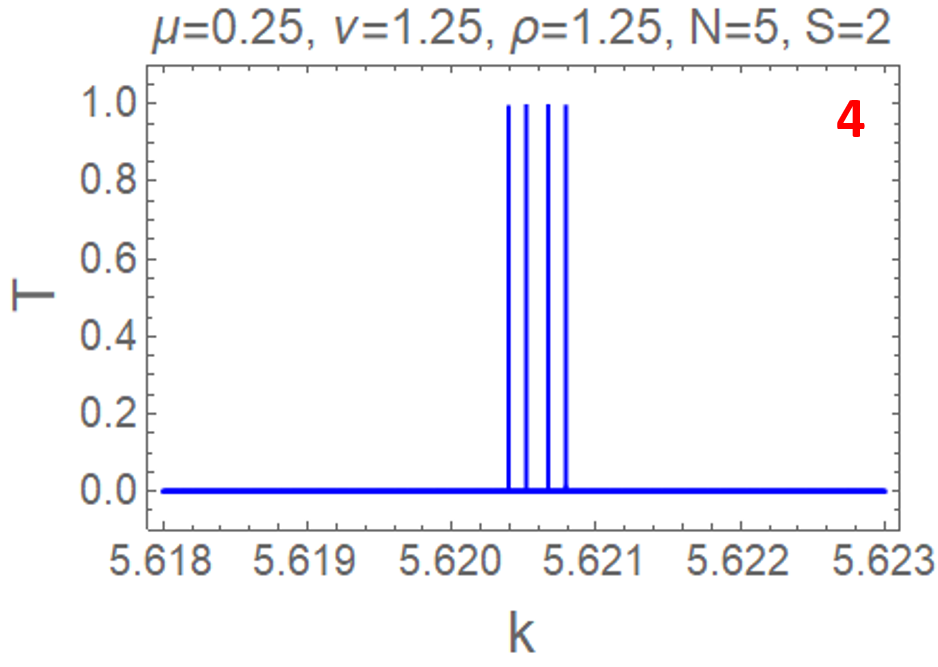} d
\caption{\it A zoomed view of the peaks identified as $1$, $2$, $3$ and $4$ in Fig. $\ref{transmission01}$ is presented, 
maintaining the same potential parameters as outlined in the previous figure. Magnified examination elucidates that these peaks are not monolithic entities as appeared in Fig. $\ref{transmission01}$. Instead, they are composed of one, two, three, and four discrete, exceedingly sharp peaks, correspondingly. This observation elucidates that the phenomena perceived as a singular, coherent transmission peak in Fig. $\ref{transmission01}$ manifest as a series of $N-1$ distinct, sharp peaks for each specified UCP-$\rho_{N}$ configuration. Additionally, magnification of peaks reveals a progressive intensification of peak acuity corresponding to ascending values of $N$.}
\label{transmission02}
\end{center}
\end{figure}
\noindent
$1$, $2$, $3$ and $4$ in Figs. \ref{transmission01}b, \ref{transmission01}d, \ref{transmission01}h and \ref{transmission01}k, respectively, are depicted in Fig. \ref{transmission02}, providing an insight into 
their exceptionally sharp characteristics. These identified peaks, labeled $1$, $2$, $3$ and $4$, correspond to the transmission resonances for systems characterized by $N=2$, $3$, $4$ and $5$, respectively. Closer inspection of these peaks, as shown in Fig. \ref{transmission02}, reveals that they are not singular but consist of one, two, three, and four distinct very sharp peaks, respectively. Thus, this enhanced perspective demonstrates that what appears to be a single, distinct transmission peak in Fig. \ref{transmission01} comprises $N-1$ distinct sharp peaks for a given UCP-$\rho_{N}$ system. 
Moreover, Fig. \ref{transmission02} demonstrates that the sharpness of the peaks enhances as $N$ progresses, signifying that the parameter $N$, which denotes the number of potential segments at stage $S=1$, plays a crucial role in determining the peak sharpness. These observation underscores the complex nature of the transmission phenomenon, highlighting the intricate layering of resonance effects within these systems. The revelation of this composite structure, comprised of multiple closely spaced sharp peaks, emphasizes the nuanced and layered complexity inherent in the transmission process, challenging simpler interpretations and necessitating a more nuanced analysis.
\begin{figure}[H]
\begin{center}
\includegraphics[scale=0.48]{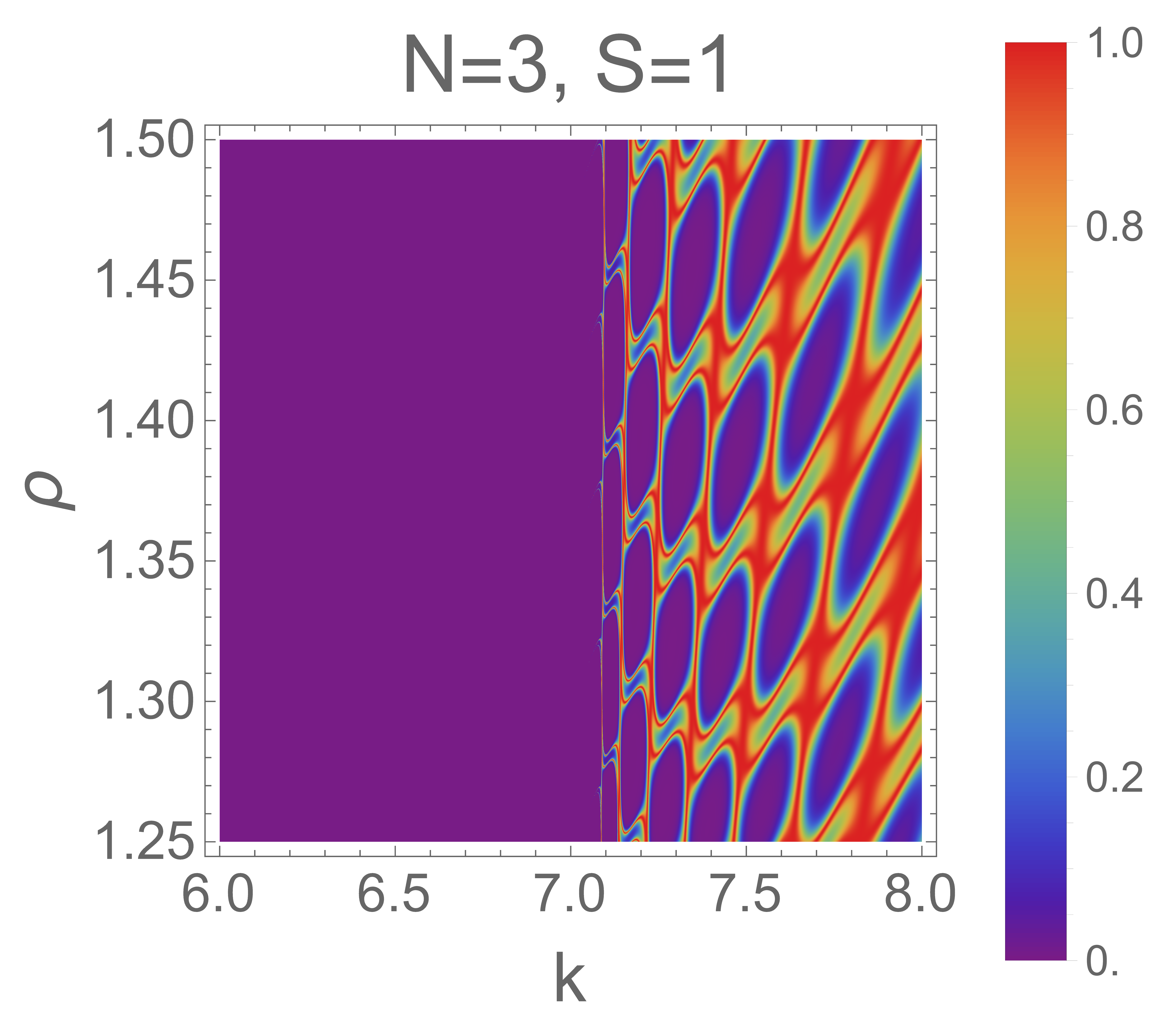} a
\includegraphics[scale=0.48]{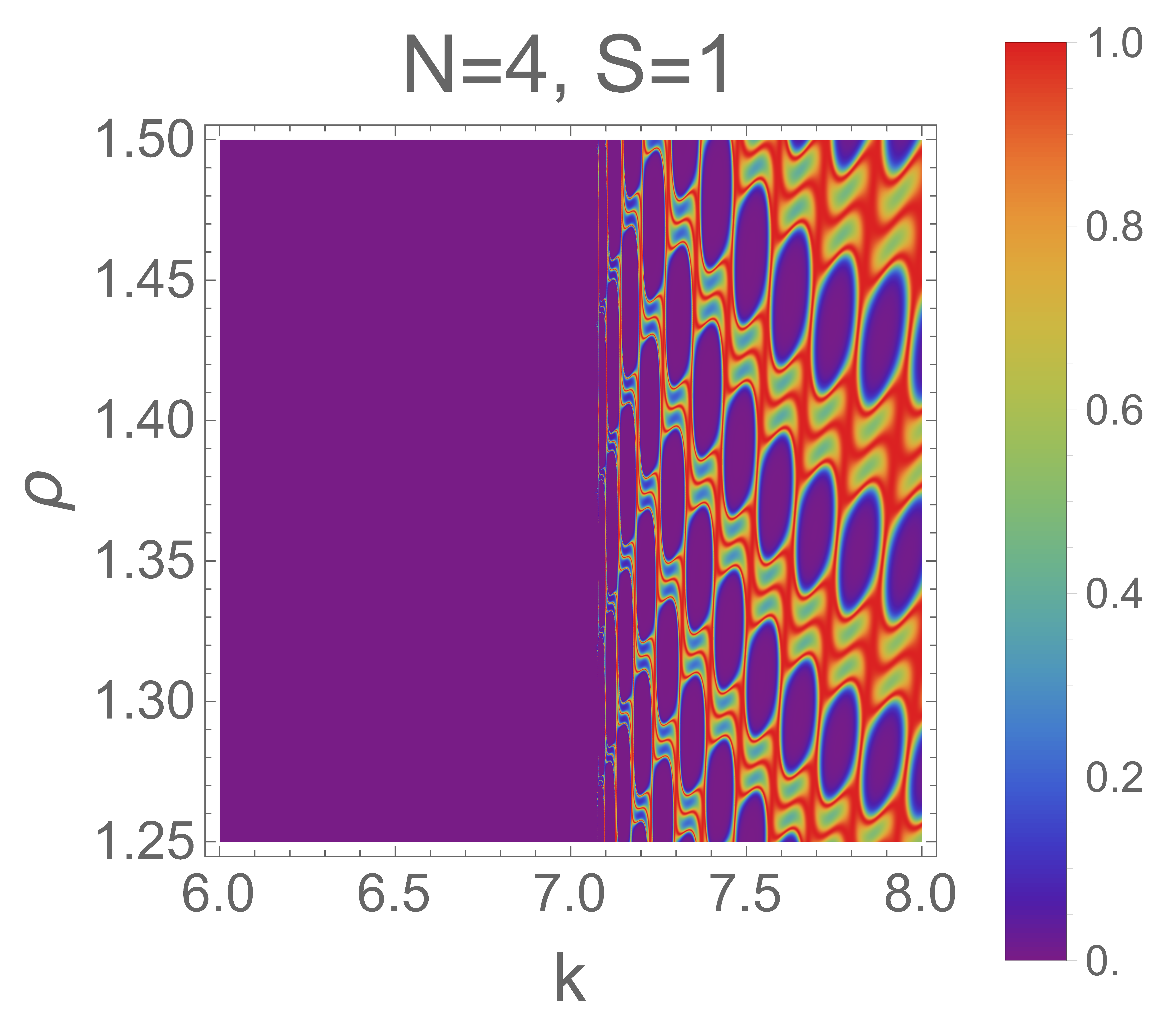} b \\
\includegraphics[scale=0.48]{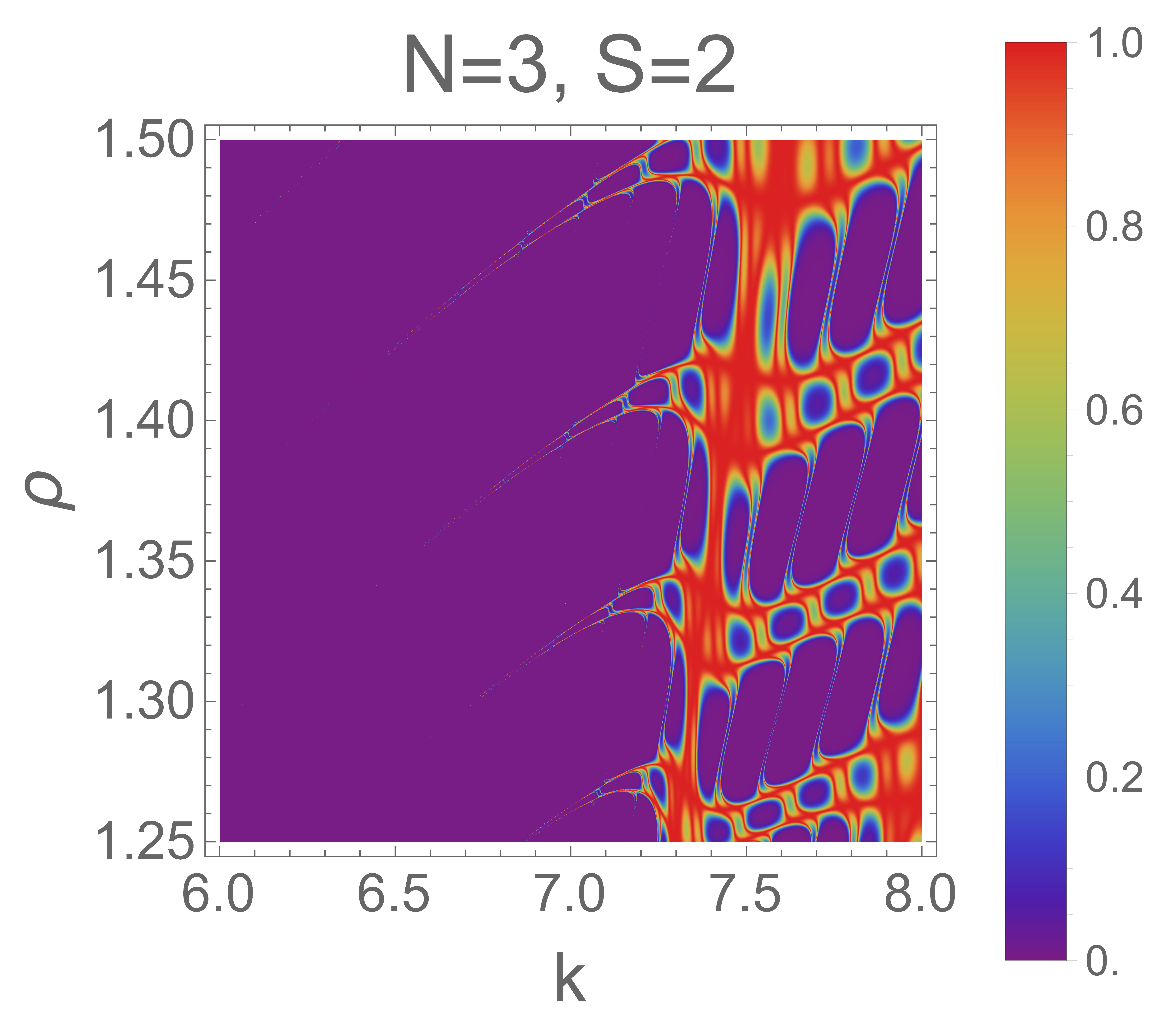} c 
\includegraphics[scale=0.48]{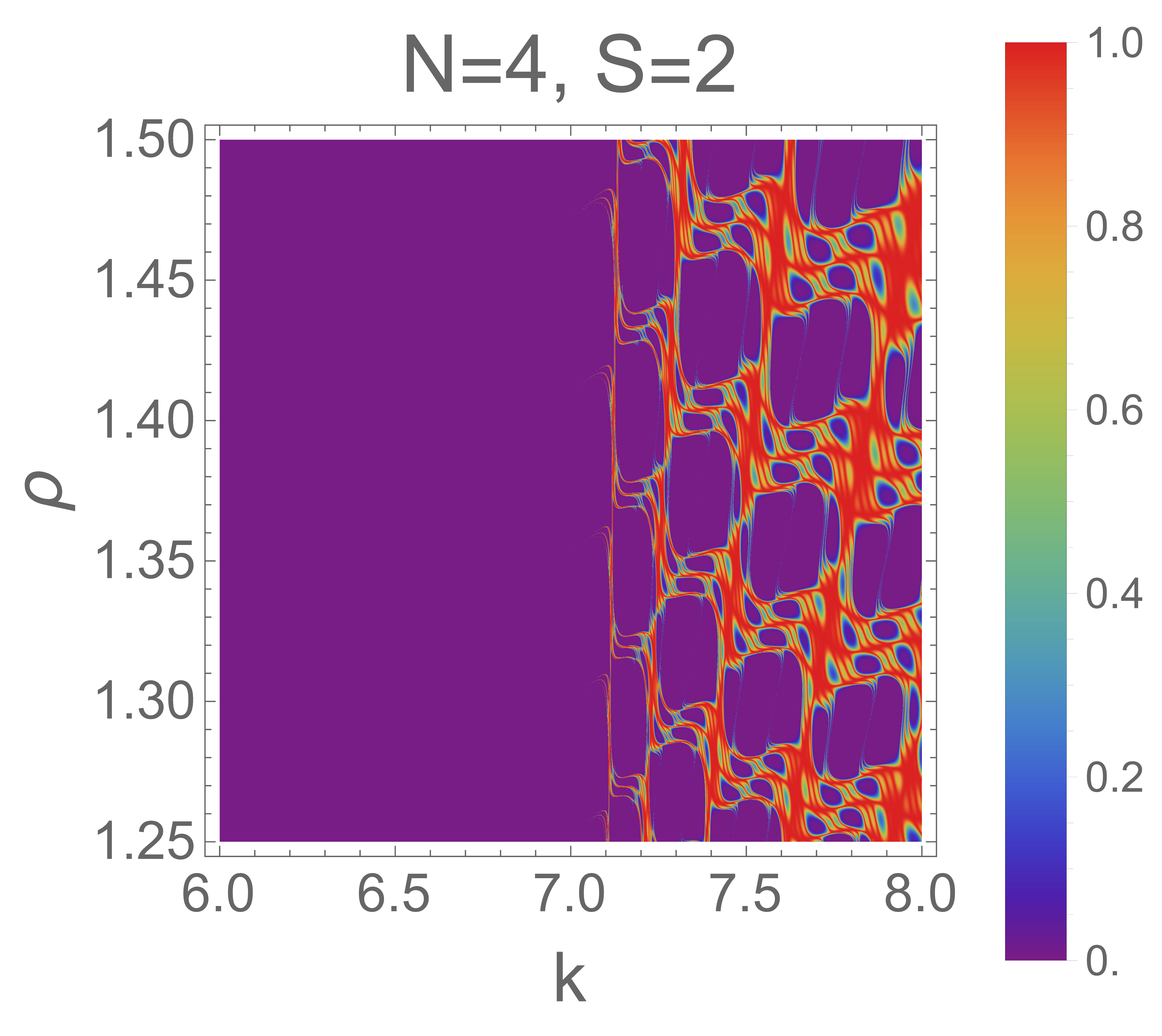} d \\
\includegraphics[scale=0.48]{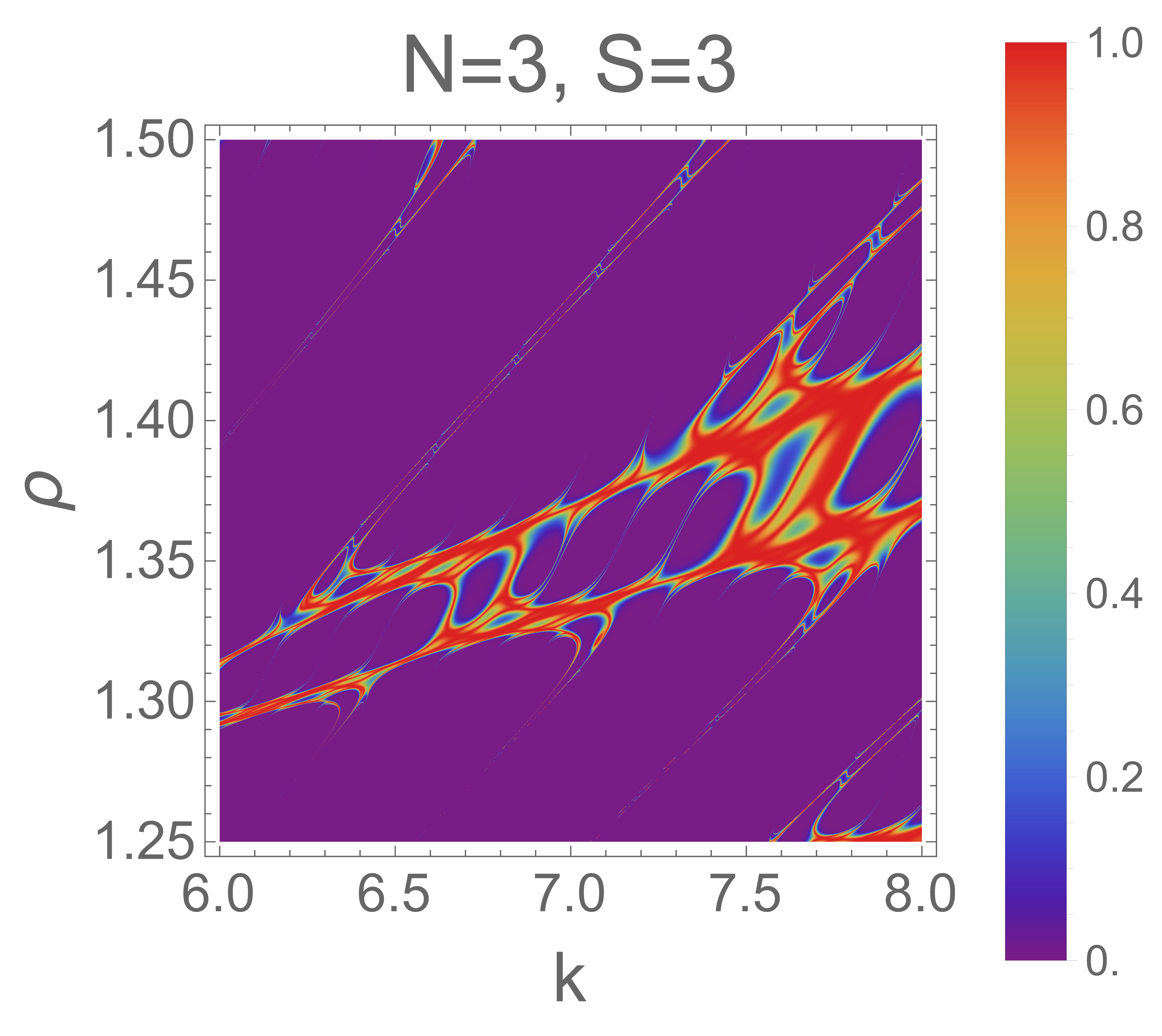} e
\includegraphics[scale=0.48]{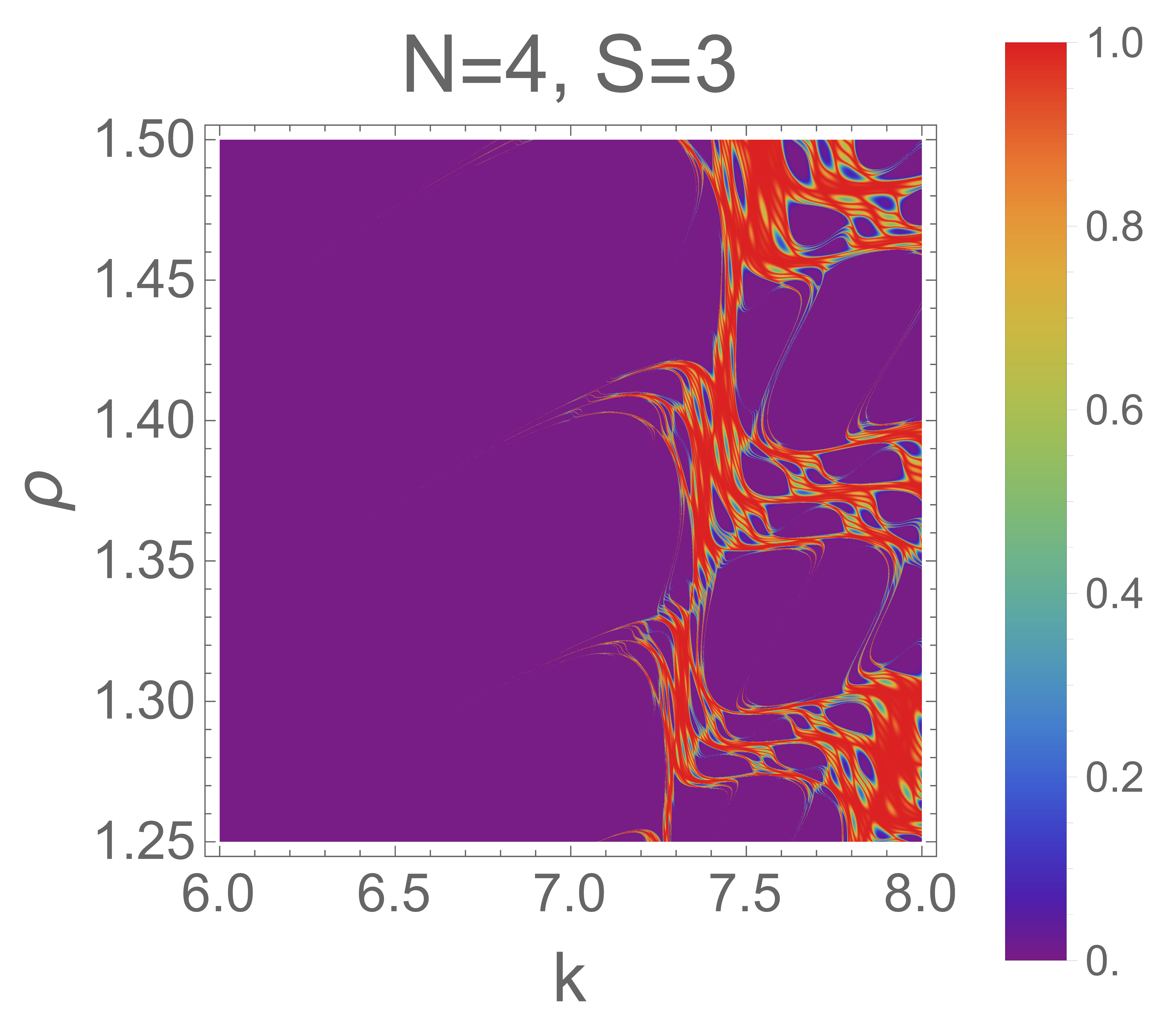} f 
\label{nu0polyadic}
\caption{\textit{Density plots showing the distribution of transmission coefficient in the $\rho-k$ plane for PCP system ($\nu=0$ case) characterized by the parameter $N=3$ (first column) and $N=4$ (second column), across various stages from $S=1$ to 3. Here the potential parameters are $\mu=0.5$, $L=25$ and $V=25$. Very sharp transmission resonances are observed between the large null region, a region in the $\rho-k$ plane where $T(k)=0$ (zero transmission is never possible for Hermitian system} \cite{christodoulides2018parity}\textit{)}.}
\label{transmission03}
\end{center}
\end{figure}
\begin{figure}[H]
\begin{center}
\includegraphics[scale=0.48]{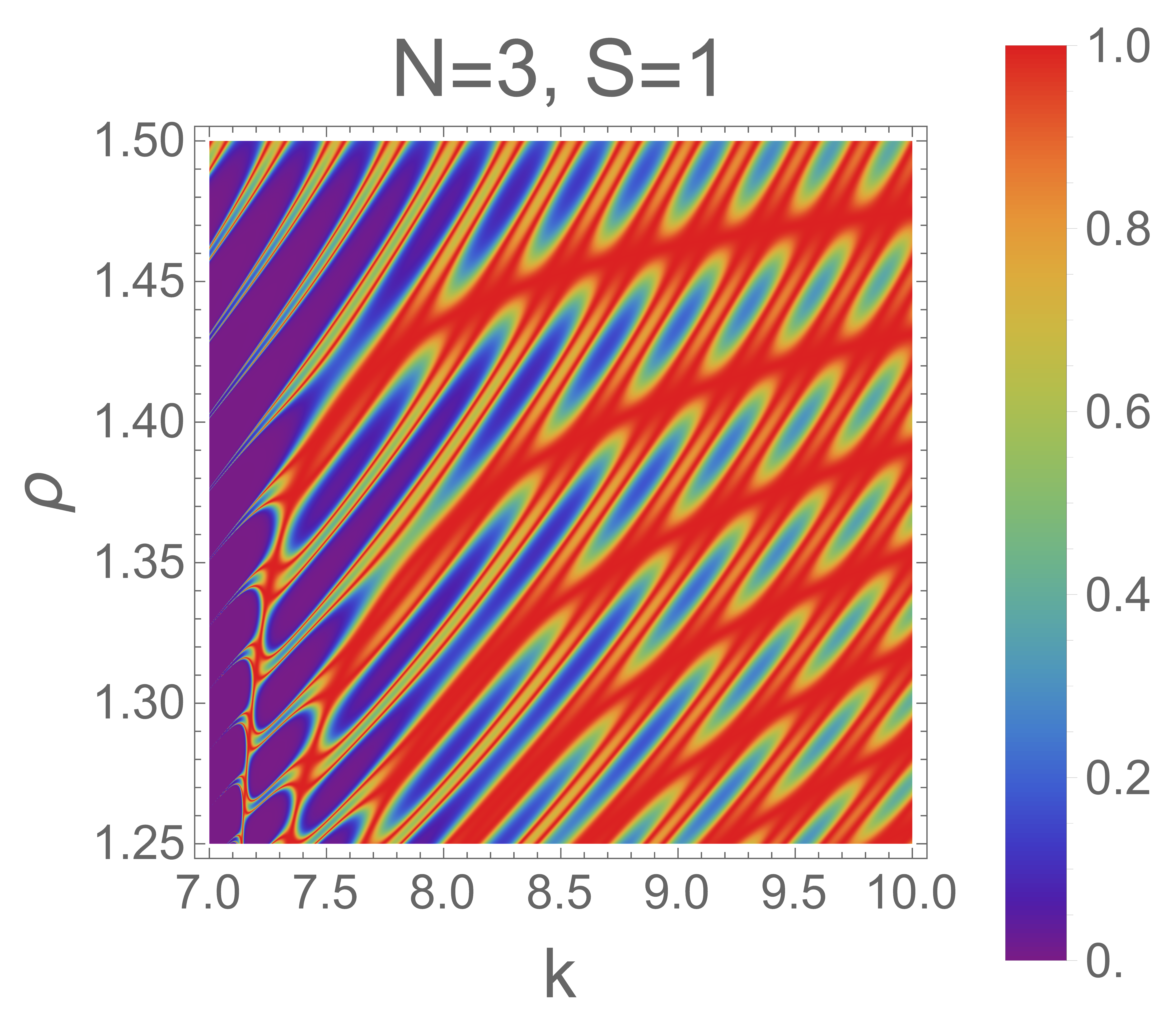} a
\includegraphics[scale=0.48]{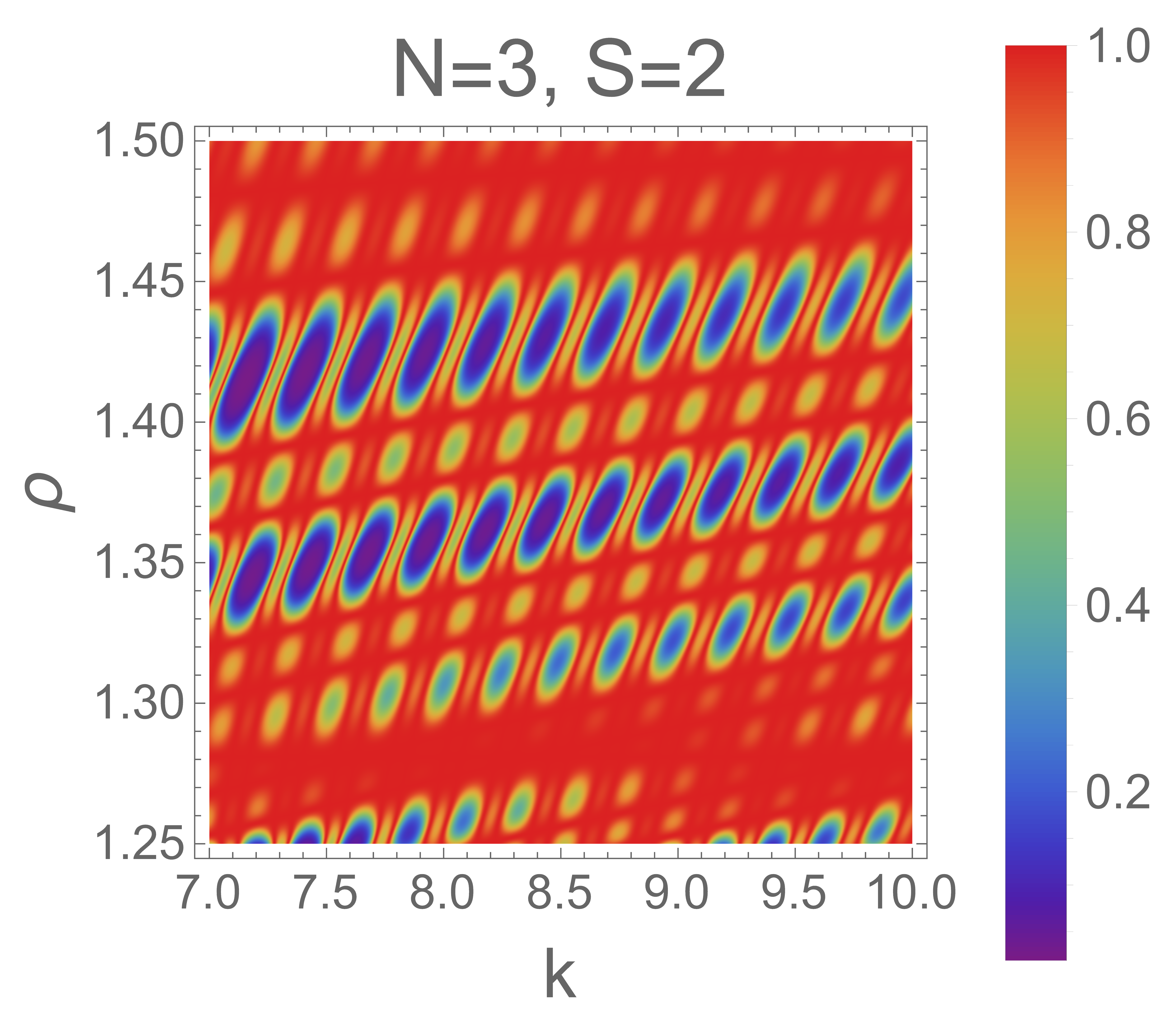} b\\
\includegraphics[scale=0.48]{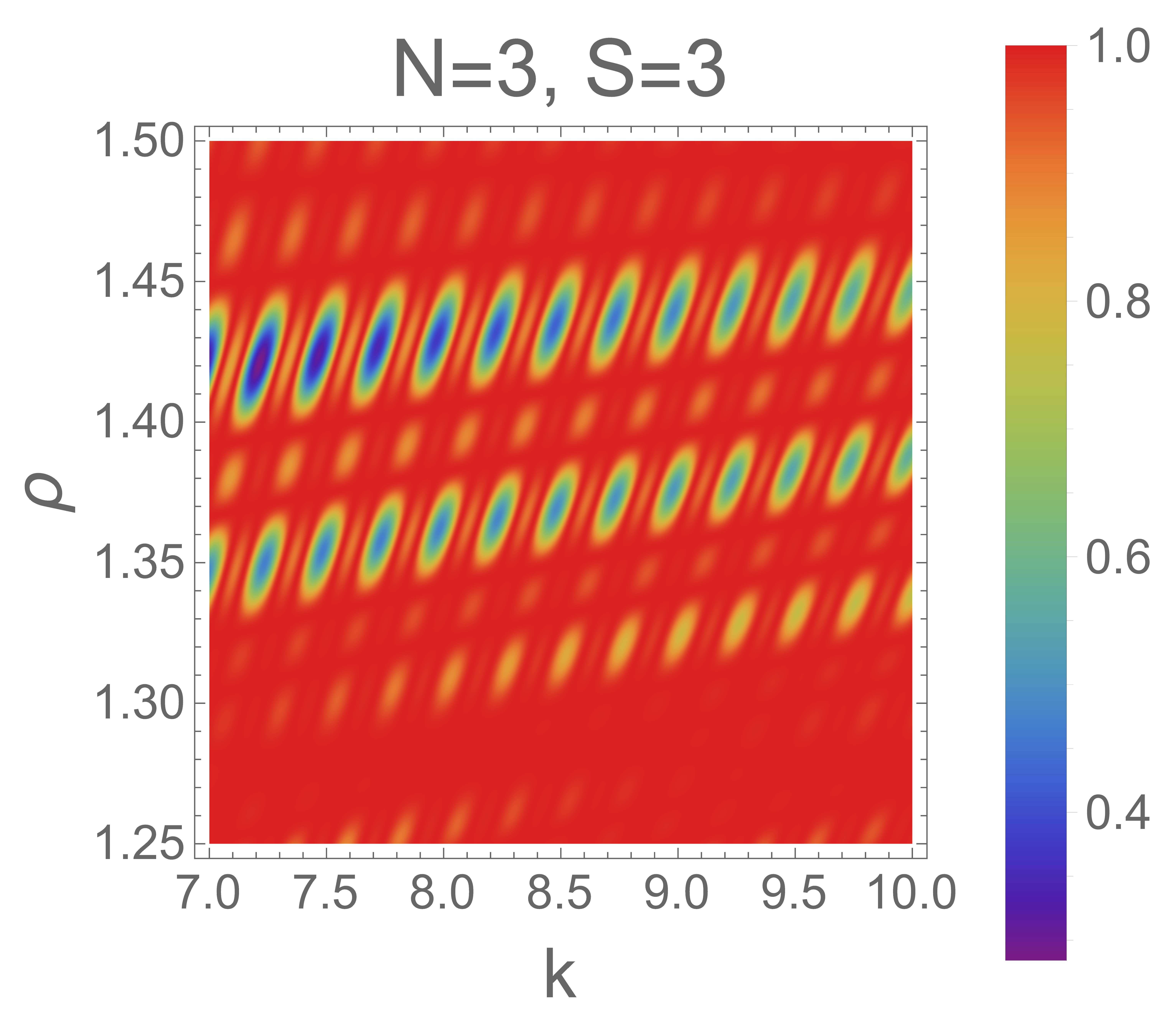} c
\includegraphics[scale=0.48]{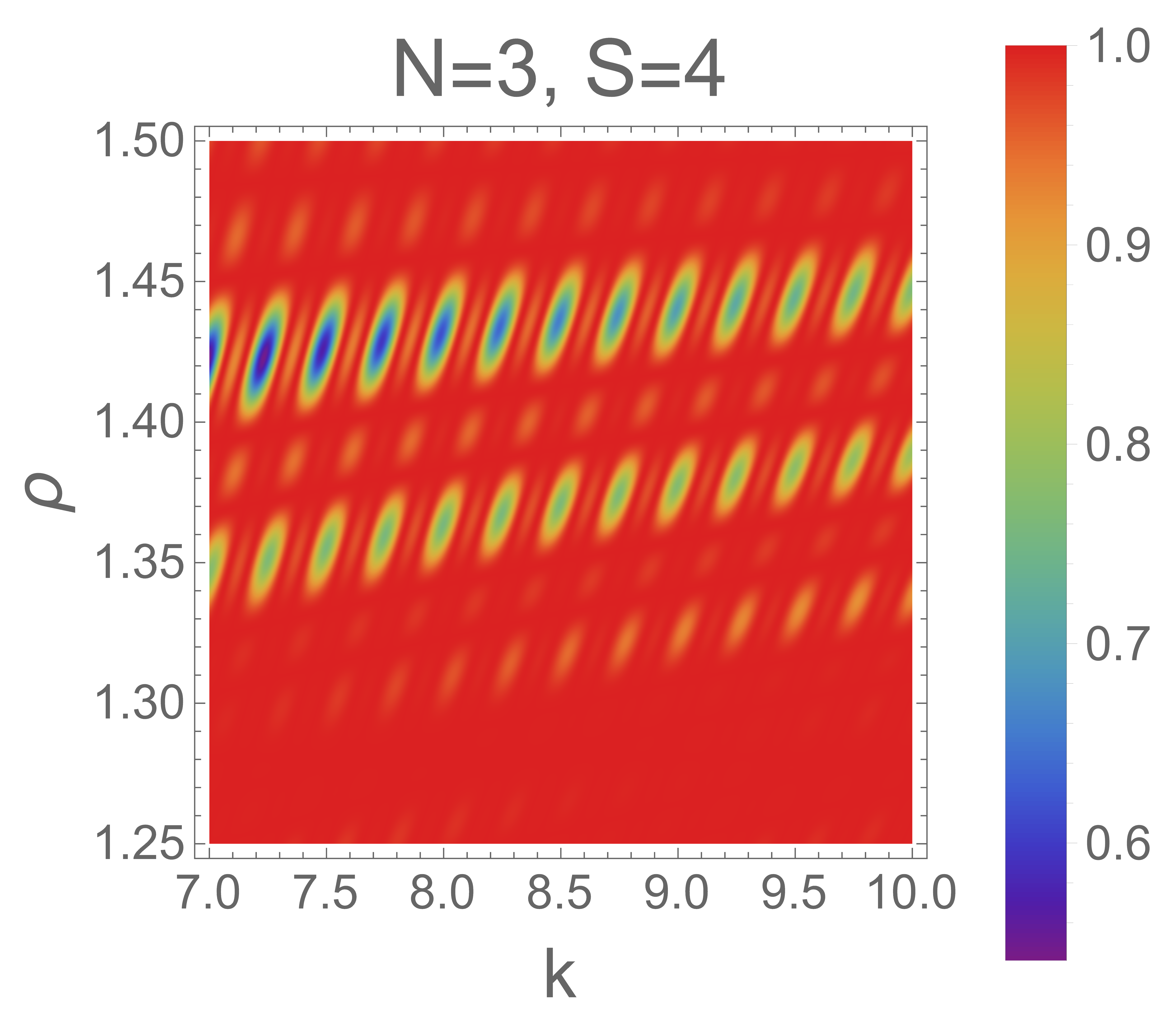} d\\
\includegraphics[scale=0.48]{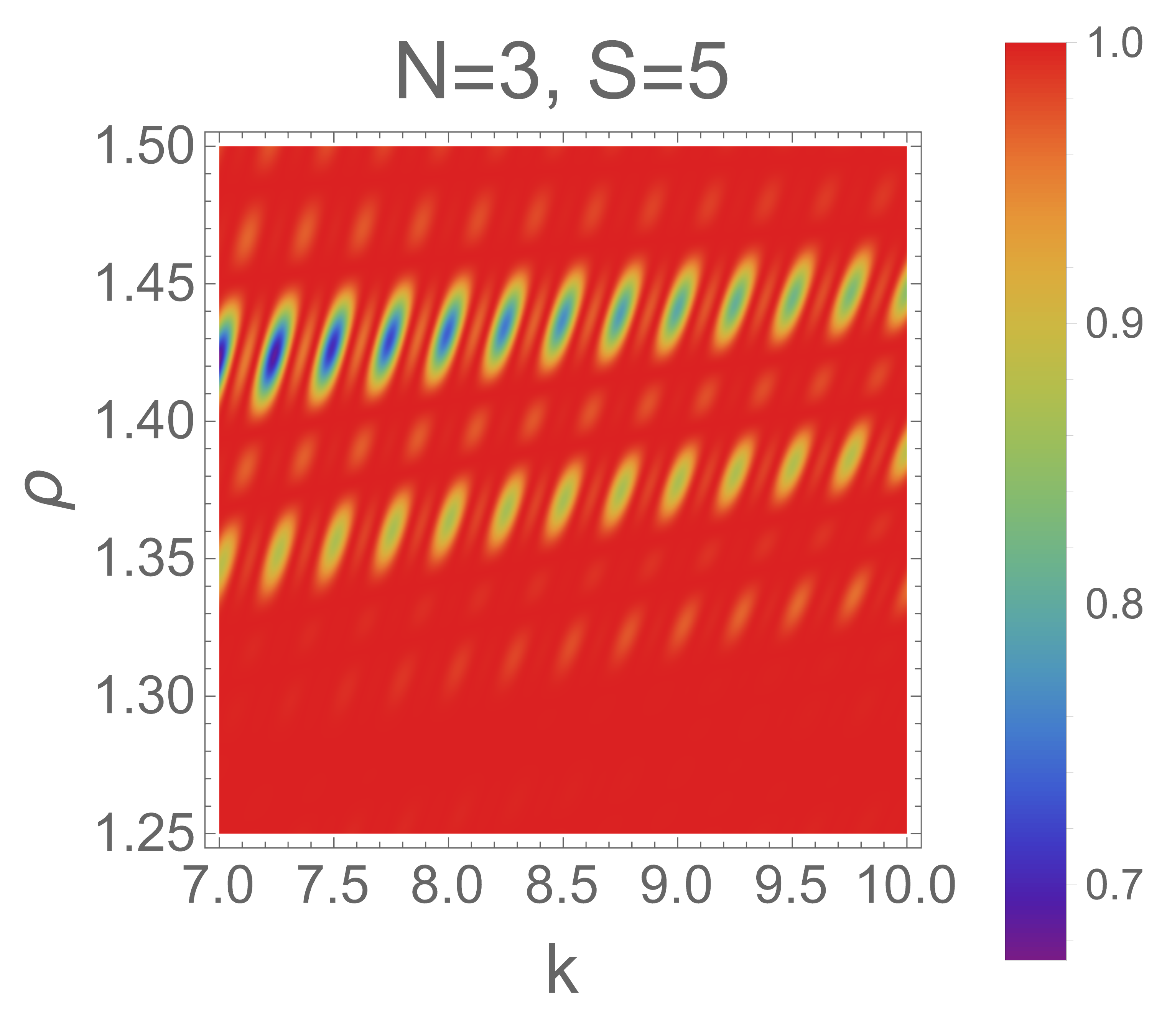} e
\includegraphics[scale=0.48]{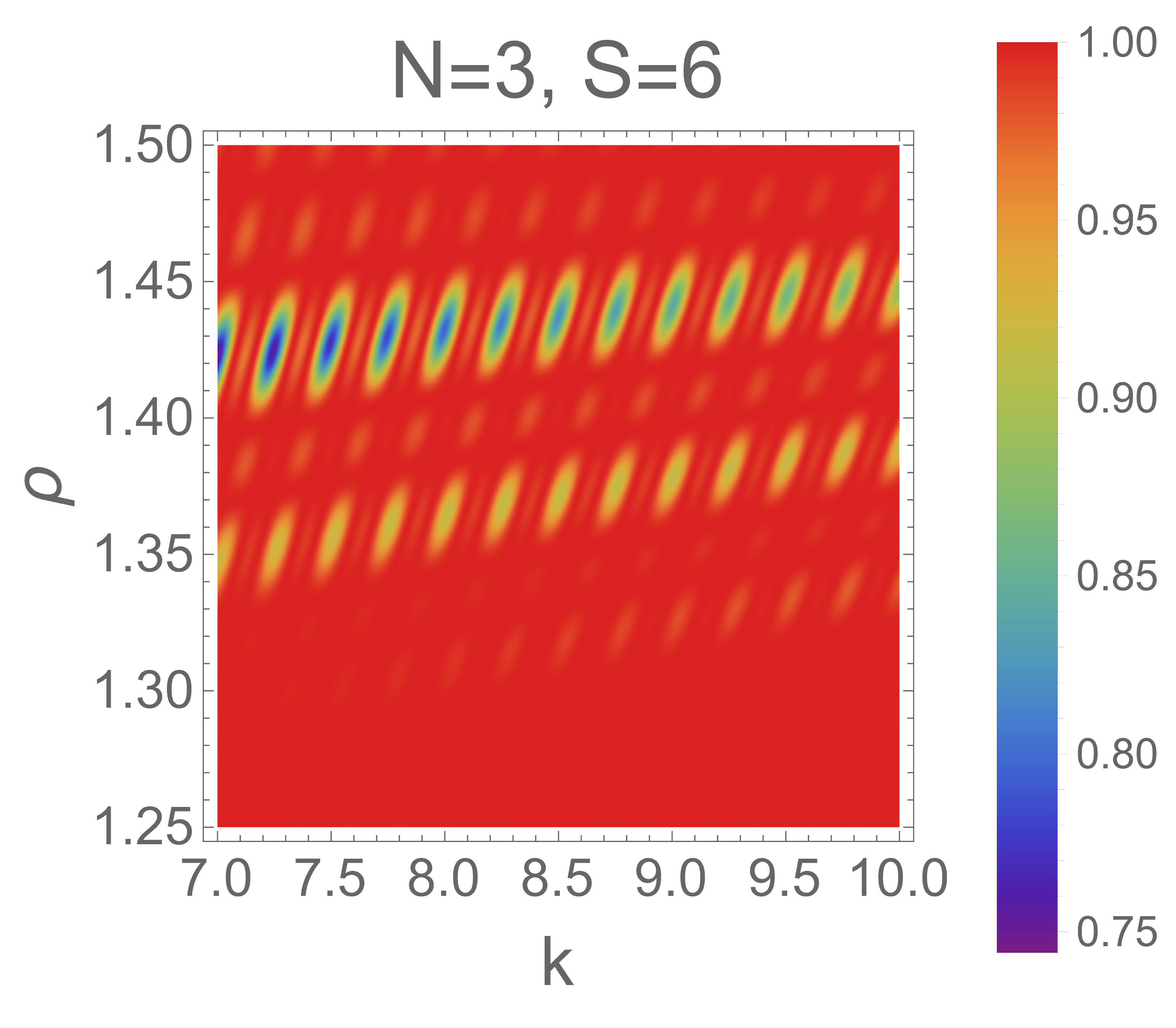} f
\caption{\it Density plots illustrating the distribution of tunneling probability within the $\rho-k$ plane are presented for the UCP-$\rho_{3}$ system across various stages spanning from $S=1$ to $S=6$. These crafted figures maintain $\mu=0.5$ and $\nu=1.15$ (non-fractal case) as fixed parameters, with $L=25$ and $V=25$ governing the potential landscape. Evident within these plots are sharply defined transmission resonances, particularly prominent at stage $S=1$. As the stage $S$ increases, an observable trend emerges: the transmission profile shows saturation and gradually occupies the red region, indicative of enhanced transmission probabilities. This indicates that advanced stages are likely to demonstrate enhanced efficiency in tunneling processes.}
\label{transmission04}
\end{center}
\end{figure}
\indent
Fig. \ref{transmission03} is systematically crafted to explore the complex tunneling phenomena demonstrated by the PCP system ($\nu=0$ case), specifically for configurations with $N=3$ (first column) and $N=4$ (depicted in the second column), thereby facilitating a comparative exploration of transmission probabilities across stages $S=1$ to $S=3$. The density plots chart a continuum of values for $\rho$ ranging from $1.25$ to $1.50$, and for $k$ between 6 to 8, with $\mu=0.5$ and both $L$ and $V$ are fixed at 25. These plots are instrumental in elucidating the dynamics of tunneling within the PCP system, serving as a pivotal visual representation of the underlying quantum mechanical tunneling. The plots highlight transmission resonances as exceedingly slender red streaks within the $\rho-k$ plane marking the locus of pronounced transmission phenomena. These streaks manifested as abrupt transitions in the transmission probability, $T(k)=0$ to $T(k)=1$. It is noted that the graphical representation precludes the visibility of the transmission spectrum for $k$ values around less than 7 due to graphic limitations. Nevertheless, beyond this hovered threshold, the transmission spectrum emerges distinctly, revealing significant null regions, areas where $T(k)=0$, which are punctuated by $N-1$ distinct red streaks indicative of sharp transmission phenomena for the given PCP system characterized by the parameter $N$. Moreover, the analysis shows that with an increment in the stage $S$, the red streaks delineating sharp transmission become more pronounced, suggesting a correlation between the intricacy of the stage and the dynamics of transmission. A comparative examination between the PCP systems of $N=3$ and $N=4$ unveils that the latter exhibits more sharp transmission resonances. The visualization and analysis also uncover expansive, well-structured null regions, highlighting the critical necessity for detailed investigations into the transmission spectrum characteristics within the PCP systems. Such in-depth studies are essential to unravel the nuanced mechanisms governing the observed tunneling phenomena. Delving into these aspects promises to shed light on the intricate interplay between the structural parameters of the PCP systems and their quantum mechanical behaviors.\\
\indent
The objective of Fig. \ref{transmission04} is to elucidate the variation in tunneling characteristics across different stages, $S$, while maintaining a constant value of $N$. Specifically, the figure presents the tunneling profiles within the $\rho-k$ plane for the UCP-$\rho_{3}$ system, where $N$ is set to 3, across a range of stages from $S=1$ through $S=6$. For these visualizations, the parameters $\mu$ and $\nu$ were set to 0.5 and 1.15, respectively, with $\nu$ distinctly non-zero. The potential parameters were held constant at $L=25$ and $V=25$. The graphical representations indicate that sharp transmission resonances are exclusively observed for the UCP-$\rho_{3}$ configuration at stage $S=1$, with subsequent stages failing to display such distinct transmission features. In the case of stage $S=1$, the system demonstrates a region of zero transmission, or a null region, when the value of $k$ hovered around 7. It is also noteworthy that with the increment in stage $S$, the transmission profile tends towards saturation, indicating a negligible variation in transmission across different stages. The saturation of the transmission profile has been comprehensively discussed in section \ref{saturation_discussion} Additionally, as the stage $S$ progresses, there is a gradual predominance of the transmission within the red region, signaling that with an increase in $S$, the UCP-$\rho_{3}$ system progressively approaches optimal transmission conditions. This observation implies that higher stages are more conducive to enhanced tunneling efficiency, suggesting a direct relationship between the stage of the system and its tunneling performance. This elucidation of the variation in tunneling profiles with stage advancement offers a valuable perspective on the optimization of tunneling efficiency through strategic stage selection.
\begin{figure}[H]
\begin{center}
\includegraphics[scale=0.48]{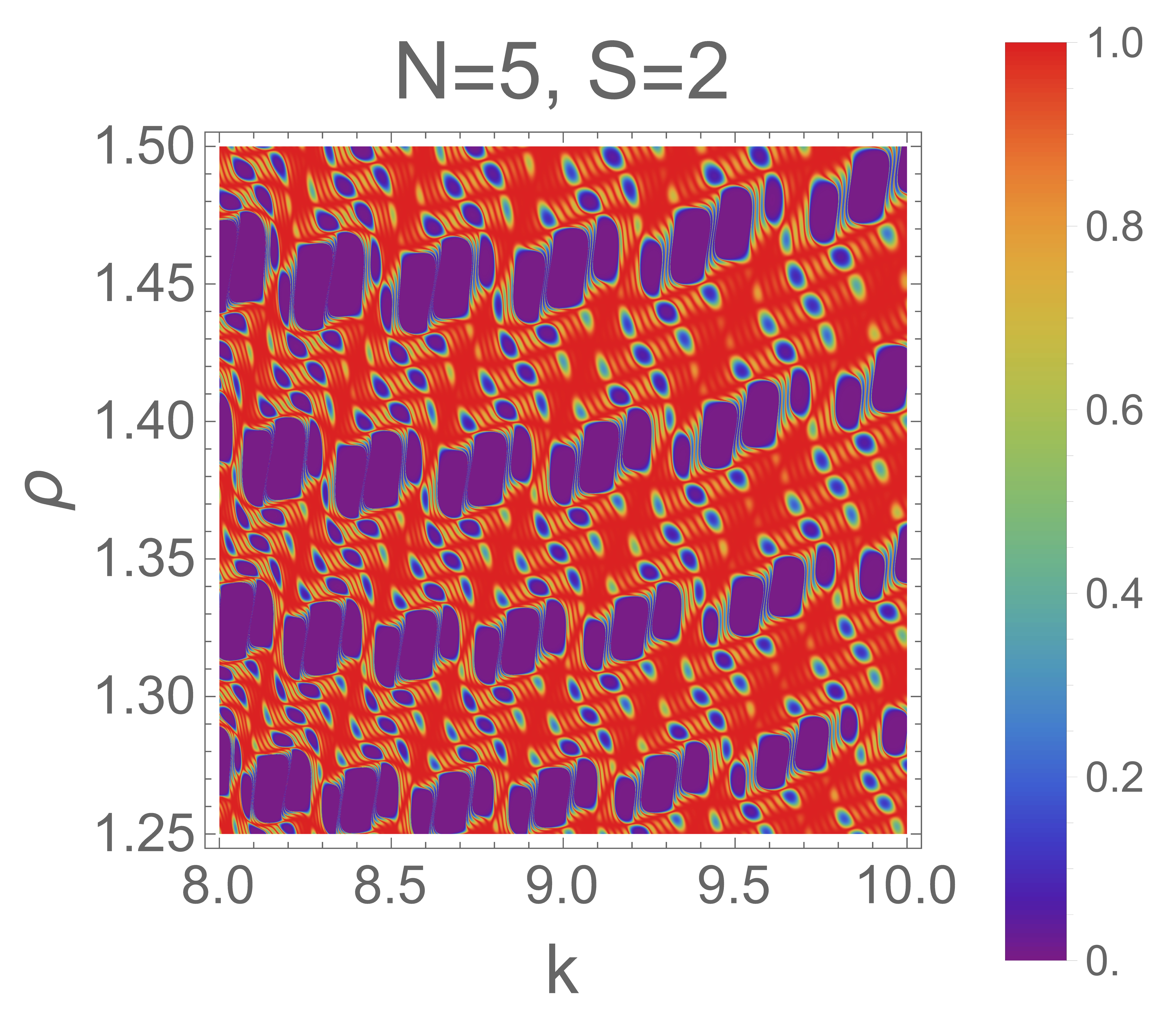} a
\includegraphics[scale=0.48]{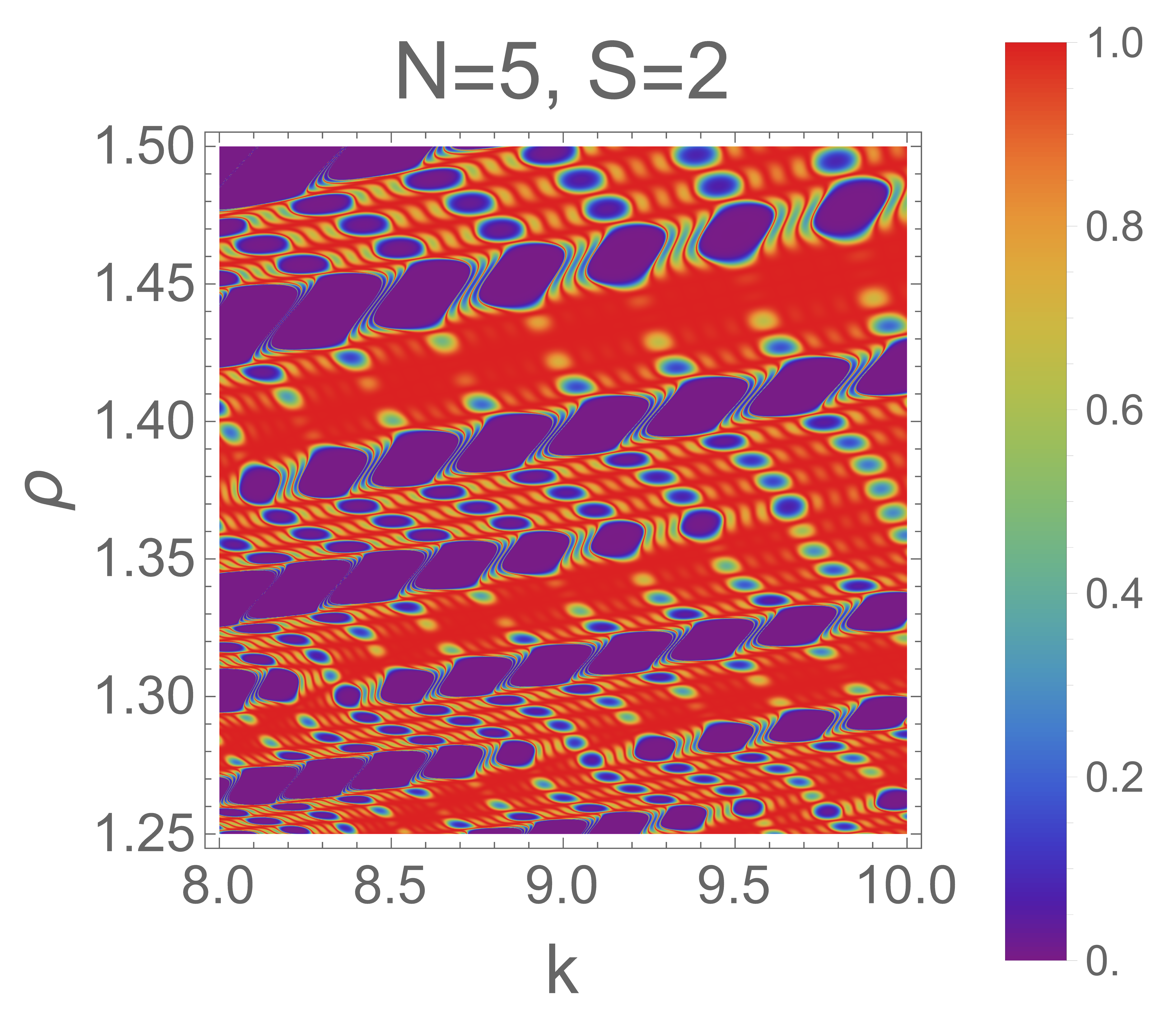} b \\
\includegraphics[scale=0.48]{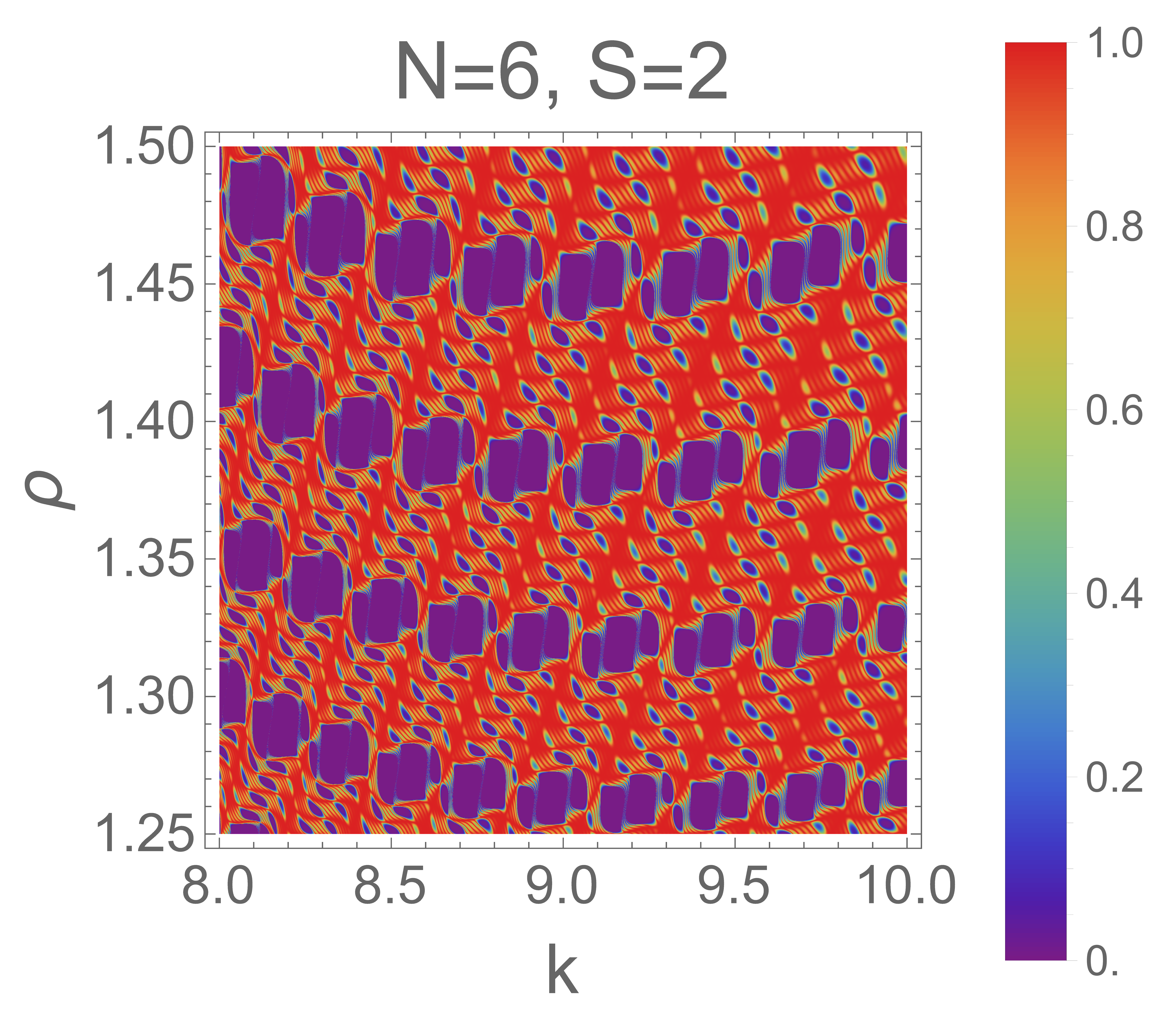} c
\includegraphics[scale=0.48]{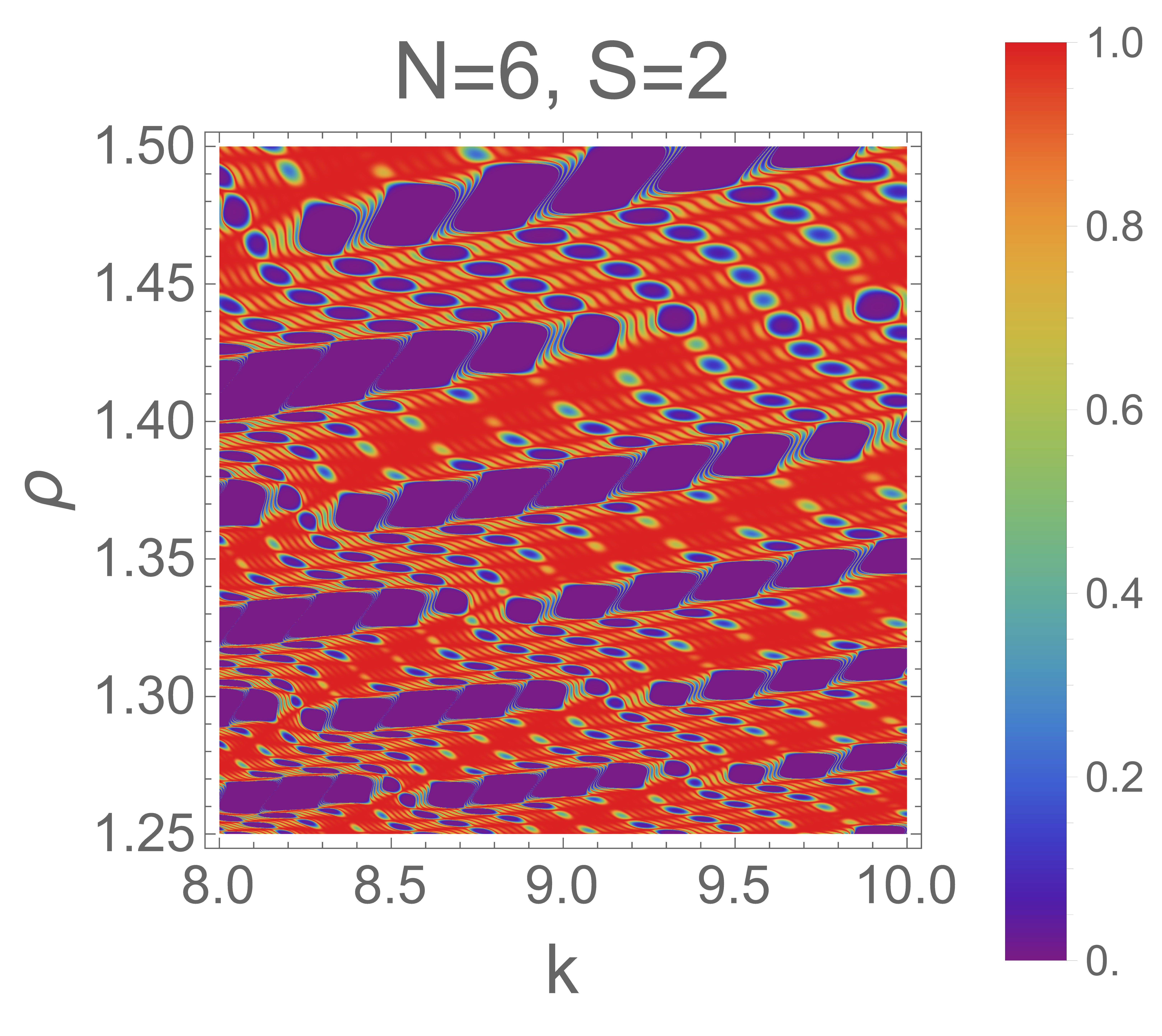} d 
\caption{\it 
Illustration of density plots showcasing the tunneling probability for two configurations: the first configuration, depicted in the first column, corresponds to the PCP system ($\nu=0$ case). The second configuration, shown in the second column, corresponds to the $\text{UCP-}\rho_{N}$ system ($\nu\neq0$ case). The first and second row is dedicated to the potential system defined by $N=5$ and $N=6$ respectively. For these plots, $\mu=0.5$ and $\nu=1.15$ ($\nu$ value is defined only for the plots located in the second column) and the other potential parameters are $L=25$ and $V=25$. This setup enables a comparison of tunneling probabilities when $\nu$ is zero and non-zero. It is observed that the PCP system exhibits sharper transmission resonances compared to the UCP-$\rho_{N=5,6}$ system. Furthermore, for the UCP-$\rho_{5}$ and UCP-$\rho_{6}$ systems, the valleys (violet region) are stretched along the $k$ axis relative to those observed in the PCP system.}
\label{transmission05}
\end{center}
\end{figure}
\indent
The comparative analysis presented in Fig. \ref{transmission05} focuses on examining the transmission profiles in $\rho-k$ space for PCP systems characterized by $N=5$ and $N=6$, along with their non-fractal counterparts, UCP-$\rho_{5}$ and UCP-$\rho_{6}$ system, for stage $S=2$. In the UCP-$\rho_{5}$ and UCP-$\rho_{6}$ systems, the parameter $\nu$, which is responsible for the transition from fractal to non-fractal behavior is held constant at $1.15$, whereas in the PCP system, $\nu=0$ by definition. The analysis reveals several key observations. Firstly, it is evident that the system with $N=6$ exhibits notably sharper transmission streaks, irrespective of the value of $\nu$ when compared to the $N=5$ system. This suggests that an increase in $N$ leads to enhanced sharpness in transmission streaks, a trend consistently illustrated across various $N$ values as depicted in the 2D plots in Fig. \ref{transmission02}. 
Further analysis of the flat valleys also called null regions, area characterized by $T(k)=0$, (this feature is never possible for any Hermitian system \cite{christodoulides2018parity} but starkly contrasting here) within the transmission profiles, which further require detailed and in-depth investigation, provides additional insights. The area occupied by the null regions in the $\rho-k$ space is larger for the system defined by $\nu=1.15$ (non-fractal case) than the system defied by $\nu=0$ (fractal case). Further, these valleys are punctuated by four (first column) and five (second column) distinct red streaks, loci of $T(k)=1$ in the $\rho-k$ plane. Also, these streaks are sharper in the plots located in the first column as compared to the plots in the second column, suggesting that the system with fractal potential exhibits sharper transmission resonances than the system with non-fractal potential. These valleys exhibit elongation along the $k$-axis in the UCP-$\rho_{5}$ and UCP-$\rho_{6}$ systems compared to the PCP system. Additionally, the area encompassed by these flat valleys is greater in the UCP-$\rho_{5}$ and UCP-$\rho_{6}$ systems compared to their PCP counterparts. These findings collectively underscore the nuanced distinctions in transmission characteristics between PCP and UCP-$\rho_{N}$ systems, shedding light on the impact of $N$ variation and the parameter $\nu$ on transmission behavior in fractal and non-fractal systems.

%%%%%%%%%%%%%%%%%%%%%%%%%%%%%%%%%%%%%%%%%%%%%%%%%%%%%%%%%%%%%%%%%%%%%%%%%%%%%%%

\subsection{Saturation of the transmission coefficient}
\label{saturation_discussion}
\begin{figure}[h! tbp]
\begin{center}
\includegraphics[scale=0.215]{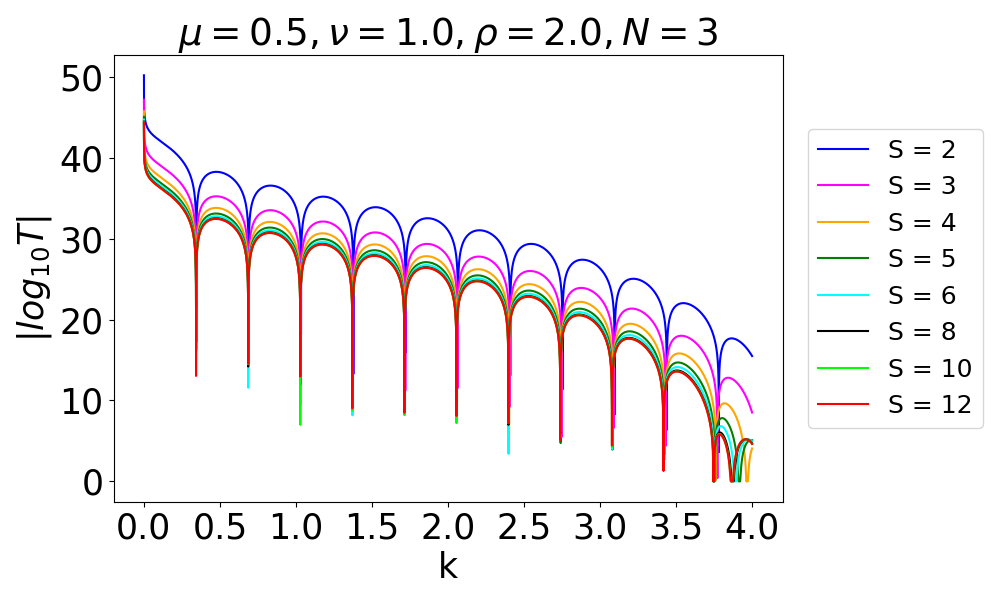} a
\includegraphics[scale=0.215]{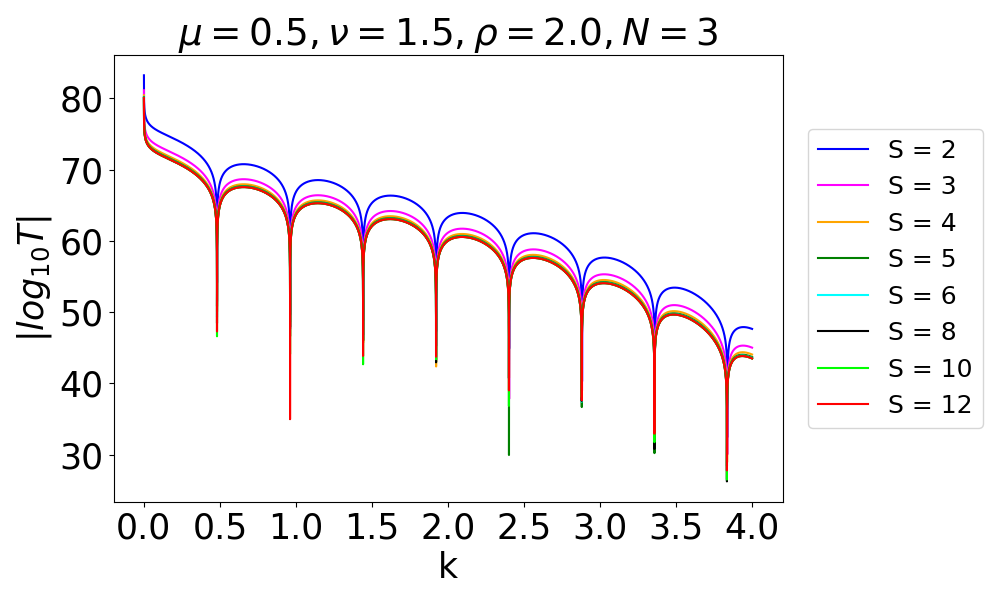} b \\
\includegraphics[scale=0.215]{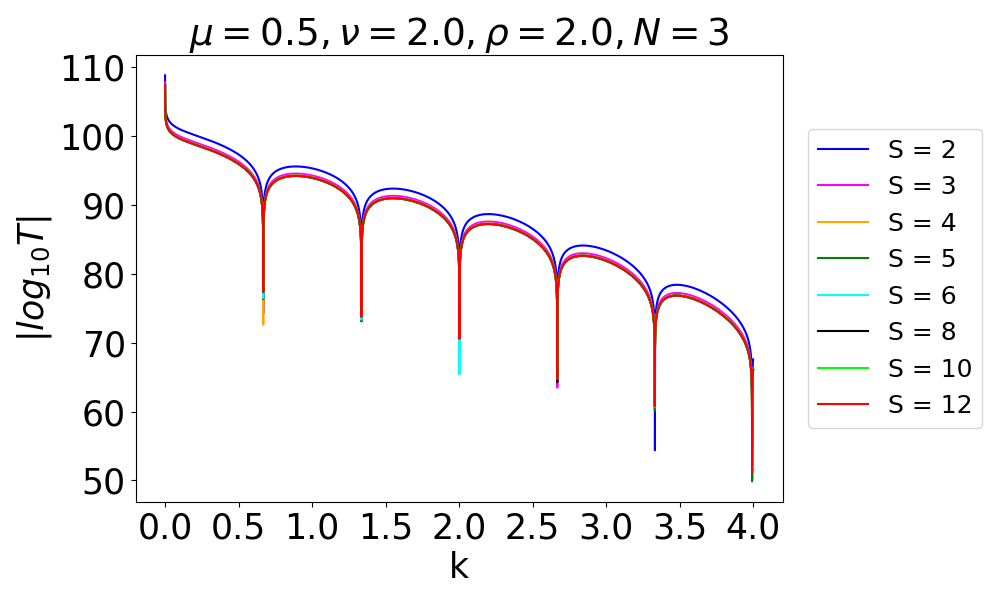} c
\includegraphics[scale=0.215]{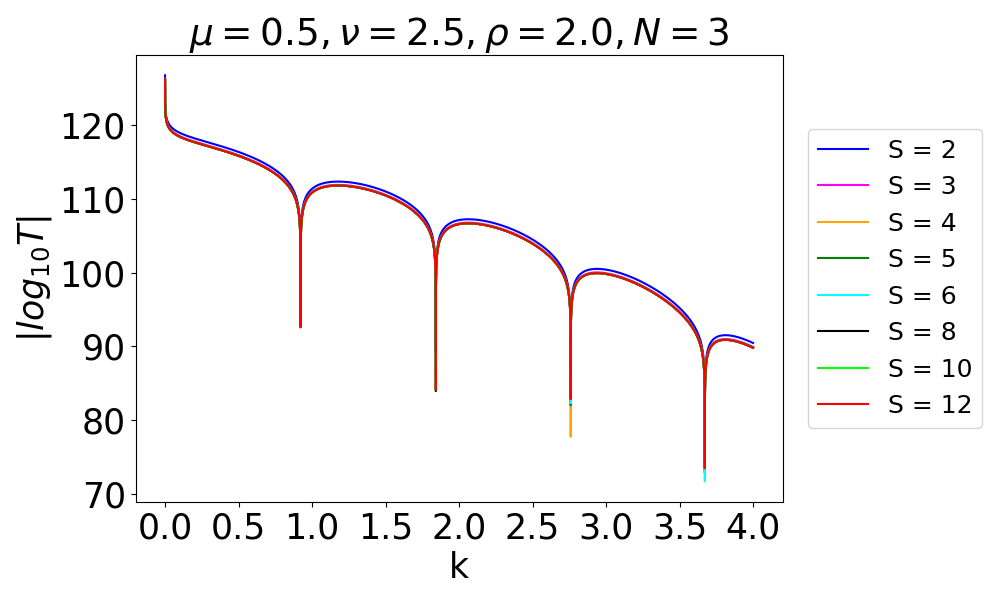} d \\
\includegraphics[scale=0.215]{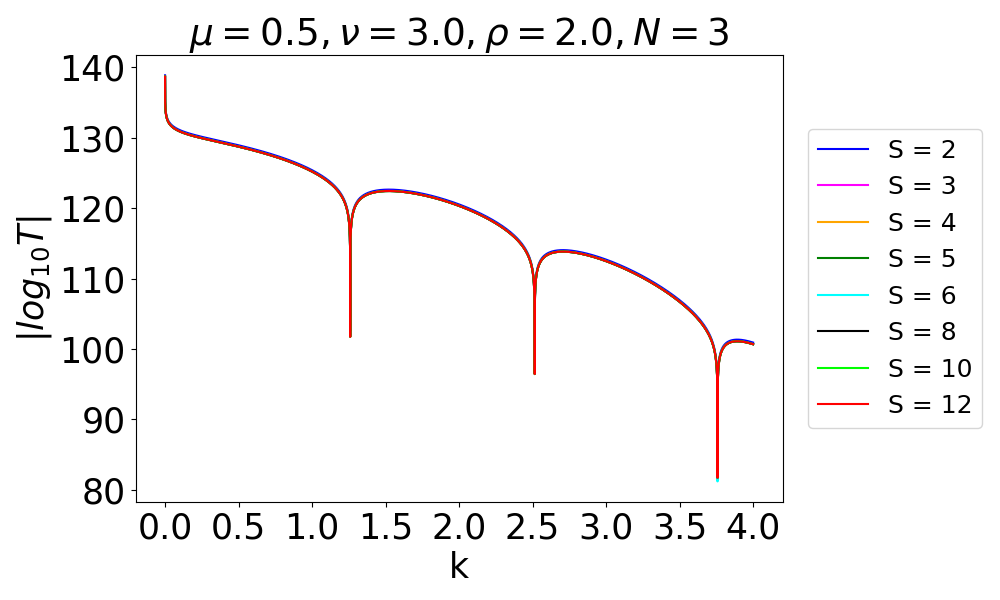} e
\includegraphics[scale=0.215]{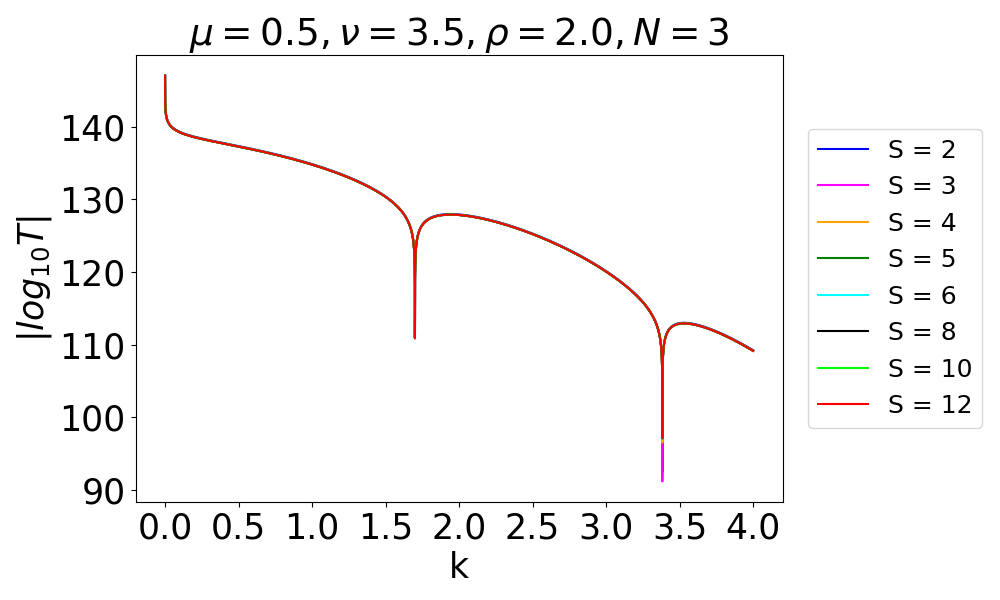} f
\caption{\it The saturation behavior in the transmission profile is depicted for a system with \(N=3\), parameterized by fixed values of $\mu=0.5$ and $\rho=2.0$, and variable \(\nu\). The potential is defined with \(L=25\) and \(V=25\). Analysis of the data reveals that increasing \(\nu\) leads to earlier saturation in the transmission profile, observable at smaller stages \(S\). This earlier saturation occurs due to the multiplication of \(\nu\) with \(S\), resulting in a higher value and consequently the removal of a thinner portion of the potential at these earlier stages. Conversely, as depicted in figure (a), a lower \(\nu\) value results in the saturation effect emerging at higher stages, such as \(S=8\), \(10\), \(12\). The differential in saturation onset between higher and lower \(\nu\) values is attributed to the varying extent of thin segment removal dictated by the \(\nu \times S\) multiplication in the term $\rho^{\mu+\nu S}$.
}
\label{saturation01}
\end{center}
\end{figure}

\begin{figure}[h! tbp]
\begin{center}
\includegraphics[scale=0.215]{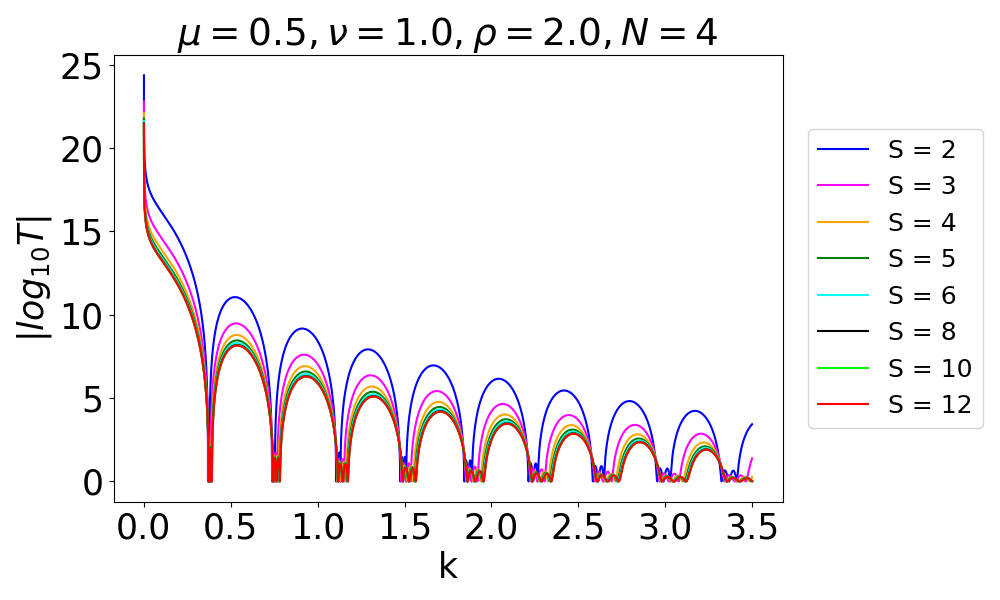} a
\includegraphics[scale=0.215]{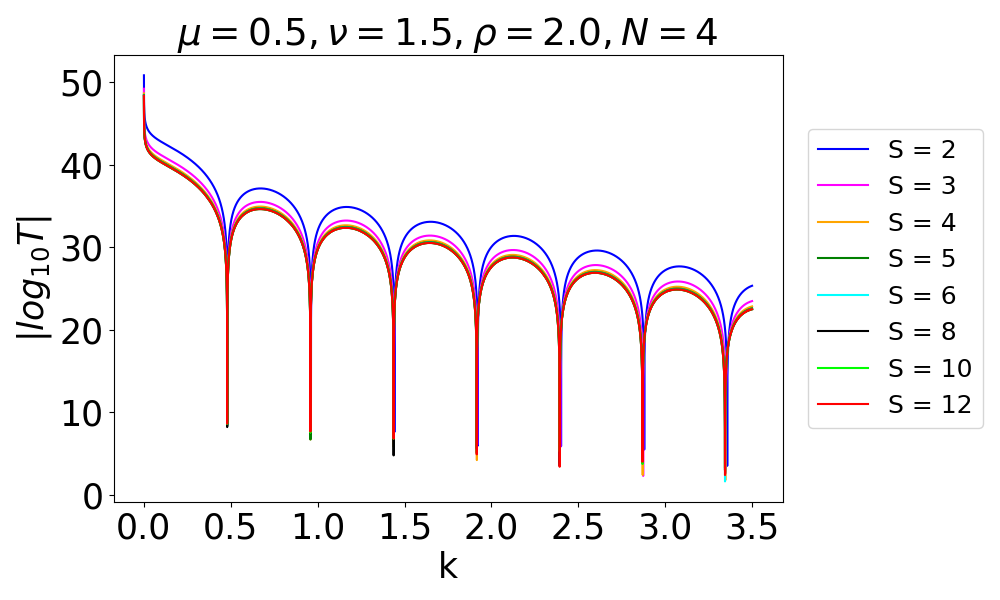} b \\
\includegraphics[scale=0.215]{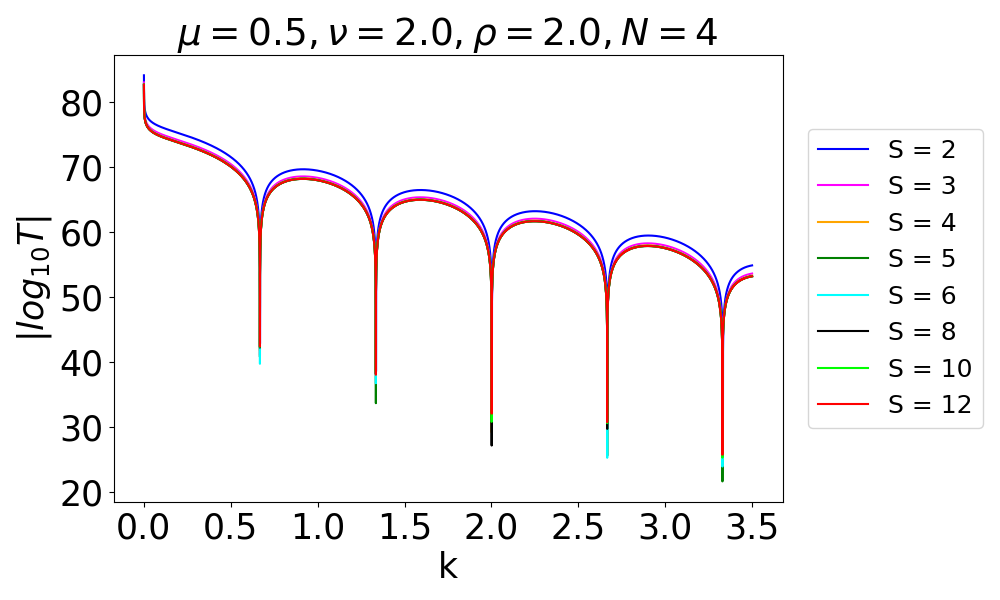} c
\includegraphics[scale=0.215]{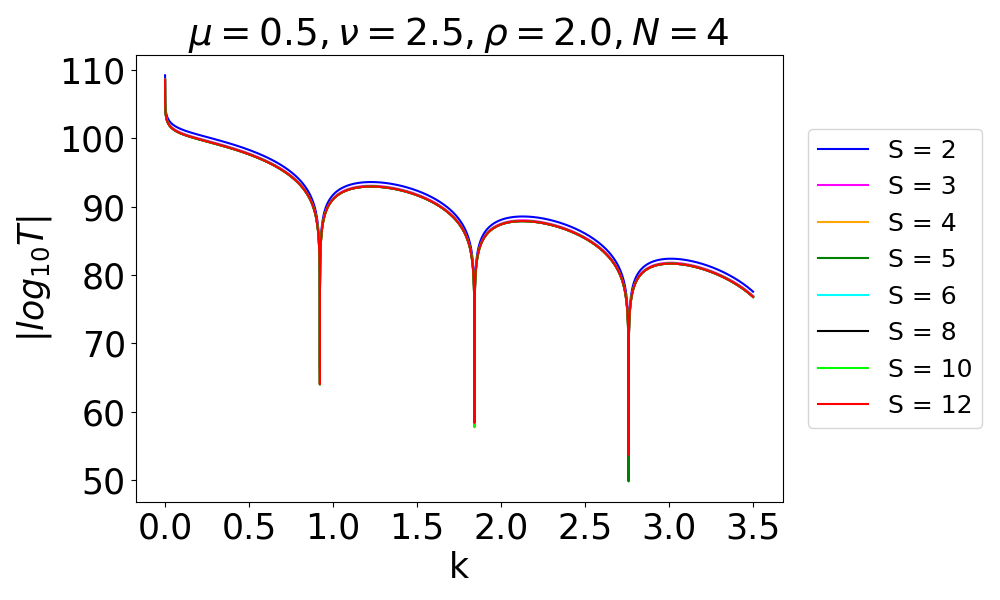} d \\
\includegraphics[scale=0.215]{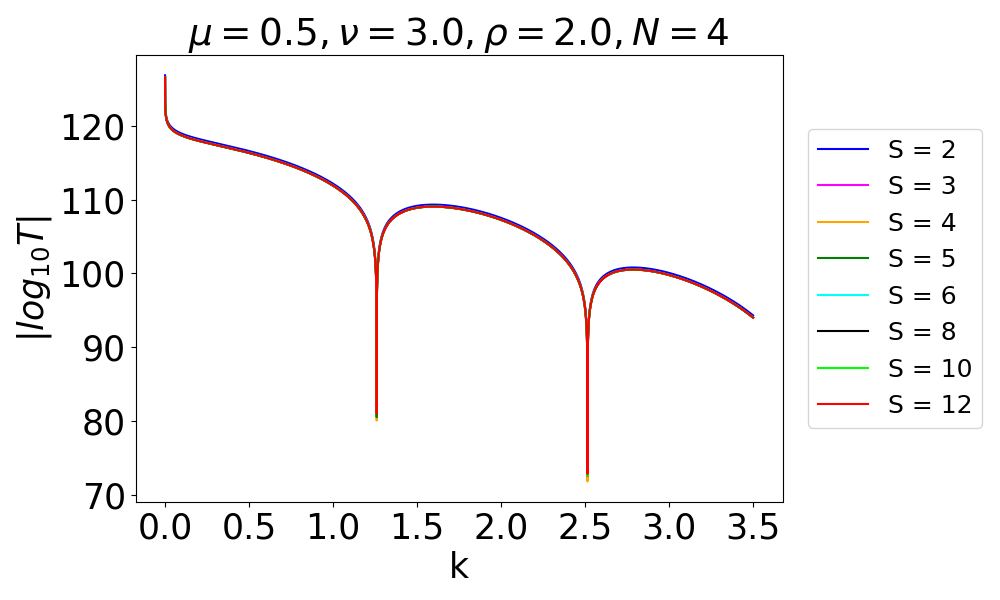} e
\includegraphics[scale=0.215]{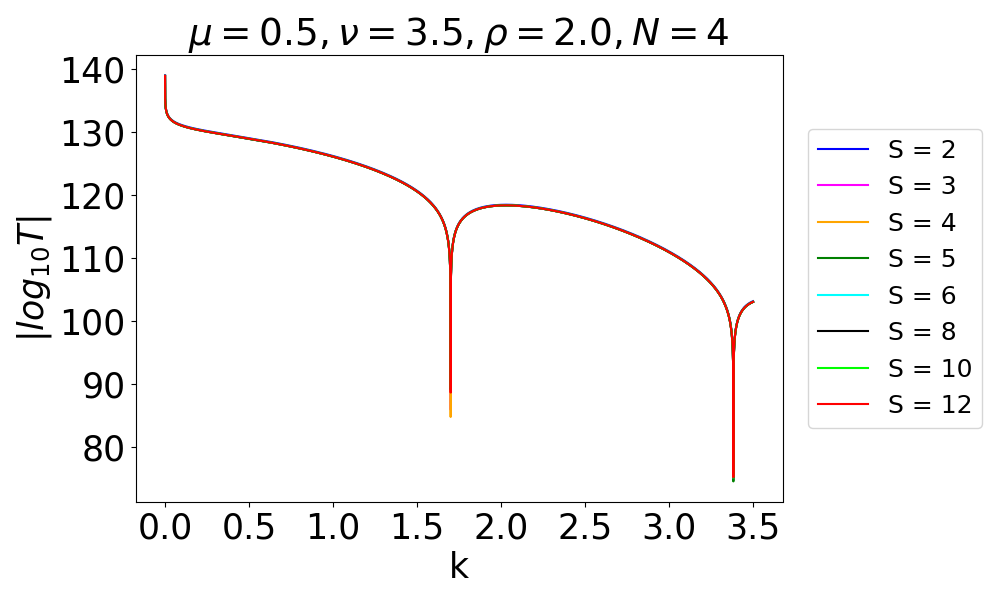} f
\caption{\it The observed saturation trends in the transmission profiles for a system characterized by \(N=4\) are illustrated, with the parameters, \(\mu\) and \(\rho\), held constant at values of $0.5$ and $2$ respectively, and \(\nu\) being varied. The potential settings are specified as \(L=25\) and \(V=25\). The analysis indicates that an increment in \(\nu\) precipitates the onset of saturation at earlier stages \(S\), a phenomenon attributable to the \(\nu \times S\) multiplication, which elevates the value leading to the excision of increasingly finer segments from the potential at these initial stages. On the contrary, depicted in Fig. (a), a diminutive \(\nu\) manifests saturation at advanced stages such as \(S=8\), \(10\), \(12\). This variance in the emergence of saturation, linked to the magnitude of \(\nu\), underscores the impact of the proportional removal of thinner potential segments, facilitated by the product of \(\nu\) and \(S\) in $\rho^{\mu+\nu S}$ term.
}
\label{saturation02}
\end{center}
\end{figure}

In the framework of UCP-$\rho_{N}$ system, the structural design involves the extraction of a fraction $\frac{1}{\rho^{\mu+\nu S}}$ at $N-1$ symmetric locations (in a defined manner) at each successive stage $S$. This design principle indicates that certain parameter combinations of $\rho$, $\mu$, $\nu$ and $S$, particularly when the denominator $\rho^{\mu+\nu S}$ assumes a large value, the extracted segment becomes increasingly slender relative to the original length of the potential segment. Consequently, as the extracted portion becomes progressively narrower, the overall potential structure undergoes a more subtle modification. In scenarios where the potential segments experience minimal structural alterations\textemdash owing to the removal of exceedingly thin segments\textemdash the transmission profile of the UCP-$\rho_{N}$ system is anticipated to exhibit characteristics of saturation. This saturation effect is reflective of the diminished impact of the structural changes on the transmission properties of the system. Further, to enhance the value of the term $\rho^{\mu+\nu S}$, multiple approaches can be applied. For example, one strategy involves fixing $\rho$ and adjusting the parameters $\mu$ and $\nu$. Alternatively, the values
of $\rho$, $\mu$ and $\nu$ can be held constant, with the term naturally escalating for higher stage $S$. We have also reported this saturation of the tunneling profile for the standard SVC potential \cite{hasan2018super} and UCP-$\rho_{2}$ system \cite{umar2023quantum}.\\
\indent
Fig. \ref{saturation01} and \ref{saturation02} illustrates the saturation of the transmission profile for the system characterized by $N=3$ and $N=4$ respectively and in both the figures the potential settings are $L=25$ and $V=25$. For crafting these figures we have fixed $\rho$ and $\mu$ at 2.0 and 0.5 respectively and varied $\nu$ and its value is labeled at each plot. This step leads to an elevation in the term $\rho^{\mu+\nu S}$ value. The progressive increase in 
$\nu$ coupled with the sequential advancement of stage 
$S$, leads to an exponential escalation in the value of $\rho^{\mu+\nu S}$. This approach effectively demonstrates how parameter manipulation, particularly the adjustment of $\nu$, influences the growth of this term and consequently, impacts the saturation trends observed in the transmission profiles of these quantum systems. For a

\begin{figure}[H]
\begin{center}
\includegraphics[scale=0.215]{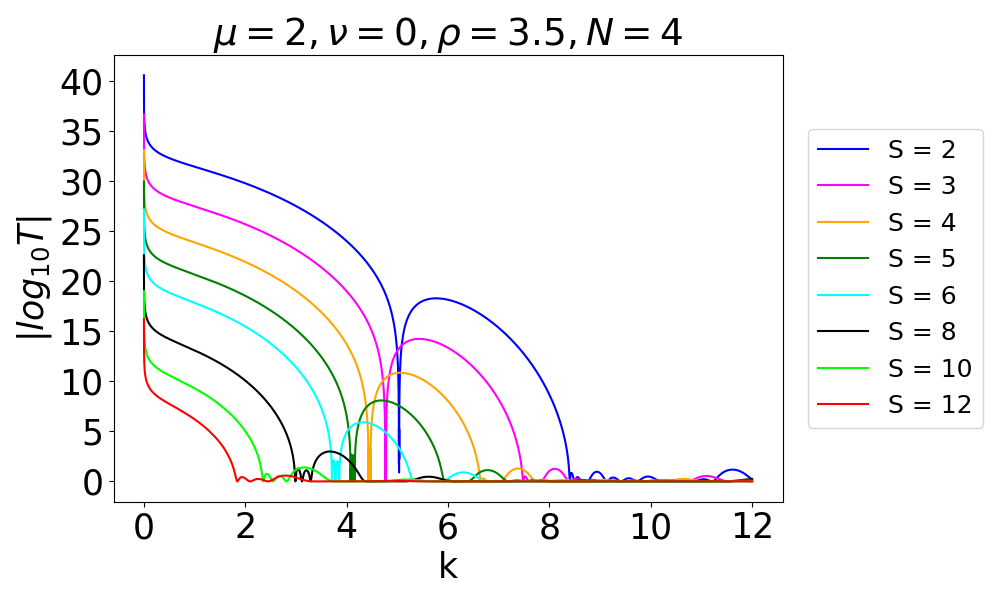} a
\includegraphics[scale=0.215]{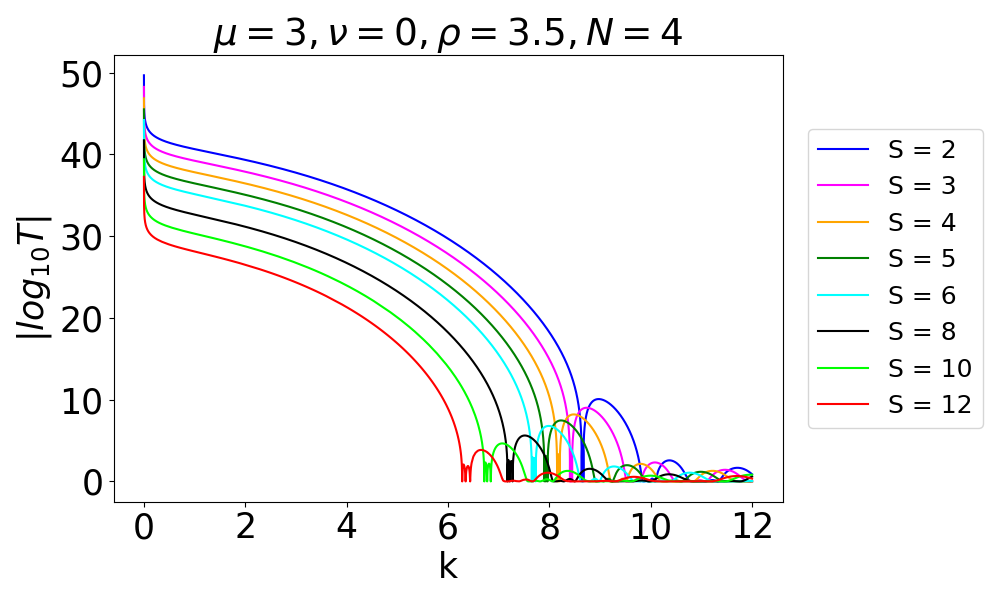} b \\
\includegraphics[scale=0.215]{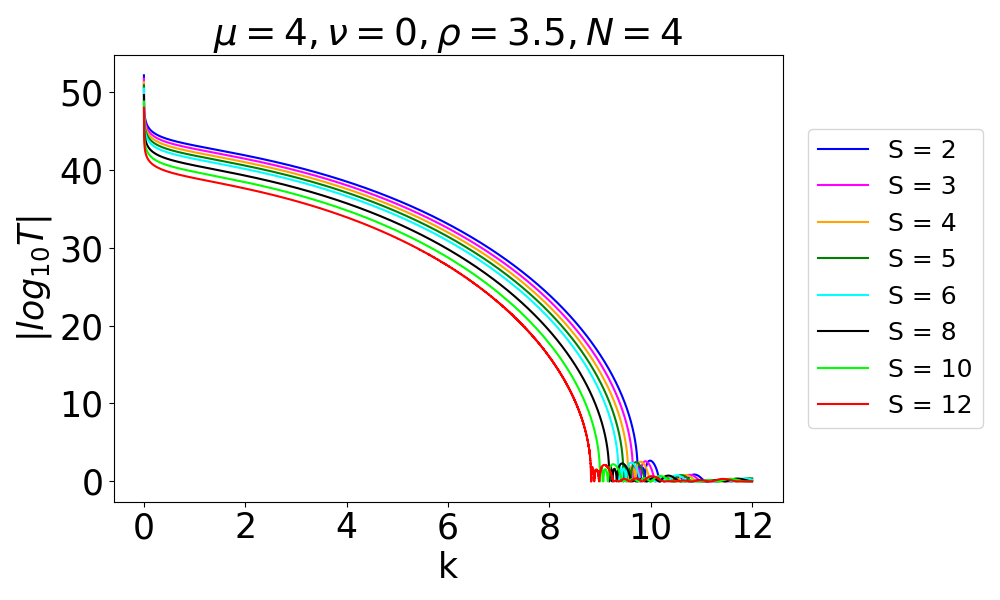} c
\includegraphics[scale=0.215]{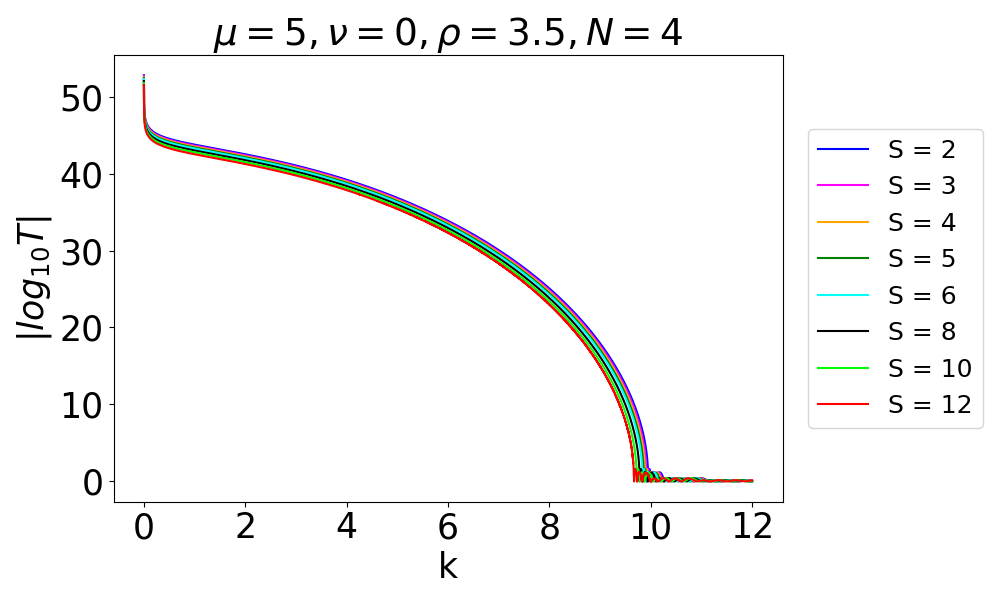} d \\
\includegraphics[scale=0.215]{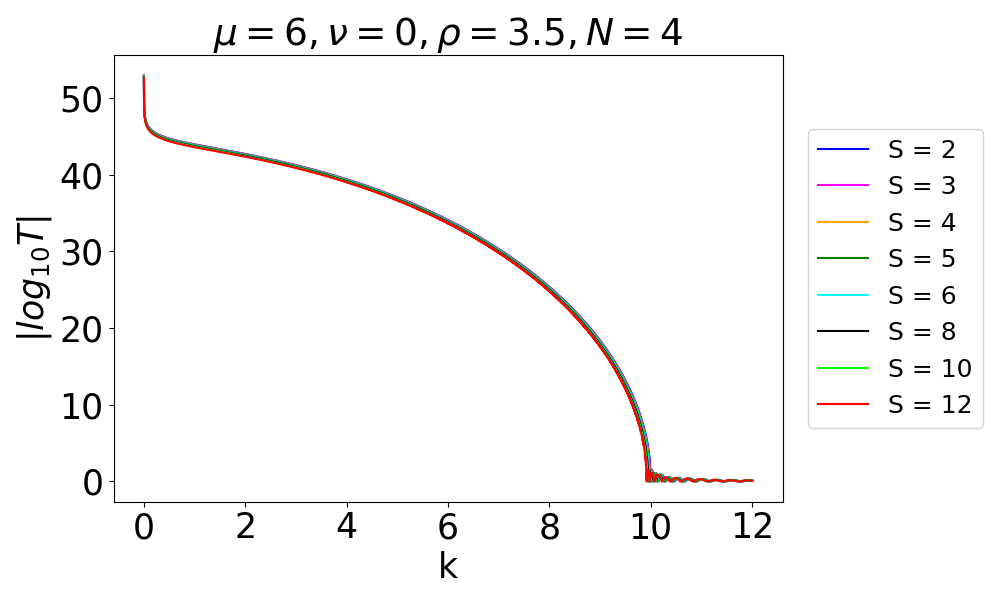} e
\includegraphics[scale=0.215]{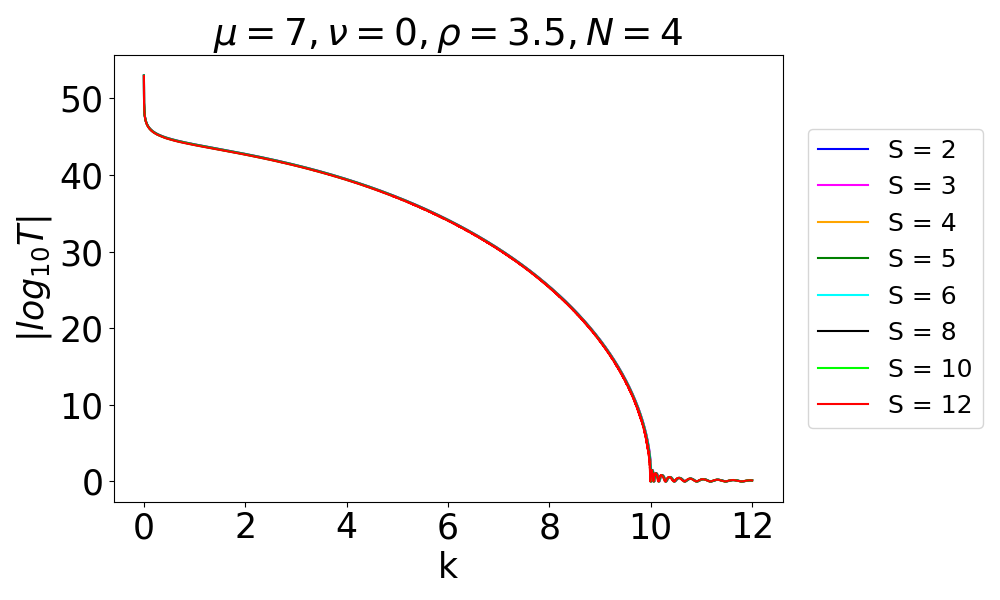} f \\
\includegraphics[scale=0.215]{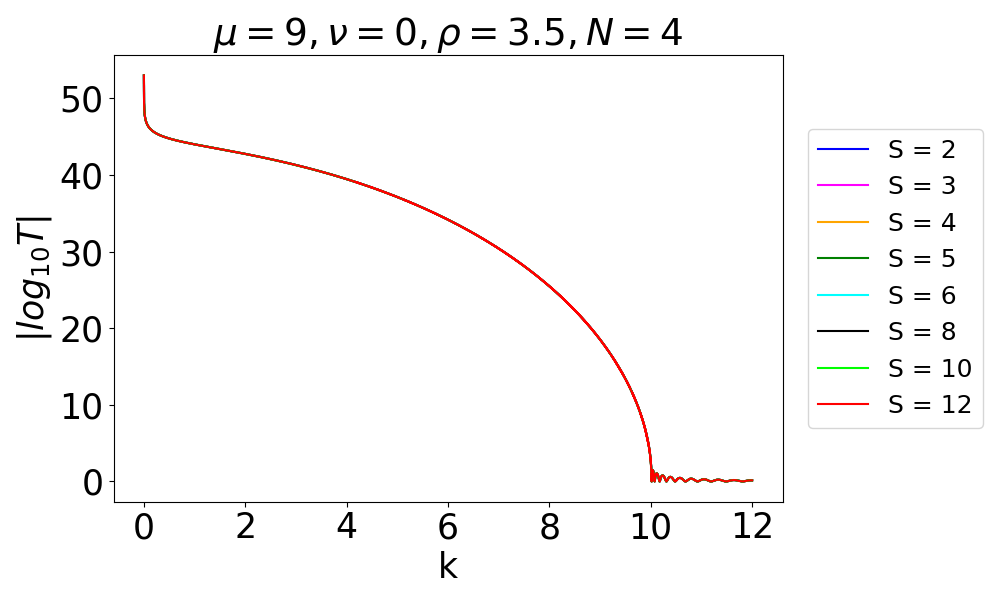} g
\includegraphics[scale=0.215]{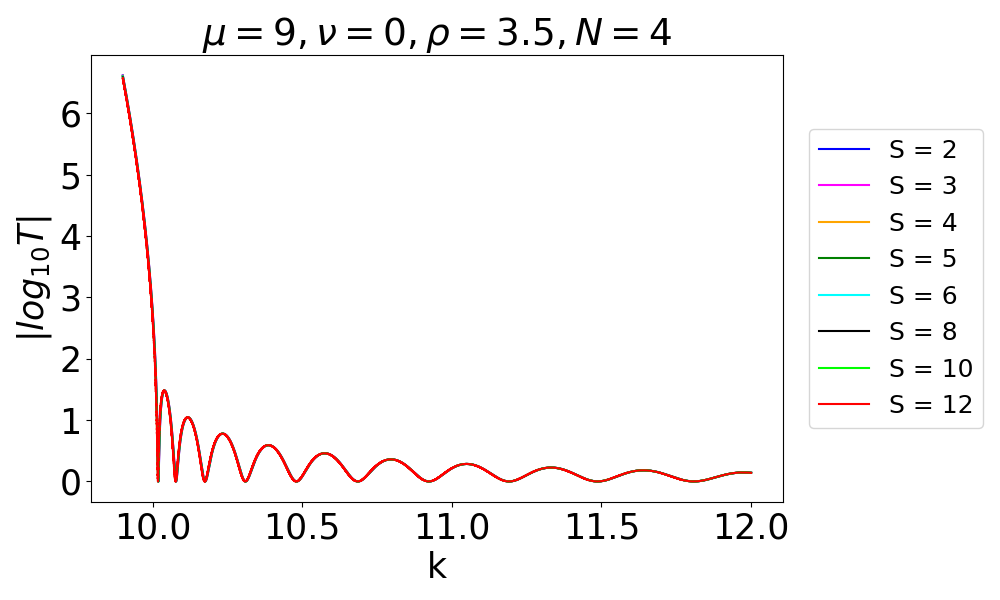} h 
\caption{\it The saturation pattern within the transmission profile of the PCP system ($\nu=0$ case) with $N=3$, is showcased. In this setup, the parameter \(\rho\) is consistently maintained at $3.5$, while \(\mu\) undergoes variation. The potential is defined with parameters \(L=25\) and \(V=25\). Given that \(\nu=0\), the stage number \(S\) does not contribute to the transmission profile, as the removal process is solely influenced by the term \(\rho^{\mu}\). Consequently, the fraction \(\frac{1}{\rho^{\mu}}\) is extracted from the potential segments. The plots reveal that lower values of \(\mu\) lead to distinct transmission profiles at each stage. However, as \(\mu\) increases, there is a noticeable trend towards the convergence of transmission profiles across different stages, which is particularly evident in Fig. g. Furthermore, Fig. h offers a magnified view of Fig. g, focusing on the range of \(k\) from $10$ to $12$ and \(\vert \log_{10}T \rvert\) from $0$ to $6.5$, providing a detailed perspective on the overlapping transmission profiles.
}
\label{saturation03}
\end{center}
\end{figure}

\noindent
better view of saturation, we have graphed $\lvert \log_{10}T \rvert$ with $k$ ranging from 0 to 4. As $T$ lies in the range 0 to 1 ($0\le T \le 1$), therefore $\log_{10}T \le 0$ and hence $f(T) =\lvert \log_{10}T \rvert \ge 0$ is a well-defined non-negative function and in our case, this function is suitable to show the saturation or complete convergence of the transmission spectrum. In Fig. \ref{saturation01}, detailing a system with \(N=3\), an initial setting of \(\nu=1.0\) presents distinct transmission profiles at early stages such as \(S=2\), \(3\), \(4\), \(5\), and \(6\). However, a notable overlap in the transmission profiles is observed at later stages, specifically \(S=8\), \(10\), and \(12\). With an increment in $\nu$, a gradual shift towards overlapping of the transmission profiles for various stages becomes apparent. This phenomenon is strikingly demonstrated in Fig. \ref{saturation01}f, which is rendered for \(\nu=3.5\). In this figure, the transmission profiles for stages \(S=2\), \(3\), \(4\), \(5\), \(6\), \(8\), \(10\), and \(12\) are observed to overlap substantially, showcasing a pronounced convergence across these various stages. This shift in saturation onset across various \(\nu\) values is attributed to the differential removal of thinner segments, as governed by the product of \(\nu\) and \(S\) in the expression \(\rho^{\mu+\nu S}\). This pattern reveals the sensitivity of the transmission characteristics of the system to the structural parameters, highlighting the crucial role of these parameters $\rho$, $\mu$ and $\nu$ in modulating the quantum tunneling behavior. Similarly, Fig. \ref{saturation02} adheres to the same principles, showcasing the transmission profile behavior for a system characterized by \(N=4\). This figure provides a parallel analysis, depicting how the transmission profiles exhibit substantial overlap across various stages, mirroring the patterns observed in Fig. \ref{saturation01}. The overarching conclusion from the analysis is that, regardless of the relatively low values of the parameters \(\rho\), \(\mu\) and \(\nu\), saturation in the transmission profiles for UCP-$\rho_{N}$ system is inevitable at higher stages. As the system evolves through successive stages, the cumulative effects of even low values of \(\rho\), \(\mu\), and \(\nu\) lead to noticeable saturation.\\
\indent
In Fig. \ref{saturation03}, the saturation trends within the transmission profile of the PCP system ($\nu=0$ case). This specific setup maintains the parameter \(\rho\) at a constant 3.5 while allowing for the adjustment of \(\mu\). The potential parameters are fixed as \(L=25\) and \(V=25\). The absence of \(\nu\) ($\nu=0$) means that the stage number \(S\) does not influence the transmission profile, as the segment removal is exclusively determined by the \(\rho^{\mu}\) term. Therefore, the fraction \(\frac{1}{\rho^{\mu}}\) is systematically excised from each potential segment. It is evident from the plots that when the value of \(\mu\) is lower, at \(\mu=2\), distinct transmission profiles for each stage are observed, demonstrating unique transmission behaviors at each stage. However, as \(\mu\) increases to a higher value, say \(\mu=9\), the transmission profiles start to exhibit a noticeable trend of overlap across different stages exhibiting saturation in the transmission profiles, which is particularly evident in Fig. \ref{saturation03}g. Furthermore, Fig. \ref{saturation03}h offers a magnified view of Fig. \ref{saturation03}g focusing on the range of $k$ from 10 to 12 and $\lvert \log_{10} T \rvert$ from 0 to 6.5, again providing a glimpse of saturation. A critical element of this phenomenon is the influence exerted by a larger value of $\mu$ on the potential structure within the system. As $\mu$ increases, the proportion of the potential segment that is removed at each iterative stage becomes progressively narrower. This more refined removal process results in modifications to the overall structure of the potential that are less pronounced. Consequently, as these structural alterations become increasingly subtle, the transmission profiles corresponding to different stages begin to show greater similarity, culminating in the observed phenomenon of profile overlapping. This pattern underscores the pivotal role that $\mu$ plays in governing the degree of structural changes effected. This dynamic is essential for understanding the modulation of transmission characteristics and their implications on the functional behavior of the system across varying stages of development.
\newpage

%%%%%%%%%%%%%%%%%%%%%%%%%%%%%%%%%%%%%%%%%%%%%%%%%%%%%%%%%%%%%%%%%%%%%%%%%%%%%%%%

\subsection{Reflection coefficient: Scaling law}
\label{scaling}
\begin{figure}[H]
\begin{center}
\includegraphics[scale=0.38]{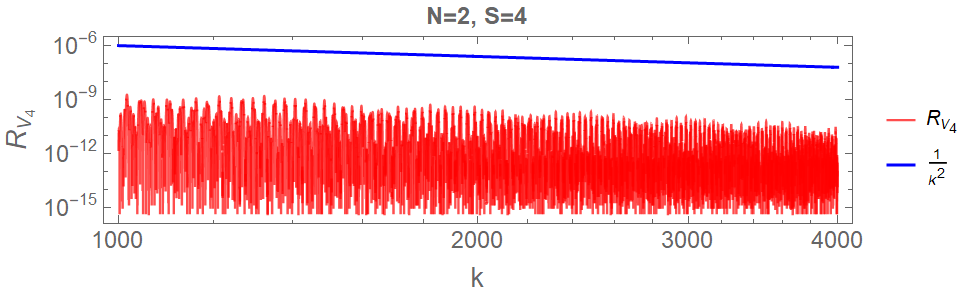} a\\
\includegraphics[scale=0.38]{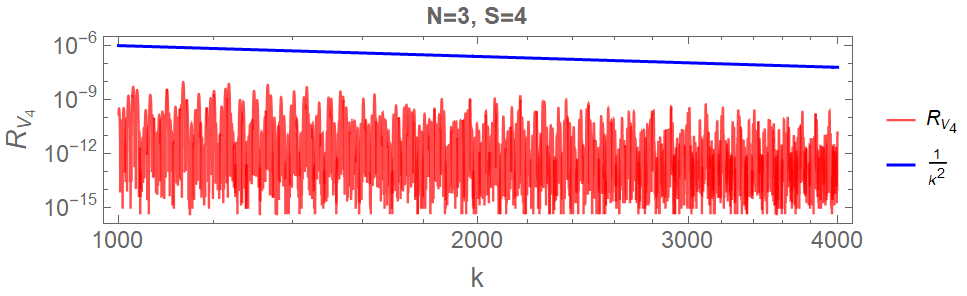} b\\
\includegraphics[scale=0.38]{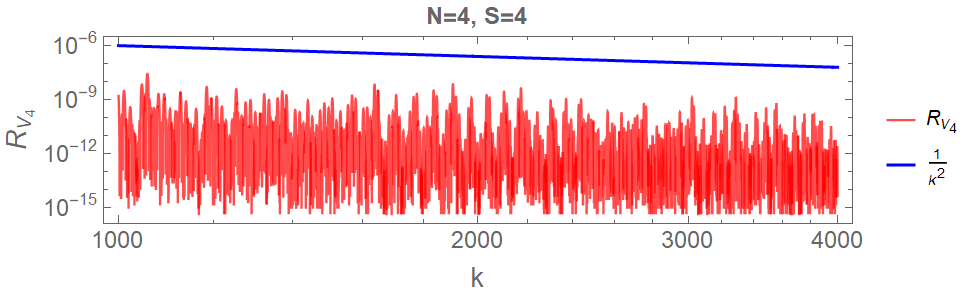} c\\
\includegraphics[scale=0.38]{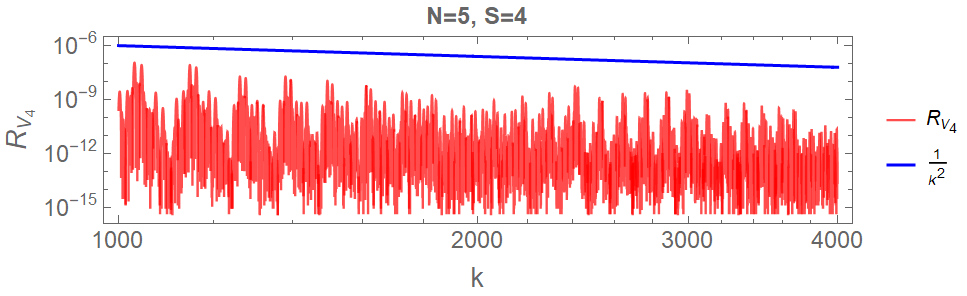} d\\
\includegraphics[scale=0.38]{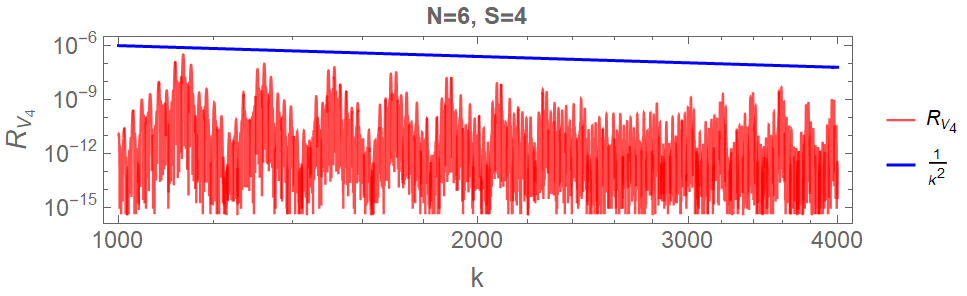} e
\caption{\it log-log plot representation of the scaling behavior of the reflection coefficient \( R_{V_{S=4}} \) is depicted, which falls off according to \( \frac{1}{k^{2}} \) (blue line). These plots are generated for a fixed stage \( S = 4 \) but with varying values of \( N \) as mentioned at the top of each figure. Notably, as the value of \( N \) increases, a change in the plot pattern is observed (lack of saturation), while maintaining the scaling behavior, governed by \( \frac{1}{k^{2}} \), consistently persists. The potential parameters employed in these plots are \( L = 1 \), \( V_{0} = 10 \), \( \rho = 3.5 \), \( \mu = 0.5 \), and \( \nu = 1.5 \).
}
\label{figure13}
\end{center}
\end{figure}

For large $k$, the reflection coefficient $R_{S}(k, N) \equiv R(k)$ is very small. With this limit, $R(k)$ can be approximated by using Eq. (\ref{T_s}) as
\begin{equation}
    R_{S}(k, N) \sim \vert  \mathbb{M}_{12} \vert ^{2} \left[\prod_{q=1} ^{S}U_{N-1}(\Gamma_{q})\right]^{2}.
    \label{r_small_value}
\end{equation}
Again for larger $k$ we have $\frac{V}{k^2} < < 1$ and upon Taylor expanding, it can be shown in the first order that
\begin{equation}
\vert \mathbb{M}_{12} \vert ^{2} \sim \left ( \frac{V b_{S}}{2k}\right )^{2}. 
\label{m12_small_value}
\end{equation}
Therefore, the expression for $R_{S}(k, N)$ becomes
\begin{equation}
    R_{S}(k, N) \sim \left ( \frac{V b_{S}}{2k}\right )^{2} \left[\prod_{q=1} ^{S}U_{N-1}(\Gamma_{q})\right]^{2}.
    \label{rr_small_value}
\end{equation}
If $V_{S}$ is the height of the potential at each stage $S$, then it can be shown that the following value of $V_{S}$ keeps the total area of the potential barrier (sum of the area of all potential segment at stage $S$) as constant as $L\times V_{0}$,
\begin{equation}
    V_{S} =\frac{L}{N^{S} b_{S}}V_{0}.
    \label{vg_value}
\end{equation}
Substituting the value of $b_{S}$ from Eq. (\ref{bsN}) in the above equation, we get
\begin{equation}
    V_{S} =\frac{V_{0}}{\prod_{j=1} ^{S}\left(1-\frac{N-1}{\rho^{\mu+\nu j}}\right)},
    \label{vg_values}
\end{equation}
where $V_{0}$ is the height of the potential barrier at $S=0$ and $b_{S}$ is the width of the unit cell rectangular barrier defined already (Eq. (\ref{bsN})). So, if $R_{V_{S}}(k, N)$ is the reflection coefficient at each stage $S$ with a potential height $V_{S}$, then it can be shown that (valid for large $k$)
\begin{equation}
    R_{V_{S}}(k, N) \sim \frac{1}{k^{2}}\left(\frac{V_{0} L}{2N^{S}}\right )^{2}\prod_{q=1} ^{S}U^{2}_{N-1}(\Gamma_{q}).
    \label{rvskngamma}
\end{equation}
Expressing the Chebyshev polynomial of second kind in terms of sinusoidal function
\begin{equation}
    U_{N}(\Gamma)=\frac{\sin{(N+1)}\gamma}{\sin{\gamma}},
\end{equation}
where $\gamma=\cos^{-1}\Gamma$, the Eq. (\ref{rvskngamma}) can be expressed as
\begin{equation}
    R_{V_{S}}(k, N) \sim \frac{1}{k^{2}}\left(\frac{V_{0} L}{2N^{S}}\right )^{2}\prod_{q=1} ^{S}\frac{\sin^{2}{(N)}\gamma_{q}}{\sin^{2}{\gamma_{q}}}=\frac{1}{k^{2}}\left(\frac{V_{0} L}{2N^{S}}\right )^{2}W^{(S)}(k, N),
    \label{rvskngammasine}
\end{equation}
where, $\gamma_{q}$ is defined as
\begin{equation}
\gamma_{q}=\cos^{-1}\Gamma_{q}.
\end{equation} The function $W^{(S)}(k, N)$, expressed as
\begin{equation}
    W^{(S)}(k, N)=\prod_{q=1} ^{S}\frac{\sin^{2}{(N)}\gamma_{q}}{\sin^{2}{\gamma_{q}}},
\end{equation}
which serves as a scaling function, is expressed as a finite product ($S$ times) of the Laue function, defined as
\begin{equation}
    L(x) = \frac{\sin^{2}(Nx)}{\sin^{2}(x)}.
\end{equation}
This function is commonly utilized in X-ray diffraction analyses. It has been documented in the literature \cite{ogawana2018transmission} that due to the multifractal properties of $W^{(S)}(k, N)$, at a sufficiently large value of $k$, the transmittance through the generalized Cantor-set also exhibits multifractal behavior. In the future, we would like to investigate the multifractality feature for our UCP case focusing on the transformation of this property as the system transitions from a fractal state to a non-fractal state ($\nu=0 \rightarrow \nu \neq 0$). Next for the case of UCP-$\rho_{2}$ system ($N=2$ case), Eq. (\ref{rvskngamma}) is modified as
\begin{equation}
    R_{V_{S}}(k, 2) \sim \frac{1}{k^{2}}\left(\frac{V_{0} L}{2^{S+1}}\right )^{2}\prod_{q=1} ^{S}U^{2}_{1}(\Gamma_{q}).
\end{equation}
Here $U_{1}(\Gamma_{q})$ represents the Chebyshev polynomial of the second kind, specifically for the first order, where $U_{1}(u)=2u$. Given this property, the above expression simplifies as
\begin{equation}
    R_{V_{S}}(k, 2) \sim \frac{1}{k^{2}}\left(\frac{V_{0} L}{2^{S+1}}\right )^{2}\prod_{q=1} ^{S}\Gamma_{q}^{2}.
\end{equation}
It is evident from Eq. (\ref{rvskngamma}) that $R_{V_{S}}(k, N)$ for UCP-$\rho_{N}$ system would scale as $\frac{1}{k^{2}}$. This is a proven result for Cantor potential in standard quantum mechanics (QM) and has been substantiated in the literature \cite{sakaguchi2017scaling}. Extending this understanding, our research has demonstrated similar scaling behaviors in other quantum potentials. We have reported this scaling phenomenon for the SVC-$\rho$ potential \cite{narayan2023tunneling} and the UCP-$\rho_{2}$ potential \cite{umar2023quantum}, thus reinforcing the universality of this scaling law across various quantum systems and now it is evident from Eq. (\ref{rvskngamma}) that this scaling feature also holds for the UCP-$\rho_{N}$ systems.  Moreover, based on the same concept, the domain of SFQM also exhibits the scaling behavior for the reflection coefficient in GC and SVC-$\rho$ potentials. In this framework, the reflection coefficient scales as $\frac{1}{k^{4\left(\frac{\alpha-1}{\alpha}\right)}}$ \cite{singh2023quantum}
 where \( \alpha \) is the Levy index, which ranges from \( 1 < \alpha \le 2 \). This finding highlights the influence of the fractional quantum mechanics framework on the scaling properties of the quantum system.\\
\indent
The objective of crafted Fig. \ref{figure13} is to elucidate the scaling behavior of the reflection coefficient \( R_{V_{S=4}} \) via a log-log plot representation. This plot demonstrates the behavior of the reflection coefficient, adhering to a scaling law of \( \frac{1}{k^{2}} \), which is represented by a blue line. The plots are consistently set for a fixed stage \( S = 4 \), yet they incorporate variations in the value of \( N \), as indicated at the top of each respective figure. An observation from these plots is the alteration in the pattern as \( N \) increases, particularly the emergence of a lack of saturation in the higher \( N \) values. Despite these pattern variations, the scaling
\begin{figure}[H]
\begin{center}
\includegraphics[scale=0.38]{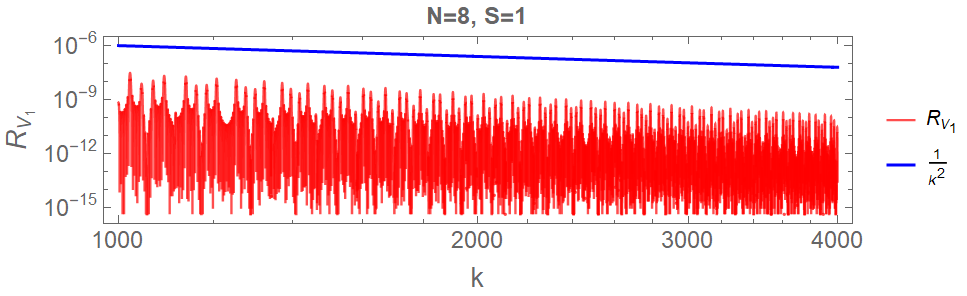} a\\
\includegraphics[scale=0.38]{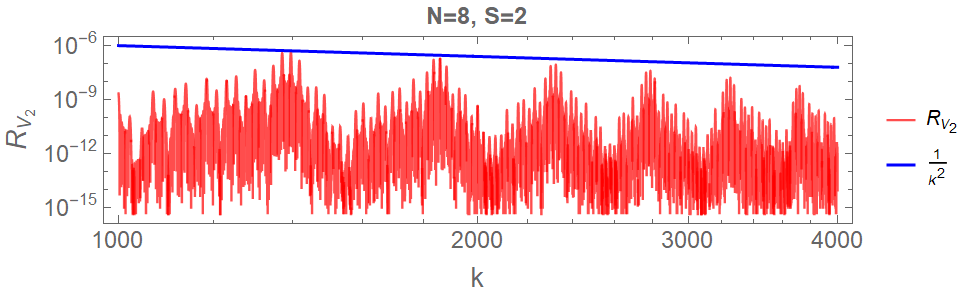} b\\
\includegraphics[scale=0.38]{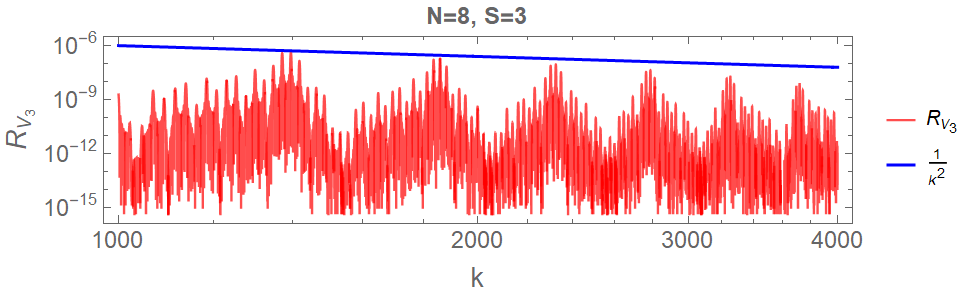} c\\
\includegraphics[scale=0.38]{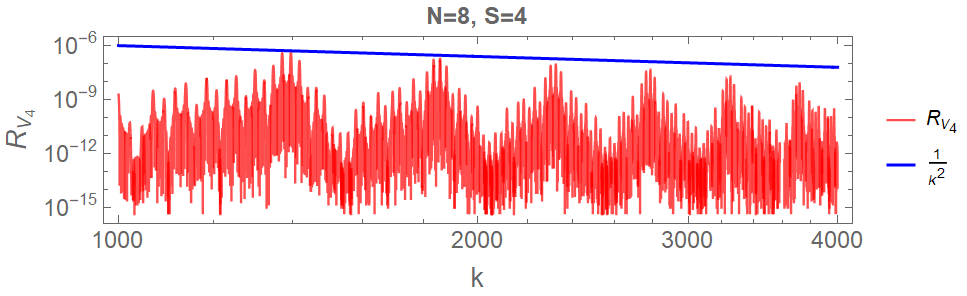} d\\
\includegraphics[scale=0.38]{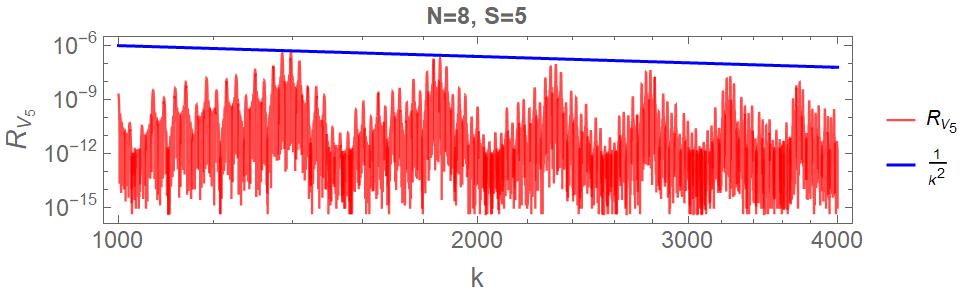} e
\caption{\it In the provided log-log plots, the scaling behavior of the reflection coefficient \( R_{V_{S}} \) as $\frac{1}{k^{1}}$ is depicted, These plots are generated for a fixed $N=8$ and different stages $S$ as mentioned at the top of each figure. These plots reveal an intriguing aspect of the reflection coefficient's behavior: a pronounced similarity (appearance of saturation of the reflection coefficient) in the patterns emerges for the stages \( S = 2, 3, 4 \) and \( 5 \) by maintaining its scaling behavior. The potential parameters employed in these plots are \( L = 1 \), \( V_{0} = 10 \), \( \rho = 3.5 \), \( \mu = 0.5 \), and \( \nu = 1.5 \).
}
\label{figure14}
\end{center}
\end{figure}

\begin{figure}[H]
\includegraphics[width=0.47\textwidth,height=0.18\textheight]{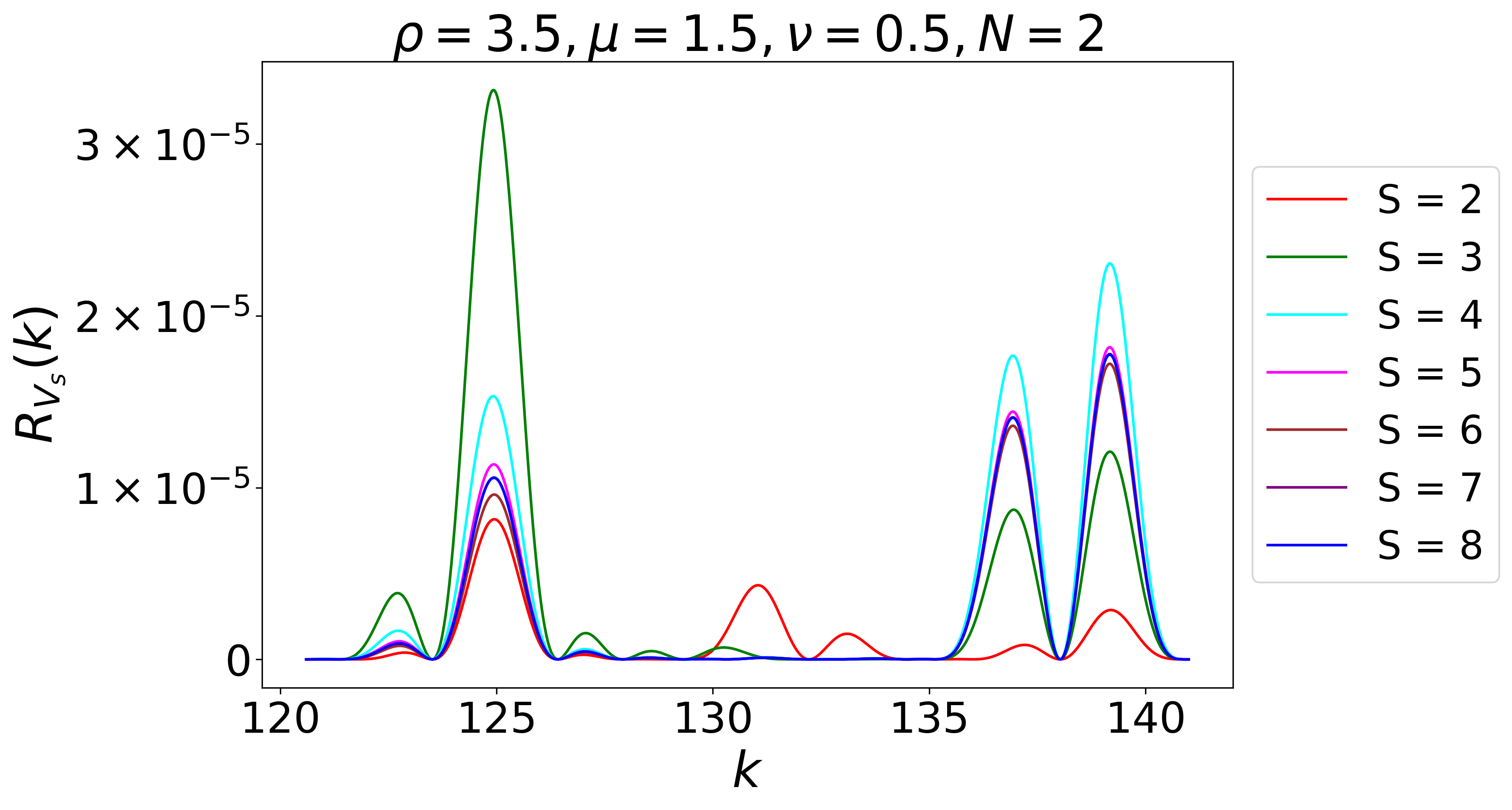} a
\includegraphics[width=0.47\textwidth,height=0.18\textheight]{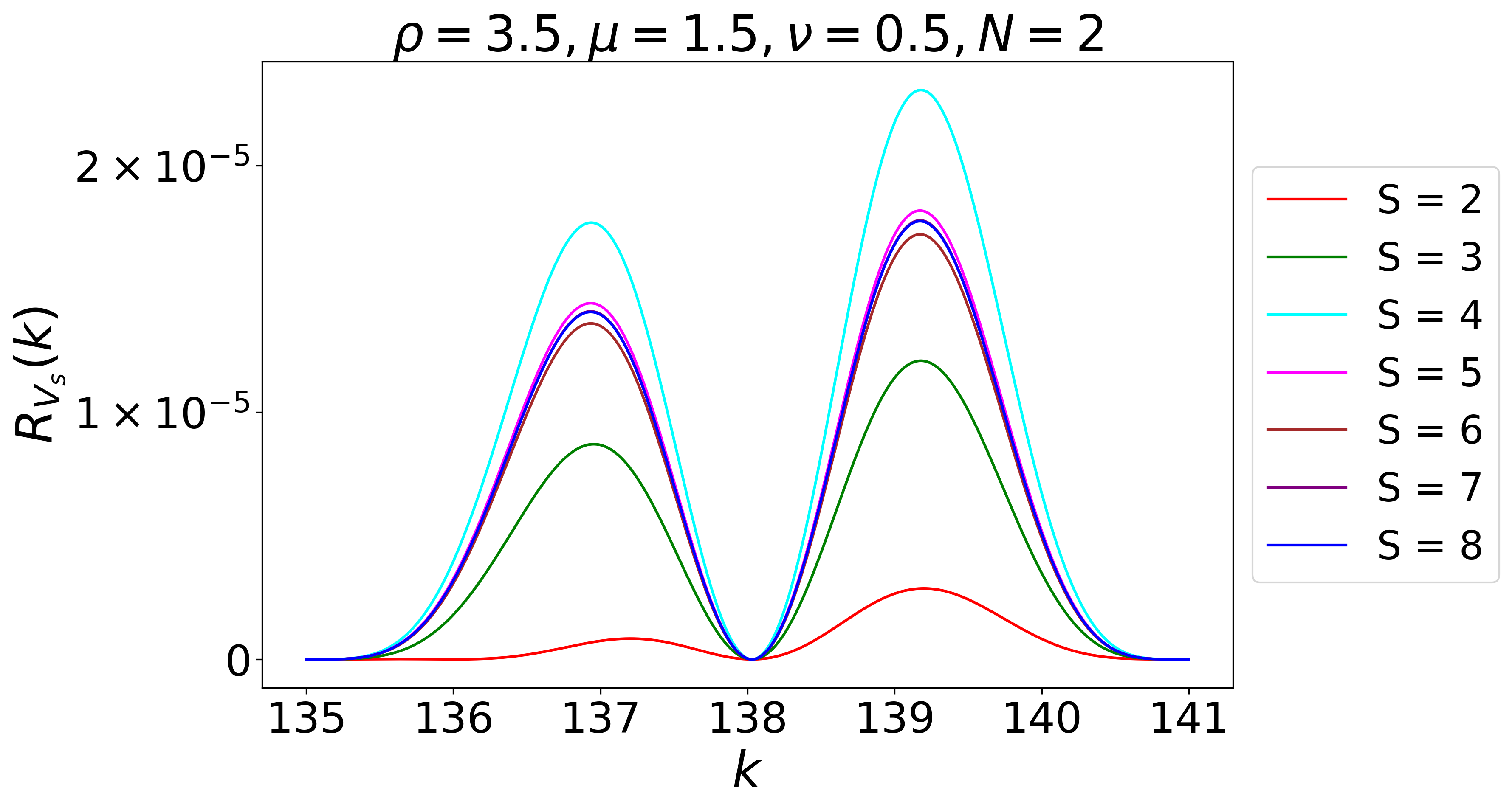} b\\
\includegraphics[width=0.47\textwidth,height=0.18\textheight]{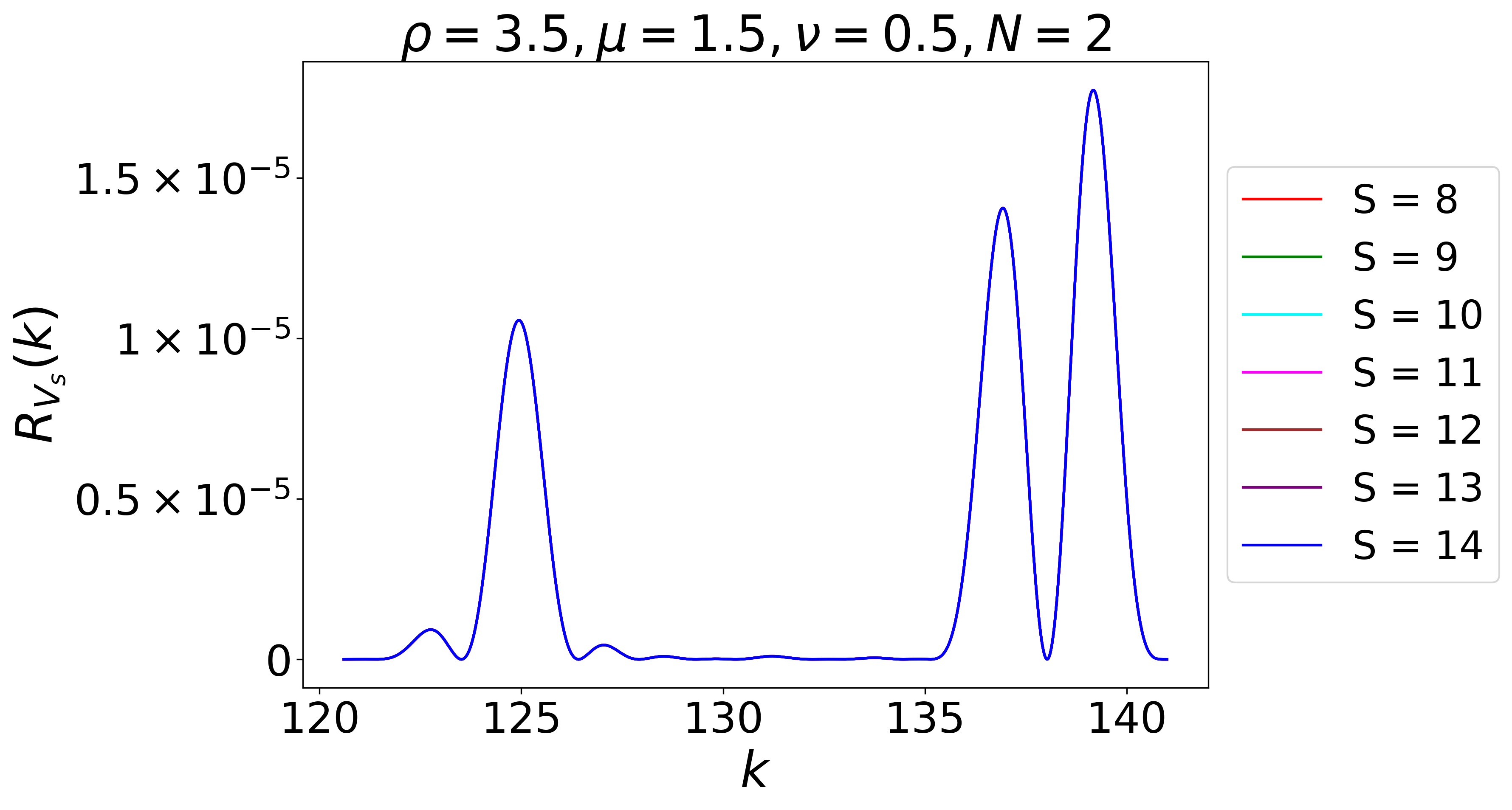} c
\includegraphics[width=0.47\textwidth,height=0.18\textheight]{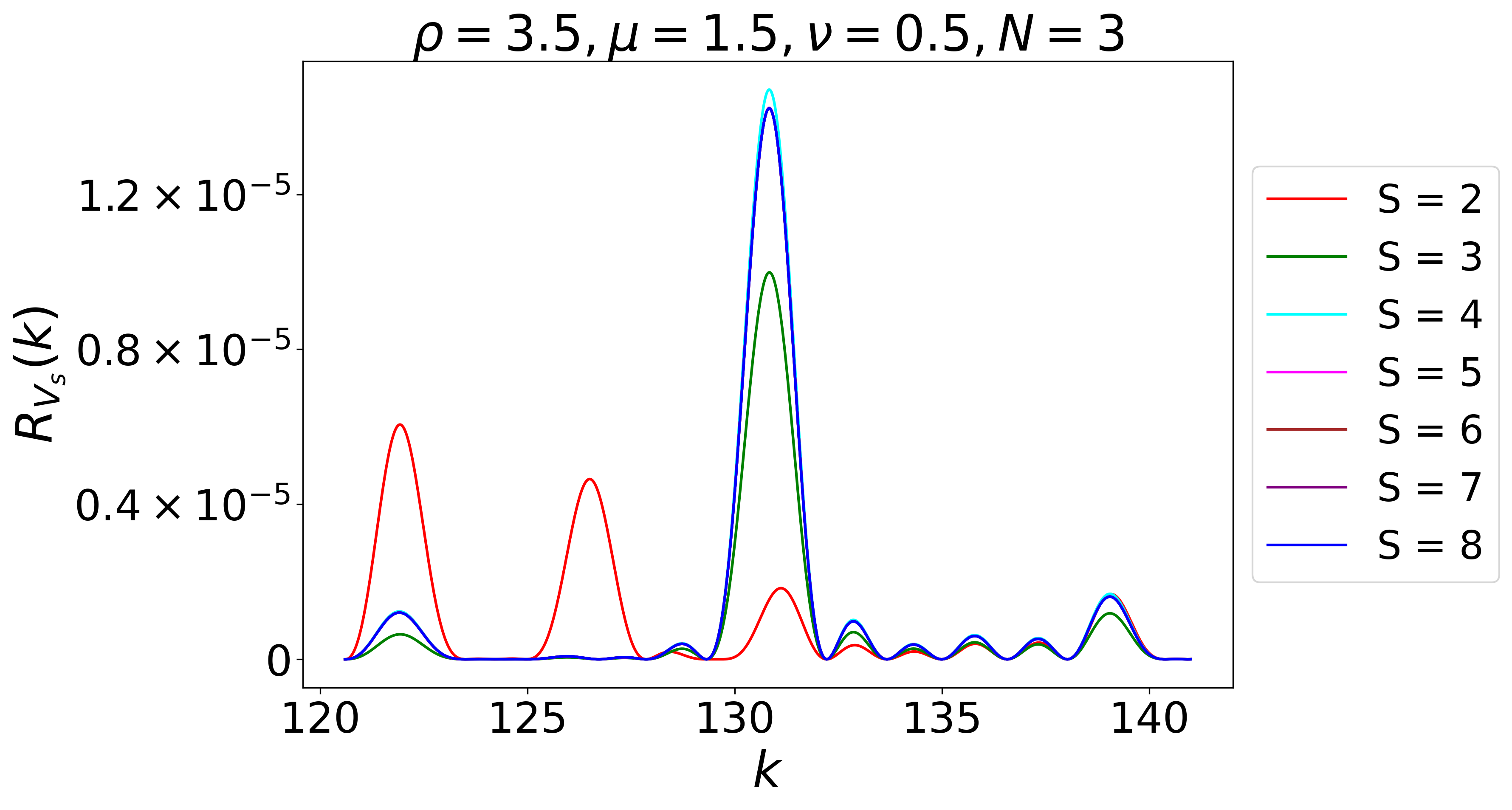} d\\
\includegraphics[width=0.47\textwidth,height=0.18\textheight]{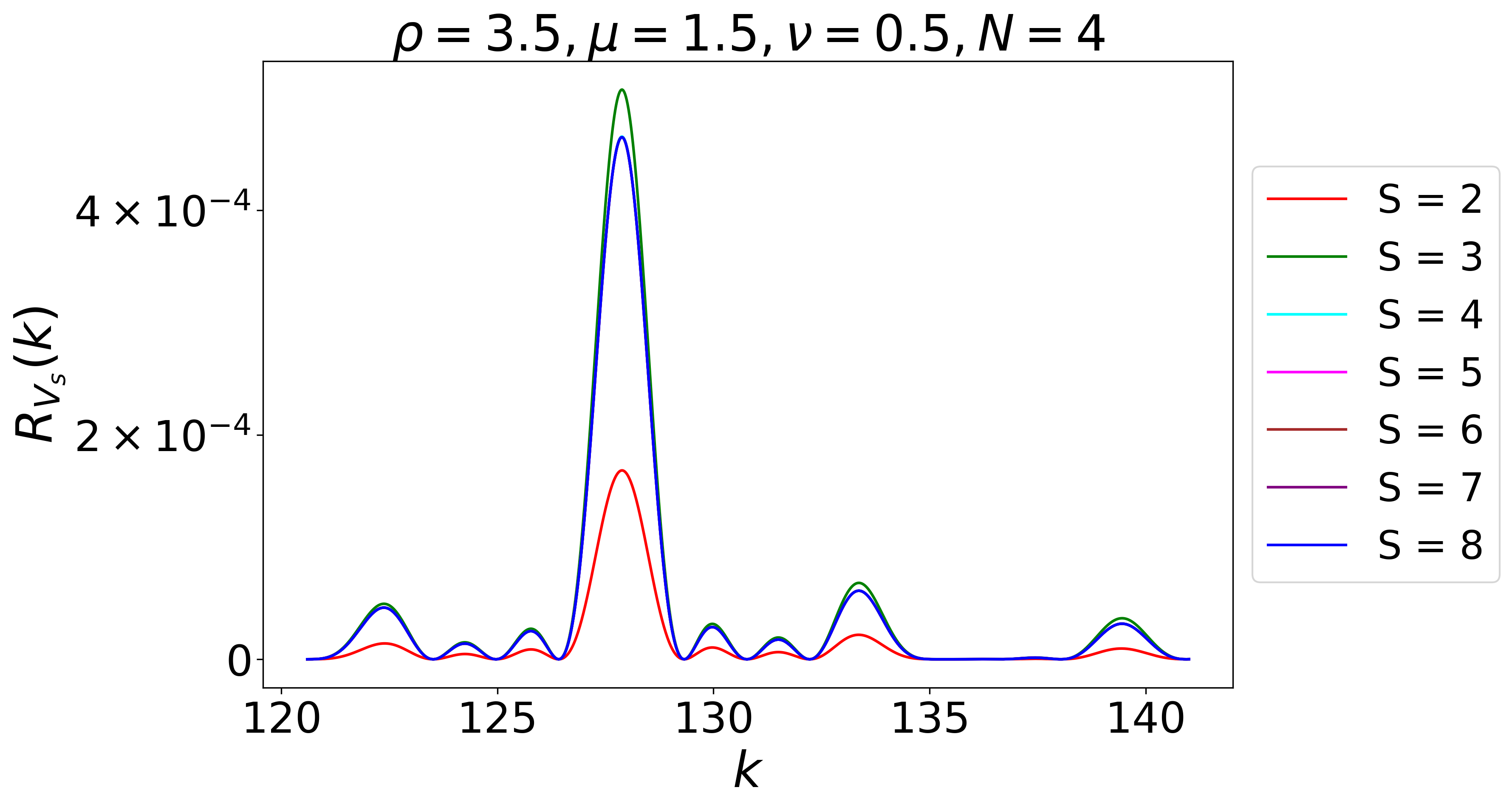} e 	\includegraphics[width=0.47\textwidth,height=0.18\textheight]{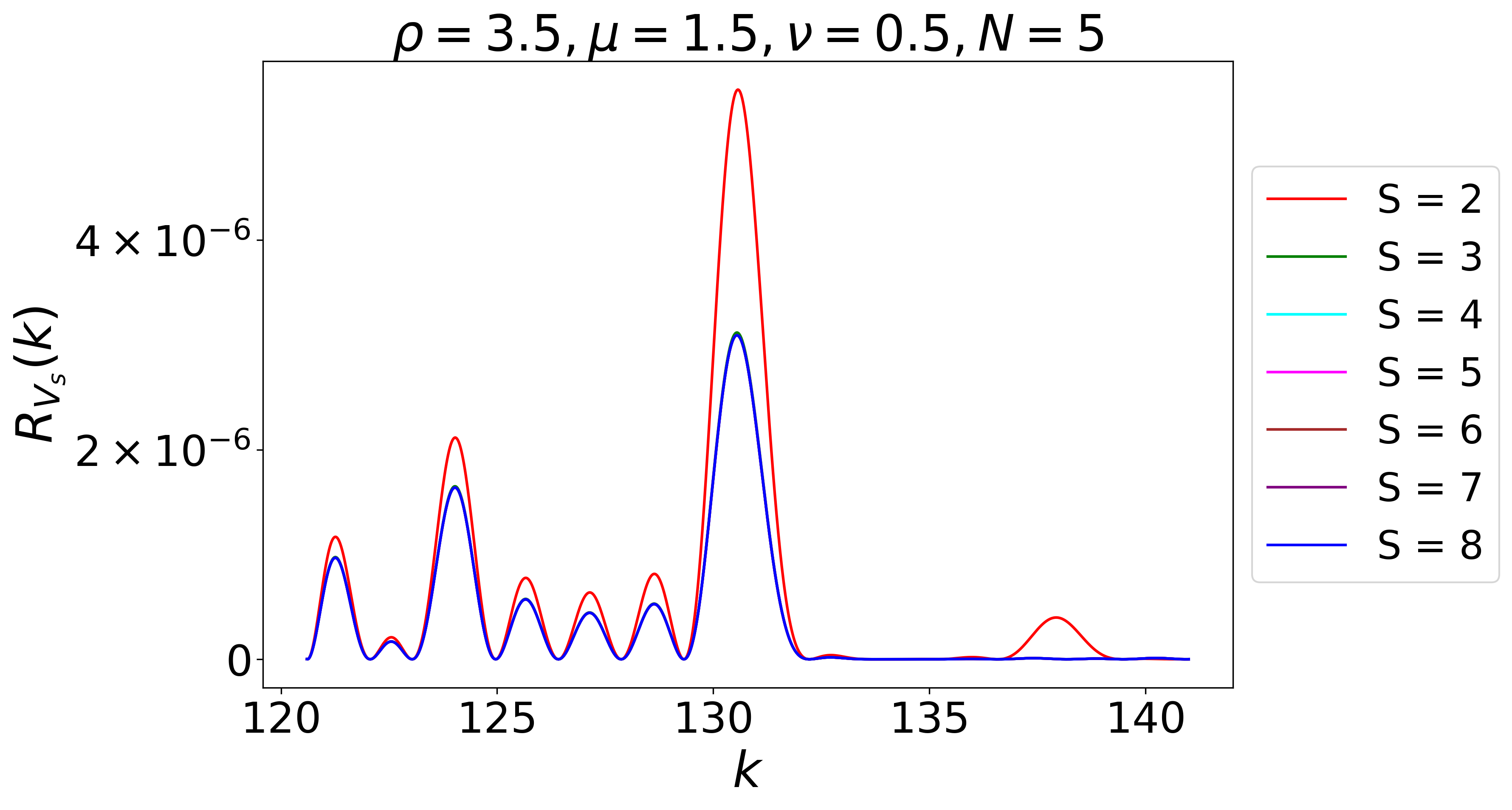} f\\
\includegraphics[width=0.47\textwidth,height=0.18\textheight]{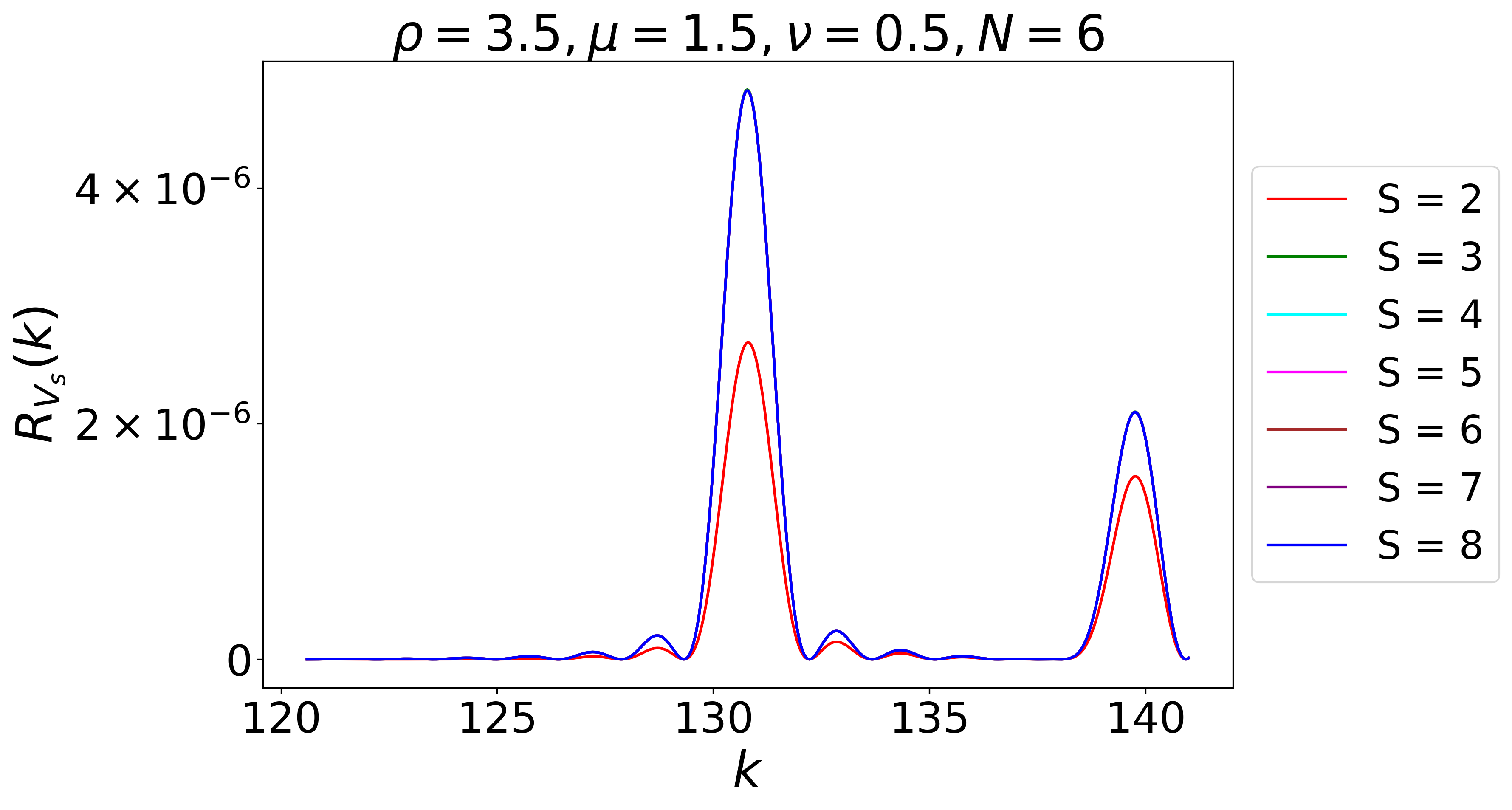} g	\includegraphics[width=0.47\textwidth,height=0.18\textheight]{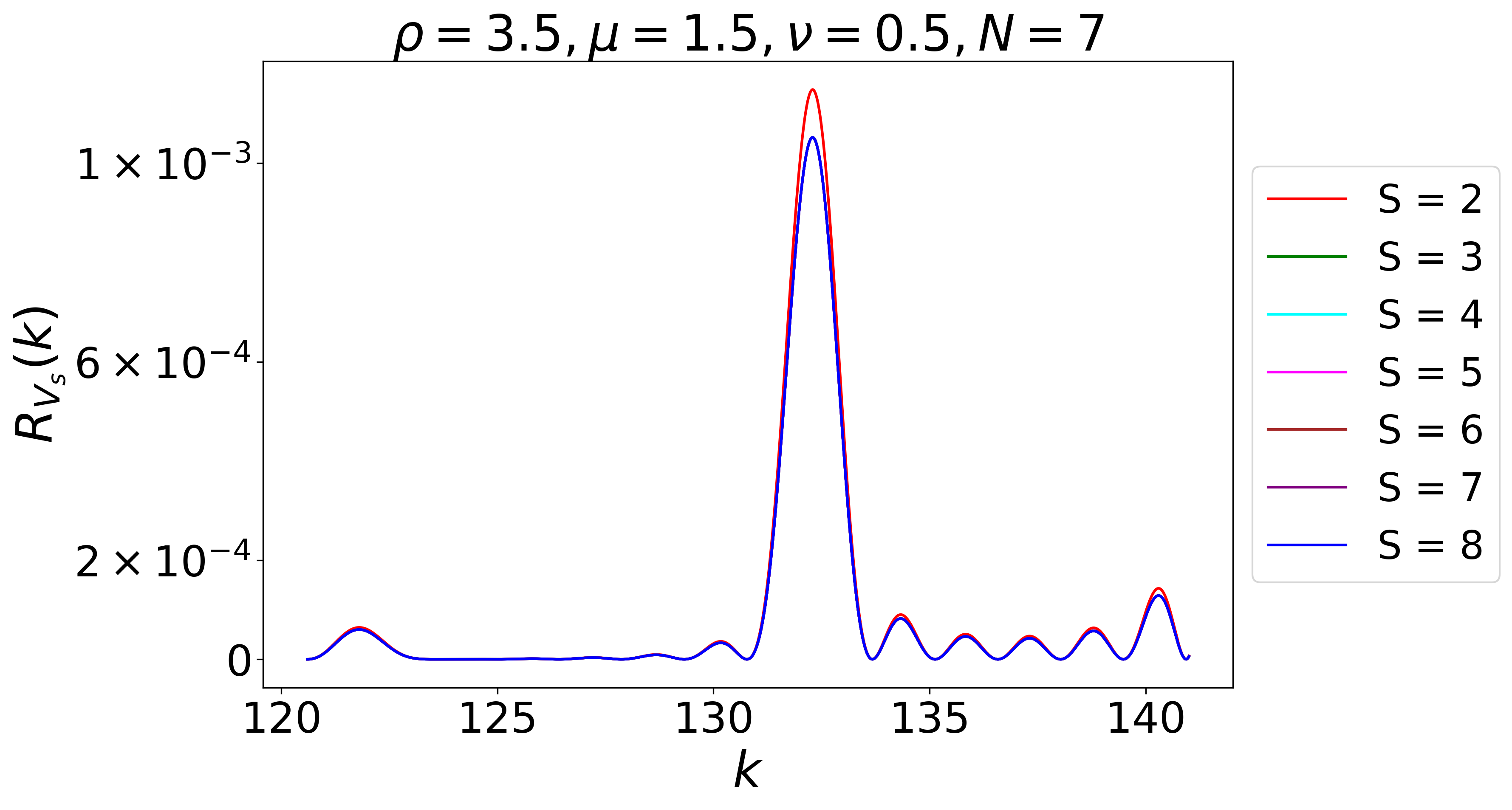} h
\caption{\it 
The plots present graphical representations to elucidate the behavior of the reflection coefficient, $R_{V_{S}}(k)$, with $k$ across various stages $S$, specifically within the high $k$ domain. Each graph corresponds to a parameter $N$, with the value of $N$ displayed atop each figure alongside other relevant parameters. The first three figures (a), (b) and (c) are dedicated to a system where $N=2$. Specifically, figure (a) delineates the variation of $R_{V_{S}}(k)$ with $k$ for stages ranging from $S=1$ to $S=8$, while figure (c) captures the variation of $R_{V_{S}}(k)$ with $k$ for a higher stage range, from $S=8$ to $S=14$. Notably, the latter scenario exhibits a convergence (complete overlap) of reflection profiles across stages, contrasting sharply with the former, where distinct reflection profiles for each stage are observable. Figure (b) provides a magnification of figure (a), focusing on the interval where $k$ spans from $135$ to $141$. Additionally, it is observed that with an increase in the value of $N$, a convergence of the reflection profiles occurs at earlier stages.}
\label{figure_convergence}
\end{figure}
\newpage
\noindent
behavior, governed by \( \frac{1}{k^{2}} \), remains steadfast across different \( N \) values. The potential parameters employed in these plots are \( L = 1 \), \( V_{0} = 10 \), \( \rho = 3.5 \), \( \mu = 0.5 \), and \( \nu = 1.5 \). The purpose of Fig. \ref{figure14} (with the same potential parameters) is also to show the scaling behavior but it is crafted for a fixed value of $N=8$ while varying stage $S$, each stage clearly labeled at the top of its corresponding figure. A salient feature of these plots is the similarity in the pattern across the stages \( S = 2, 3, 4 \) and \( 5 \) showing the appearance of saturation in the reflection profile. Notably, this saturation does not deviate from established $\frac{1}{k^{2}}$ law.\\
\indent
Fig. \ref{figure_convergence} offers a comprehensive study of graphical analyses aimed at delineating the intricate behavior of the reflection coefficient, denoted as $R_{V_{S}}(k)$, as a function of wave number $k$ across a spectrum of stages, $S$, with a particular focus on the higher $k$ regime. The notation $V_{S}$ signifies that with each progressive stage, the height of the potential barrier is modulated in such a manner that the cumulative area of all potential segments at any given stage remains invariant, precisely quantified as $L\times V_{0}$. Each graphical representation is associated with a distinct parameter $N$, with the respective value of $N$ and other pertinent parameters displayed on the top of each figure with $L=2$ and $V_{0}=10$. The subsets of Fig. \ref{figure_convergence}, specifically Figs. \ref{figure_convergence}a, \ref{figure_convergence}b, and \ref{figure_convergence}c, are allocated to elucidate the system characterized by $N=2$. Fig. \ref{figure_convergence}a expounds on the variation of $R_{V_{S}}(k)$ as a function of $k$ across a range of stages from $S=1$ to $S=8$, revealing distinct reflection profiles for each stage within this interval. Conversely, Fig. \ref{figure_convergence}c extends this analysis to a higher range of stages, from $S=8$ to $S=14$, wherein a pronounced convergence, or a complete overlap, of reflection profiles across these stages is observed, marking a stark contrast to the divergent profiles elucidated in Fig. \ref{figure_convergence}a. Fig. \ref{figure_convergence}b serves to magnify the interval within Fig. \ref{figure_convergence}a, specifically targeting the $k$ span of $135$ to $141$, providing an enhanced view of the reflection profiles within this narrow range. For a system characterized by $N=3$ depicted in Fig. \ref{figure_convergence}d, the reflection profile remains distinctive up to stage $S=4$. Post this stage, a merging of reflection profiles for subsequent stages is evident, indicating a convergence trend. Progressing to $N=5$, $N=6$ and $N=7$, as showcased in Fig. \ref{figure_convergence}f, \ref{figure_convergence}g and \ref{figure_convergence}h respectively, a distinct reflection profile is discernible exclusively for stage $S=2$, with stages extending from $S=3$ to $S=8$ demonstrating a convergence of reflection profiles. This progression highlights a notable trend where an increment in the value of $N$ precipitates the convergence of reflection coefficient at earlier stages, a phenomenon meticulously captured through the graphical representations in Fig. \ref{figure_convergence}.

%%%%%%%%%%%%%%%%%%%%%%%%%%%%%%%%%%%%%%%%%%%%%%%%%%%%%%%%%%%%%%%%%%%%%%%%%%%%%%

\section{Conclusion and Discussions}
\label{conclusion}
In the present investigation, the concept of the UCP has been broadened to introduce the GUCP or UCP-$\rho_{N}$ framework. This advancement is distinguished by the introduction of a critical parameter, \(N\), which denotes the count of potential segments at the inaugural stage ($S=1$). Additionally, it is elucidated that the PCP, characterized by its minimal lacunarity, constitutes a specific incarnation within the UCP-$\rho_{N}$ spectrum. The UCP-$\rho_{N}$ paradigm not only augments the repertoire of Cantor-derived potentials but also facilitates a methodical pathway to navigate the continuum between fractal $(\nu=0)$ and non-fractal $(\nu \neq 0)$ domains. This framework thereby significantly contributes to a more nuanced comprehension of the quantum mechanical properties inherent in these complex systems. By incorporating this systematic modulation between fractal and non-fractal characteristics, the UCP-$\rho_{N}$ system offers a versatile tool for exploring the quantum mechanical behavior across a diverse array of Cantor-type potentials. This system facilitates a deeper exploration of quantum behavior within these complex potential landscapes, advancing our comprehension of how fractal structures can influence quantum mechanics.\\
\indent
The fundamental basis of our investigation lies in the application of the SPP framework, a comprehensive extension of the concept of locally periodic potentials. This approach has been instrumental in enabling the derivation of a closed-form analytical expression for the transmission coefficient $T_{S}(k, N)$ utilizing the $q$-Pochhammer symbol for its representation. A critical and noteworthy discovery within our investigation is the identification of exceptionally sharp transmission resonances manifested within the UCP-$\rho_{N}$ potential framework. These resonances are particularly significant, as they underscore the aptitude of potential for the development of quantum filters characterized by exceedingly narrow bandwidths, demonstrating precision in control and effectiveness in selective filtering.\\
\indent
This investigation has demonstrated the phenomenon of transmission profile overlap, or saturation, across varying stages within the UCP-$\rho_{N}$ system. Employing a structural protocol, the system iteratively excises a specific fraction, \(\frac{1}{\rho^{\mu+\nu S}}\), from \(N-1\) symmetrically distributed points at each stage \(S\). This methodical process illuminates that with certain parameters, \(\rho\), \(\mu\), \(\nu\), and \(S\), settings especially when the expression \(\rho^{\mu+\nu S}\) assumes larger values, the resultant excised segments significantly diminish relative to the original segment dimensions. Consequently, this selective removal process induces subtle structural evolutions within the potential landscape. Such minimal structural shifts, particularly with the excision of exceedingly fine segments, lead to a saturation effect in the transmission profile of the potential system. Various strategies can be employed to increase the magnitude of \(\rho^{\mu+\nu S}\). Specifically, the analysis extends to the transmission profile saturation within the UCP-\(\rho_{3}\) and UCP-\(\rho_{4}\) frameworks, as well as the PCP system (\(\nu=0\)) for \(N=4\). In the UCP-\(\rho_{3}\) and UCP-\(\rho_{4}\) architecture, by holding \(\rho\) and \(\mu\) constant and varying \(\nu\), an exponential increase in the value of \(\rho^{\mu+\nu S}\) is observed as \(\nu\) progresses alongside stage \(S\) leading to the saturation in transmission spectrum. Furthermore, in the polyadic Cantor system (\(\nu=0\)), the absence of \(\nu\) and hence stage \(S\) in the segment removal implies a consistent fraction of \(\frac{1}{\rho^{\mu}}\) is removed at each stage $S$, thereby only \(\rho\) and \(\mu\) parameters participated in the potential segment removal exercise. This has resulted in the observation of transmission profile for a system characterized by $N=4$ with a fixed \(\rho\) and high magnitude of \(\mu\).\\
\indent
In the continued pursuit of our research, we have directed our focus towards the analysis of the reflection coefficient \(R_{V_{S}}(k, N)\) within a framework where the cumulative potential barrier area is preserved at a constant value of \(L \times V_{0}\) at each stage $S$. Here, \(L\) delineates the comprehensive extent of the system, ensuring that each progressive stage of the UCP-\(\rho_{N}\) system is encapsulated within this specified length, while \(V_{0}\) signifies the initial potential height at stage \(S=0\). Our investigations have yielded significant evidence suggesting that \(R_{V_{S}}(k, N)\) exhibits a scaling behavior proportional to \(\frac{1}{k^{2}}\) in the regime of large \(k\) values. This relationship has been validated through both graphical representations and analytical methodologies. This particular scaling law, previously identified in diverse potential systems such as SVC-$\rho$ \cite{narayan2023tunneling},  SVC(\(\rho, n\)) \cite{singh2023quantum01} and specifically the UCP-\(\rho_{2}\) \cite{umar2023quantum} potential systems in QM, has now been extended to encompass the UCP-\(\rho_{N}\) potential framework in the current study.\\
\newpage
\indent
In summarizing, the innovative blending of fractal and non-fractal polyadic elements by the UCP-$\rho_{N}$  potential heralds novel avenues for both theoretical exploration and empirical inquiry. The potential invites further scholarly investigation into the intricate implications of its newly introduced parameters, signaling the emergence of a more expansive spectrum of potentials that may redefine our comprehension and application of quantum potentials derived from the systematic segmentation of real lines. An intriguing prospect for extending this research encompasses examining a wider variety of potentials, conceptualized by the strategic excision of a $\frac{1}{\rho^{f(S)}}$ fraction from the center of each segment in successive stages, where $f(S)$ is a stage $S$ dependent function. Envisioning $f(S)$ as a polynomial function, such as $f(S)=a_{0}+ a_{1}S+ a_{2}S^{2}+\ldots+ a_{n}S^{n}$, proposes a framework for synthesizing a diverse array of potential systems under a unified Cantor system. Specifically, setting $a_{0}=\mu$ and $a_{1}=\nu$ with the remaining coefficients at zero, delineates the UCP-${\rho_{N}}$ architecture. Moreover, exploring $f(S)$ as trigonometric functions like as $f(S)=\sin{S}$, $f(S)=\cos{S}$, $f(S)=\cos^{3}S$, exponential function as $f(S)=e^{S}$ etc could also be the future direction of work and could significantly broaden the Cantor (fractal and non-fractal) potential family, offering a rich domain for employing the SPP formalism to dissect transmission phenomena. In our recent investigations, we have delved into a more comprehensive formulation of the SVC potential, denoted as SVC($\rho, n$) \cite{singh2023quantum01}, where $n$ represents a real number. Within this framework, $n=0$ and $n=1$ are indicative of the GC potential and the SVC-$\rho$ potential, respectively. This model is characterized by the removal of a fraction $\frac{1}{\rho^{S^{n}}}$ from the center of the potential segment at the preceding stage, $S-1$. Notably, the parameter $n$ plays a pivotal role in modulating the transmission profile. Looking ahead, we aim to broaden the scope of this system to encompass the polyadic domain, introducing a novel formulation as UCP($\rho_{N}, n$), thereby expanding the potential applicability and theoretical underpinnings of this intriguing quantum mechanical model. \\

%%%%%%%%%%%%%%%%%%%%%%%%%%%%%%%%%%%%%%%%%%%%%%%%%%%%%%%%%%%%%%%%%%%

\noindent
{\it \bf{Acknowledgements}}: MU extends gratitude towards the Optics and Photonics Centre (OPC), Indian Institute of Technology (IIT) Delhi for fostering research activities through their supportive environment and encouragement. 
MH is grateful for the support from SPO-ISRO HQ, which encouraged research activities. VNS acknowledges the assistance from the BHU Physics Department and BPM, who provided invaluable support. BPM, in turn, appreciates the support received through the research grant under the IoE scheme (Number - 6031) from BHU, funded by the UGC Government of India.

%%%%%%%%%%%%%%%%%%%%%%%%%%%%%%%%%%%%%%%%%%%%%%%%%%%%%%%%%%%%%%%%%%%%

\newpage
\bibliographystyle{elsarticle-num}
\bibliography{Ref}

\end{document}